%% file: happrox-paper.tex
\newif\ifdraft\draftfalse
\newif\iffinal\finaltrue
\newif\iflong\longfalse
\documentclass[twocolumn,pdftex]{svjour3}
\pdfoutput=1

\titlerunning{
Safe collision avoidance
for horizontal turns}
\authorrunning{Y. Kouskoulas, et al.}

\title{
Envelopes and Waves: Safe Multivehicle Collision Avoidance for Horizontal Non-deterministic Turns

\thanks{This research was partially funded under the sponsorship of the Federal Aviation Administration Traffic Alert \& Collision Avoidance System (TCAS) Program Office (PO) AJM-42 under contract number DTFAWA-11-C-00074 as well as internal funds from the Johns Hopkins University Applied Physics Laboratory.}}
\author{Yanni Kouskoulas\and
T. J. Machado \and
Daniel Genin \and
Aurora Schmidt \and
Ivan Papusha \and
Joshua Brul\'{e} 
}

\institute{Y. Kouskoulas \at
No Affiliation \\
\email{yxkous@gmail.com}
\and T. J. Machado \at
New Mexico State University \\
\email{tjm@nmsu.edu}
\and D. Genin \and A. Schmidt \and I. Papusha \and J. Brul\'{e}
\at Johns Hopkins University Applied Physics Laboratory\\
11100 Johns Hopkins Rd, Laurel, MD 20723, USA\\
\email{daniel.genin@jhuapl.edu}
}

\input{preamble.tex}

\input{macros.tex}

\begin{document}

\maketitle

\begin{abstract}
We present an approach to analyzing the safety of asynchronous, independent, non-deterministic, turn-to-bearing horizontal maneuvers for two vehicles. Future turn rates, final bearings, and continuously varying ground speeds throughout the encounter are unknown but restricted to known ranges. We develop a library of formal proofs about turning kinematics, and apply the library to create a formally verified timing computation. Additionally, we create a technique that evaluates future collision possibilities that is based on waves of position possibilities and relies on the timing computation. The result either determines that the encounter will be collision-free, or computes a safe overapproximation for when and where collisions may occur.
\end{abstract}

\section{Introduction} 
\input{horiz-introduction}

\section{Reasoning Foundations for Turn-to-bearing Maneuvers}

\input{horiz-proof}

\section{Reasoning about the Timing of Intersecting Turns}
\input{horiz-analysis_of_horizontal_motion}

\section{Quantifying collision timing computations over the conflict area}
\input{happrox-introduction}

\input{happrox-approach}

\section{Sound Approximation of Collision Timing Equations}
\input{happrox-approx}

\input{horiz-conclusion}

\newcommand{\doi}[1]{}
\bibliographystyle{spmpsci}
\bibliography{horiz-paper}

\end{document}

%% file: preamble.tex
\usepackage{amsmath} 

\usepackage{mathtools}
\usepackage{amssymb} 
\usepackage{graphicx} 
\usepackage[labelfont=bf]{caption}
\usepackage{subcaption}
\usepackage{epstopdf}
\usepackage{hyperref}

\usepackage{fancyvrb}
\usepackage{tabularx}
\usepackage{color}
\usepackage{xspace}
\usepackage{mdwlist}

\usepackage{cuted} 

\usepackage{rotating}
\usepackage{datetime}

\usepackage{tikz}
\usetikzlibrary{arrows}

\usepackage{multirow}

\usepackage{pgf,tikz}
\usetikzlibrary{arrows}

\usepackage[group-separator={,}]{siunitx}


%% file: macros.tex
\DeclareMathOperator{\pathsegment}{path\_segment}
\DeclareMathOperator{\straight}{straight}
\DeclareMathOperator{\turning}{turning}

\DeclareMathOperator{\atan}{atan}
\DeclareMathOperator{\sign}{sign}
\DeclareMathOperator{\sinc}{sinc}

\newcommand{\ra}{\ensuremath{r_{\alpha}}}
\newcommand{\rb}{\ensuremath{r_{\beta}}}
\newcommand{\tha}{\ensuremath{\theta_{\alpha}}}
\newcommand{\thb}{\ensuremath{\theta_{\beta}}}
\newcommand{\sa}{\ensuremath{s_{\alpha}}}
\renewcommand{\sb}{\ensuremath{s_{\beta}}}

\newcommand{\KeYmaera}{Ke\kern-0.1emYmaera\xspace}
\newcommand{\KeYmaeraX}{\KeYmaera~X\xspace}

\newcommand{\nondeterministic}{non-de\-ter\-min\-is\-tic}
\newcommand{\turntobearing}{turn-to-bear\-ing}
\newcommand{\twoargument}{two-ar\-gu\-ment}



%% file: horiz-introduction.tex
Autonomous and semi-autonomous systems that control ground vehicles,
boats, and aircraft all need to reason about horizontal turns in order
to create plans for future motion that meet system objectives.

We are specifically motivated by aircraft collision avoidance maneuvers that
combine vertical and horizontal advice to ensure multi-aircraft
encounters are safely separated. These maneuvers advise aircraft to
turn at the same time they change vertical velocity---the
objective being to keep the aircraft separated in altitude during
periods when their positions might coincide horizontally. This
requires correctly computing the time interval that describes when in
the future both aircraft might come into horizontal conflict.

This paper develops a formalization of \emph{\nondeterministic{}}
turn-to-bearing motion, where a vehicle turns following a circular
arc until reaching a certain bearing, and then follows a straight path
thereafter. Turn-to-bearing motion is the building block for
Dubins trajectories used in many different techniques in the
literature (see Section~\ref{section:literature_review}), but here we 
consider that the parameters that describe our future path are 
non-deterministic and uncertain at the beginning of the turn.

The formalization is embodied in a 
library of proofs that are detailed descriptions of these kinematics, 
and are machine-checked to guarantee correctness. Each theorem in this 
paper corresponds to a proof in the formalization.\footnote{Coq proofs
are at  \href
{https://bitbucket.org/ykouskoulas/ottb-foundation-proofs}
{https://bitbucket.org/ykouskoulas/ottb-foundation-proofs}.}
We believe that the library can serve as a foundation for formal
reasoning about horizontal turns in the Coq proof assistant,
supporting the development of insight and correct reasoning for a wide
variety of path planning and collision avoidance algorithms. Furthermore, we 
hope that it helps guarantee a high level of correctness and robustness for
robotic systems' horizontal motion, and that it provides the basis for
certification artifacts (i.e., proofs) that can be used to establish system
algorithm and software correctness.

We develop an approach to evaluating collision possibilities during
an encounter. We apply the library to develop and 
formally verify an exact pointwise timing computation, use it to 
create a sound approximation of the timing over an area, and quantify
the timing computation over the reachable area of future motion.

This paper is an extension of \cite{kouskoulas2020}, which presents the novel contributions of the original work, but also new ideas.
The contributions of the original work are: the development of a Coq library
for reasoning about \nondeterministic{} Dubins-style paths; an
additional Coq library defining a variety of \twoargument{} arctangent
functions with different branch cuts that are each sensitive to the
quadrant and sign of their arguments; a new expression for computing
the appropriate angle necessary for connecting Dubins paths to a
destination waypoint\footnote{There exist alternate expressions for
this angle, but to our knowledge, the formulation in this paper is
new.}; and formally verified expressions of the timing constraints of uncertain turn-to-bearing motion.
The contributions that are unique to this extension are: development of novel,
sound approximations for the location of a vehicle within the reachable area using waves as moving boundaries; and a simple, efficient, piecewise
approach to calculating the range of possible collision times between two vehicles
each of which followings turn-to-bearing kinematics. It also improves the presentation of conjectures from the original paper that enhance ease of comprehension for readers who want to use the equations in practical applications.

The rest of this paper is organized as follows: Section \ref{section:literature_review}
considers prior work in formalizing horizontal motion and analyzing potential collisions;
Section 3 describes the details of our library and how we formalized
non-deterministic, turn-to-bearing paths in Coq; Section 4 applies the library
to derive exact, formally verified solutions for the timing of intersecting turns
at a \emph{given} point. The rest of the sections present new ideas that were not in the original paper: Section 5 develops a strategy for quantifying the
collision timing computations over \emph{all}
points in the conflict area, providing a bookkeeping framework for managing the
solution to different pieces of the exact timing equations separately;
Section 6 develops formally verified approximations to the timing equations that are
appropriate for quantification within their polygonal region;
Section 7 presents a method for quantifying the timing equations over each piece of the domain, devising
a sound solution for collision timing between two turning vehicles; and finally Section 8 discusses our conclusions and future work.

\section{Literature Review}\label{section:literature_review}

A number of efforts have gone on to formalize horizontal motion and
prove properties about it, but all have characteristics that
distinguish them from our work. For instance, 
\cite{jeyaraman2005} develops an approach for
maneuvering and coordinating vehicles following Dubins paths utilizing
Kripke models which is verified via a model checker, but does not
incorporate non-determinism in the turning models, and does not consider
timing characteristics of the turns. Examples such as \cite{abhishek2020} 
use differential dynamic logic with \KeYmaeraX to model collision
avoidance in automobiles with skidding, but unlike our work
they are concerned mainly with geometric properties of paths and do
not consider timing. The work in \cite{mitsch2017} is an excellent
treatment of collision avoidance in a wide variety of uncertain
turning scenarios for ground robots. It assumes obstacles
characterized by maximum velocity bounds, is not focused on timing
analysis, and is not tailored for use in mixed vertical and horizontal
collision avoidance. We develop a new expression for calculating allowable
tangents to a turn; an alternate solution to this problem is reported 
in~\cite{platzer2017}.

In \cite{Wu2012}, authors develop an algorithm for safe trajectories for robots and dynamic
obstacles following constant speed trajectories with an upper and lower
bounds on curvature, which encompasses turn-to-bearing maneuvers we consider.
However, it uses a rather coarse approximation of the collision region in
the velocity obstacle space, taking a union of all reachable regions over a
time window, where we show how to compute a much tighter conflict region
approximation over time. This has the advantage of allowing greater
maneuverability which is particularly important in adversarial scenarios.

Closely related to this work is \cite{platzer2009}, which considers
curved, horizontal aircraft avoidance maneuvers, but without combining
them with vertical maneuvers; and \cite{jeannin2017}, which considers
vertical maneuvers, but with straight line horizontal kinematics, and
does not allow combination with horizontal maneuvers.

Also closely related to this work is \cite{kouskoulas2017}, which
analyzes vertical maneuvers, but contains timing parameters that can
be set to ensure safety for simultaneous horizontal maneuvers. Our timing computation
can be used to set parameters that safely compose turn-to-bearing
horizontal maneuvers with arbitrary bounded-acceleration vertical
maneuvers.

Dubins paths, constructed of circular arc segments and straight lines,
are used to model horizontal motion in many path planning and
collision avoidance algorithms, such
as \cite{cons2014,pantelis2014,ma2006,mcgee2006,song2017,zhao2019}.
These examples are not formally verified, and although some are
created with aircraft in mind, they are not designed for timing analysis or adversarial collision avoidance assumptions in our work.

Many years of work have gone into the tools and libraries that we
used for our development, including the Coq proof assistant \cite{coq}
and the Coquelicot extensions for its real
library \cite{boldo2015}. Our libraries are intended to contribute to
this toolbox.

%% file: horiz-proof.tex
The first step in reasoning about the safety of turn-to-bearing maneuvers
is to formalize the definition of a turn-to-bearing trajectory in a manner
suitable for use in the Coq proof assistant. We consider the
mathematical definition of turn-to-bearing kinematics, a library interface,
and the formalization of geometric properties that are necessary for our
safety analysis.

\begin{figure}[ht]
  \centering
  \begin{subfigure}[t]{0.47\textwidth}
        \centering
        \includegraphics[width=\textwidth]{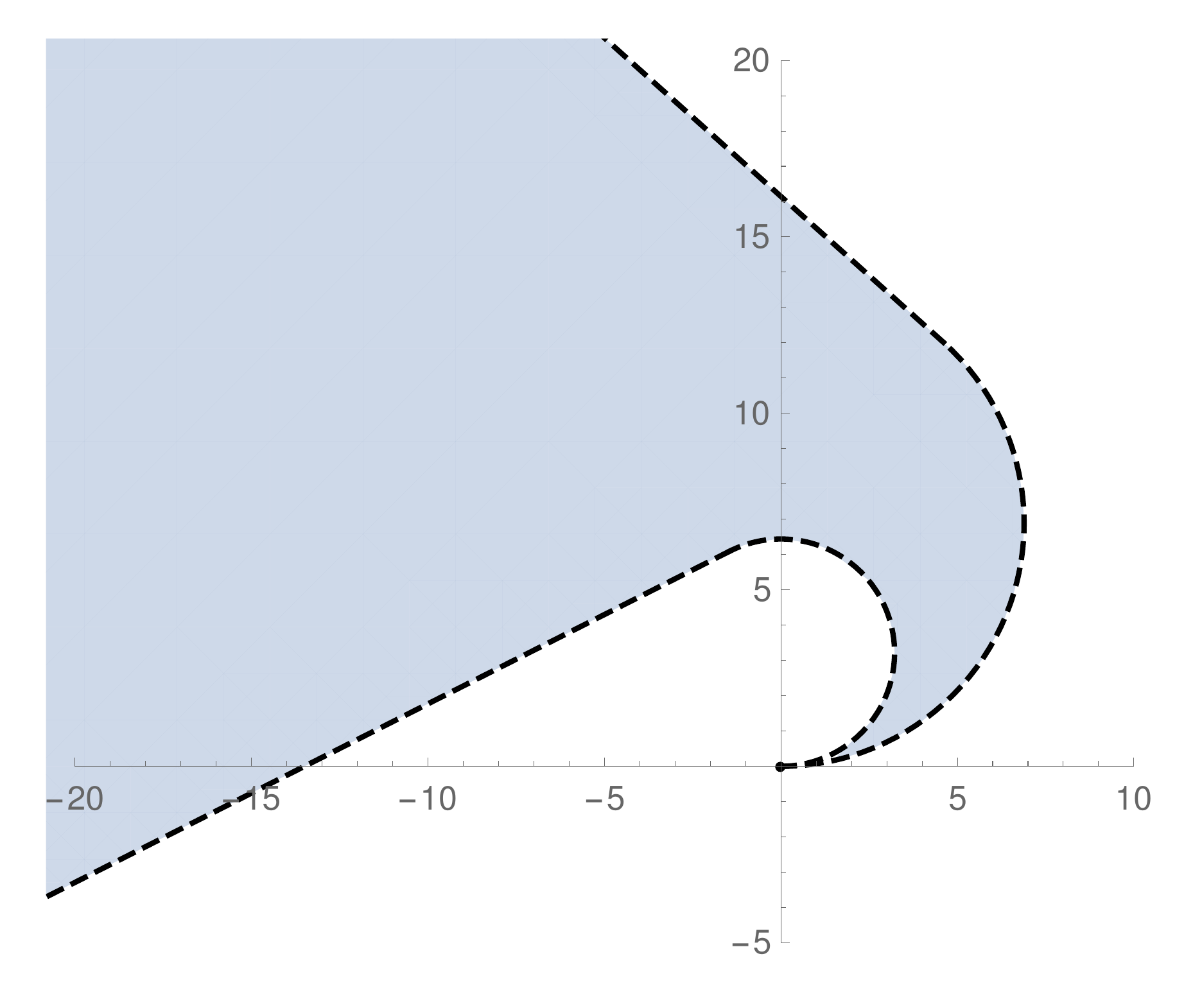}%
        \raisebox{55pt}[0pt][0pt]{
          \makebox[0pt][l]{
            \hspace*{-98pt}
            \includegraphics[angle=-147,scale=.013]{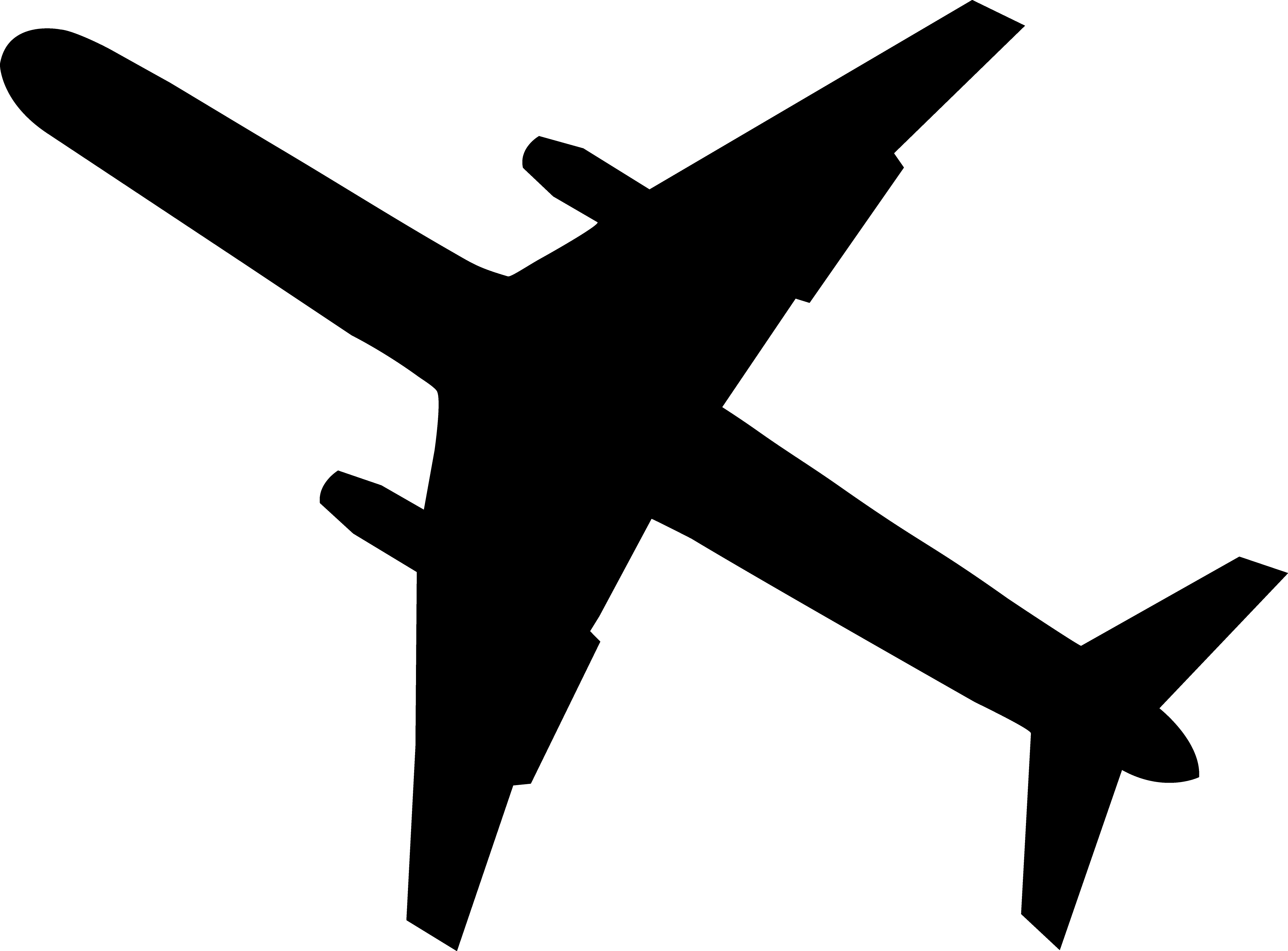}}}
        \caption{Shaded area is reachable in the future.}
    \end{subfigure}\hfill
    \begin{subfigure}[t]{0.47\textwidth}
        \centering
        \includegraphics[width=\textwidth]{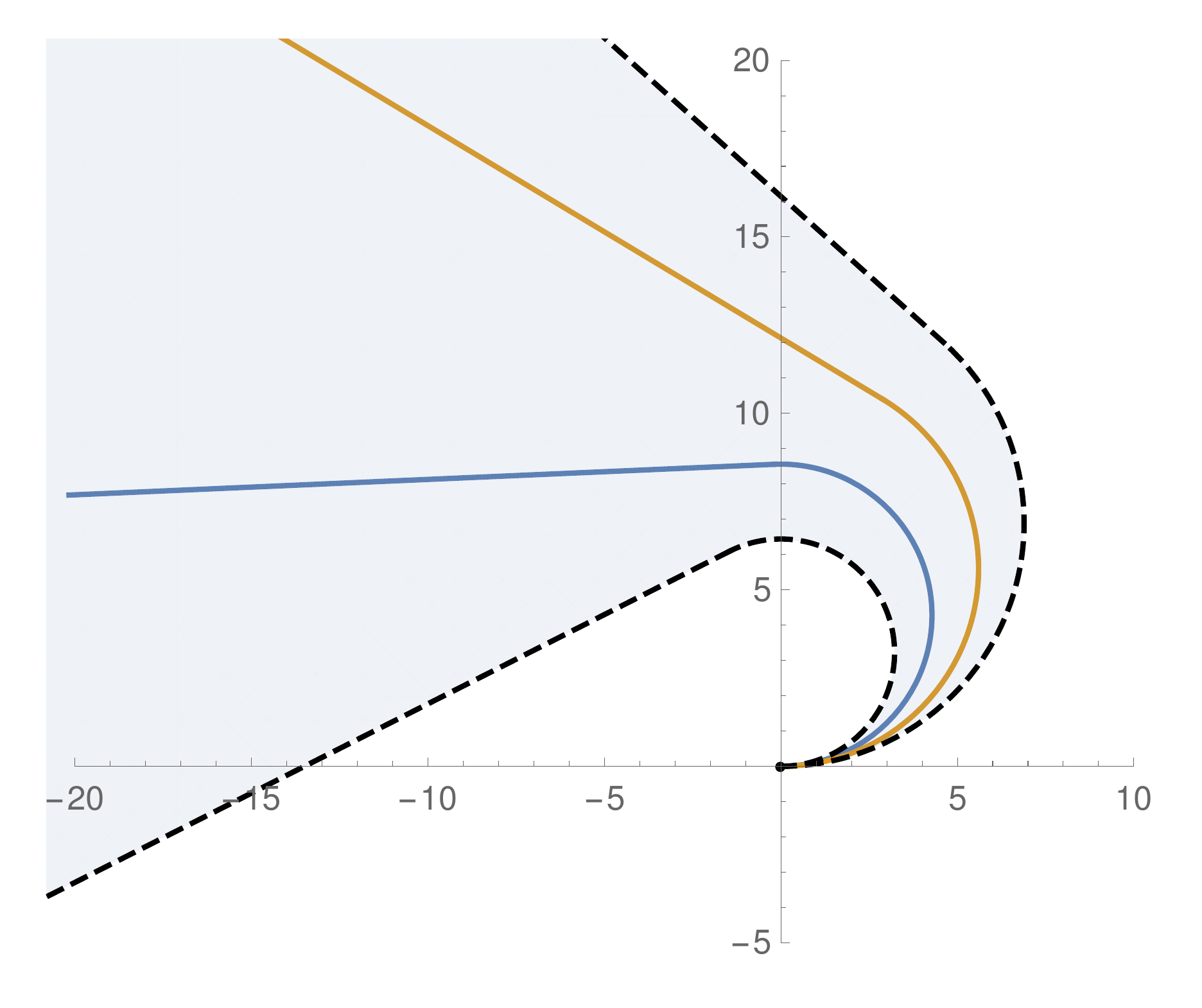}
        \caption{Paths show two possible trajectories.}%
        \raisebox{117pt}[0pt][0pt]{
          \makebox[0pt][l]{
            \hspace*{-54pt}
            \includegraphics[angle=36,scale=.013]{figs/airplane_topview_CC.png}}}%
        \raisebox{157pt}[0pt][0pt]{
          \makebox[0pt][l]{
            \hspace*{4pt}
            \includegraphics[angle=0,scale=.013]{figs/airplane_topview_CC.png}}}
    \end{subfigure}
  \caption{Visualizing turn-to-bearing motion.}  \label{fig:ottb}
\end{figure}

\subsection{Non-deterministic Turn-to-Bearing Kinematics}

We define \emph{non-deterministic one-turn-to-bearing} motion as a set
of trajectories representing a range of future motion possibilities
that might be followed by the vehicle. We characterize this motion
with a tuple that represents the set of future trajectories that are
possible $(x_0, y_0, \theta_0, \ra, \rb, \tha, \thb, \sa,\sb)$ where:
$(x_0, y_0)$ and $\theta_0$ are initial position and orientation of
the vehicle; \ra, \rb, \tha, \thb\ are bounds on the fixed turn radius and
cumulative change in orientation after completing the circular turn, respectively;
and \sa, \sb\ are bounds on the speed throughout the encounter, which
unlike the other parameters is assumed to vary continuously as a
function of time. We adopt the convention of using positive radii and
bearing offsets to represent counterclockwise (left) turns, and
negative radii and bearing offsets to represent clockwise (right)
turns. Left turns are represented by $0<\ra\leq \rb$ and $0<\tha<\thb
< 2\pi$, while right turns by $\ra\leq \rb < 0$ and $-2\pi<\tha<\thb <
0$.  In all cases, we assume $0<\sa\le\sb$. Realizing a specific
future trajectory requires drawing from this sample space. Each
trajectory has parameters $r, \theta_c, s(t)$ satisfying
the constraint predicate $\chi(r, \theta_c, s)
= \theta_c\in[\tha,\thb]\land r\in[\ra ,\rb]\land(\forall u,
s(u)\in[\sa,\sb])$ which represents a path with initial turn that we
model using a circular arc of radius $r$, followed by a linear path
tangent to the turn whose bearing is offset by $\theta_c$ from
$\theta_0$. The path is traversed with continuously varying speed
$s(t)$. Figure~\ref{fig:ottb}a
plots a visualization of the turn-to-bearing
envelope for $(x_0,
y_0, \theta_0, \ra, \rb, \tha, \thb, \sa,\sb) =
(0,0,0,3.22,6.89,2.41,3.62,1,2)$, while Fig. \ref{fig:ottb}b shows example trajectories consistent with that envelope.

Components of the vehicle's trajectory for these kinematics are given by
\begin{align}
  &  J_x(t) =
		 \begin{cases}
          r \sin\left(\frac{d(t)}{r} + \theta_0\right) - r \sin(\theta_0) + x_0
          &  d(t)\leq r\theta_c \\
			 \begin{aligned}
				 &(d(t)-r\theta_c) \cos(\theta_c + \theta_0) \\
				 &\quad + r \sin(\theta_c+\theta_0) - r \sin(\theta_0)  + x_0 
			 \end{aligned}
			&  d(t)>r\theta_c \\
         \end{cases} \label{jx}\\
  & J_y(t) = 
         \begin{cases}
         - r\cos\left(\frac{d(t)}{r}+\theta_0\right) + r \cos(\theta_0) + y_0
         & d(t)\leq r\theta_c \\
			\begin{aligned}
				& (d(t)-r\theta_c) \sin(\theta_c + \theta_0) \\
				&\quad - r\cos(\theta_c+\theta_0) + r \cos(\theta_0) + y_0 
		    \end{aligned}
			&  d(t)> r\theta_c \\
  \end{cases}\label{jy}
\end{align}
for overall trajectory $J(t) = J_x(t)\hat{x}+J_y(t)\hat{y}$. The distance
traveled on the path is related to speed during the trajectory in the
usual way, $d(t) = \int_0^t s(\gamma) d\gamma$.

\subsection{Library Interface}

The library we have developed is organized around the representation of
a path in $\mathbb{R}^2$ and a predicate
\begin{equation}
\pathsegment(D, f_x(d), f_y(d), (x_0,y_0), (x_1,y_1))
\end{equation}
which, when true, asserts: that $f_x(d)$ and $f_y(d)$ are parameterized
functions describing the $x$ and $y$ positions of the path in the
coordinate plane; that the resulting path is continuous and
integrable; and that $f_x(d)$ and $f_y(d)$ are parameterized by the
path distance, i.e.
$\int_0^d\sqrt{(f'_x(\alpha))^2+(f'_y(\alpha))^2}d\alpha = d$;
that $(f_x(0),f_y(0)) = (x_0,y_0)$; and that
$(f_x(D),f_y(D)) = (x_1,y_1)$.
Parameterizing our path representation by path distance creates a
canonical representation of the geometry for each path, isolating it
from timing considerations associated with variations in speed during
the maneuver. This allows us to analyze each aspect separately and
combine them in the end.

Note that although the turn-to-bearing paths in the library define a 
starting and ending point separated by distance $D$, the paths continue
indefinitely.

The library also contains piecewise functions parameterizing the $x$
and $y$ positions for turn-to-bearing paths
$H_x(r,\theta_0,x_0,\theta_c,rtp,d)$ and
$H_y(r,\theta_0,y_0,\theta_c,rtp,d)$, meant to be used with the
$\pathsegment$ predicate. The functions are equivalent to
Eqs.~\eqref{jx} and \eqref{jy}, differing only in that they are
parameterized by distance $d$ instead of time $t$. The functions are
curried before being used in $\pathsegment$, instantiated with
starting point $(x_0,y_0)$, initial orientation $\theta_0$, the turn
radius $r$, and the angular offset for the final bearing
$\theta_c$. They also require an argument named $rtp$, which must be a
proof object showing that $0 < r\theta_c < 2\pi|r|$, ensuring the
signs of $r$ and $\theta_c$ to be identical, and enforcing an upper
bound on $\theta_c$. The files \emph{ttyp.v} and \emph{tdyn.v} define
the $\pathsegment$ predicate, the parameterized turn-to-bearing paths,
and prove lemmas about path continuity, differentiability, and
path-length parameterization of $H_x$ and $H_y$ so they can be used
with the $\pathsegment$ predicate. Along with the parameterization,
the library contains predicates $\straight$ and $\turning$ which
indicate whether the parameters describing a path reach the final
destination point while traveling in a straight line, or turning on a
circular arc, respectively.

The rest of the library includes trigonometric definitions and identities
that are missing from the Coq standard library
(\emph{atan2.v}, \emph{strt.v} and \emph{strt2.v}), lemmas that help
the user introduce \turntobearing{} $\pathsegment$ predicates into
the context (\emph{tlens.v}), lemmas that derive consequences and
mathematical relationships from \turntobearing{} $\pathsegment$
assumptions (\emph{tlens.v}), lemmas about timing intervals
(\emph{ttim.v}), and theorems about the computation of timing properties
 based on pathlength (\emph{dtlen.v}). The size of the development is 
significant, around 40k lines of proof scripts.

Because Coq allows expression in a higher order logic, it permits
quantification over any variable. This means we can hold the starting
and ending points of the path fixed and quantify over the other
parameters to reason about waypoints, or fix ranges of parameters and
quantify over the radii and angles to reason about ranges of
non-deterministic possibilities in turn radius and final bearing.

In this paper, for clarity, we present lemmas from the library in a
standard position and orientation such that $(x_0,y_0)=(0,0)$,
$\theta_0=0$, and $(x_1,y_1)=(x,y)$. To analyze intersecting paths
that are oriented and positioned arbitrarily with respect to one
another, the more general form can be recovered by assuming that
\begin{align}
  x &= (x_1 - x_0)\cos(\theta_0) + (y_1 - y_0)\sin(\theta_0) \\
  y &= - (x_1 - x_0)\sin(\theta_0) + (y_1 - y_0)\cos(\theta_0)
\end{align}
The library itself contains the translations and rotations to allow
full generality when working with more than one path.

\subsection{Trigonometric Properties}

Geometric intuition which might seem simple does not always translate
naturally to formal analysis in a proving environment.

\begin{figure}[htbp]
  \begin{center}
    \begin{subfigure}[t]{.42\textwidth}
      \includegraphics[width=\textwidth]{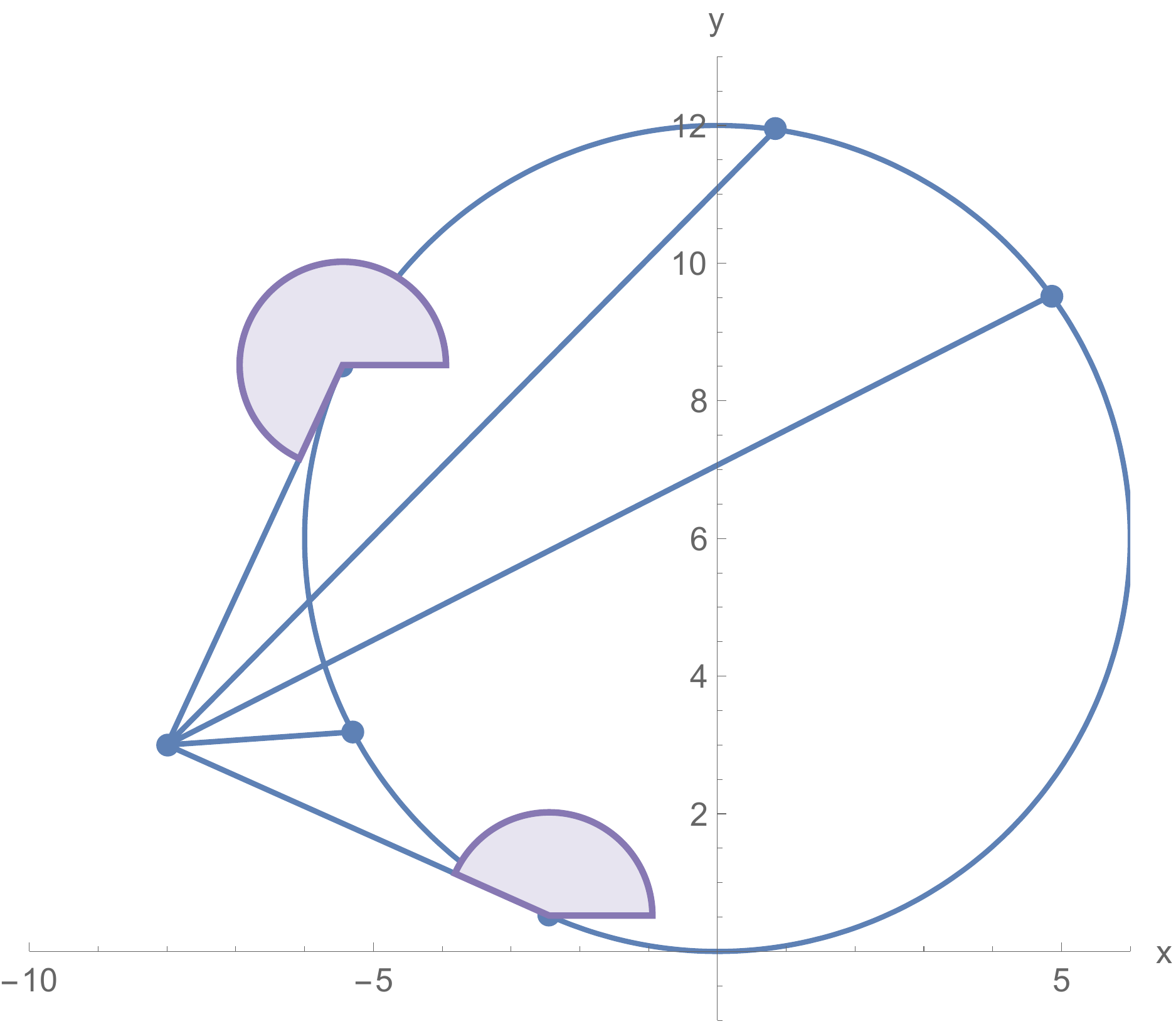}
      \caption{Angle of each line segment is $\kappa$}
    \end{subfigure}\hfill
    \begin{subfigure}[t]{.47\textwidth}
      \includegraphics[width=\textwidth]{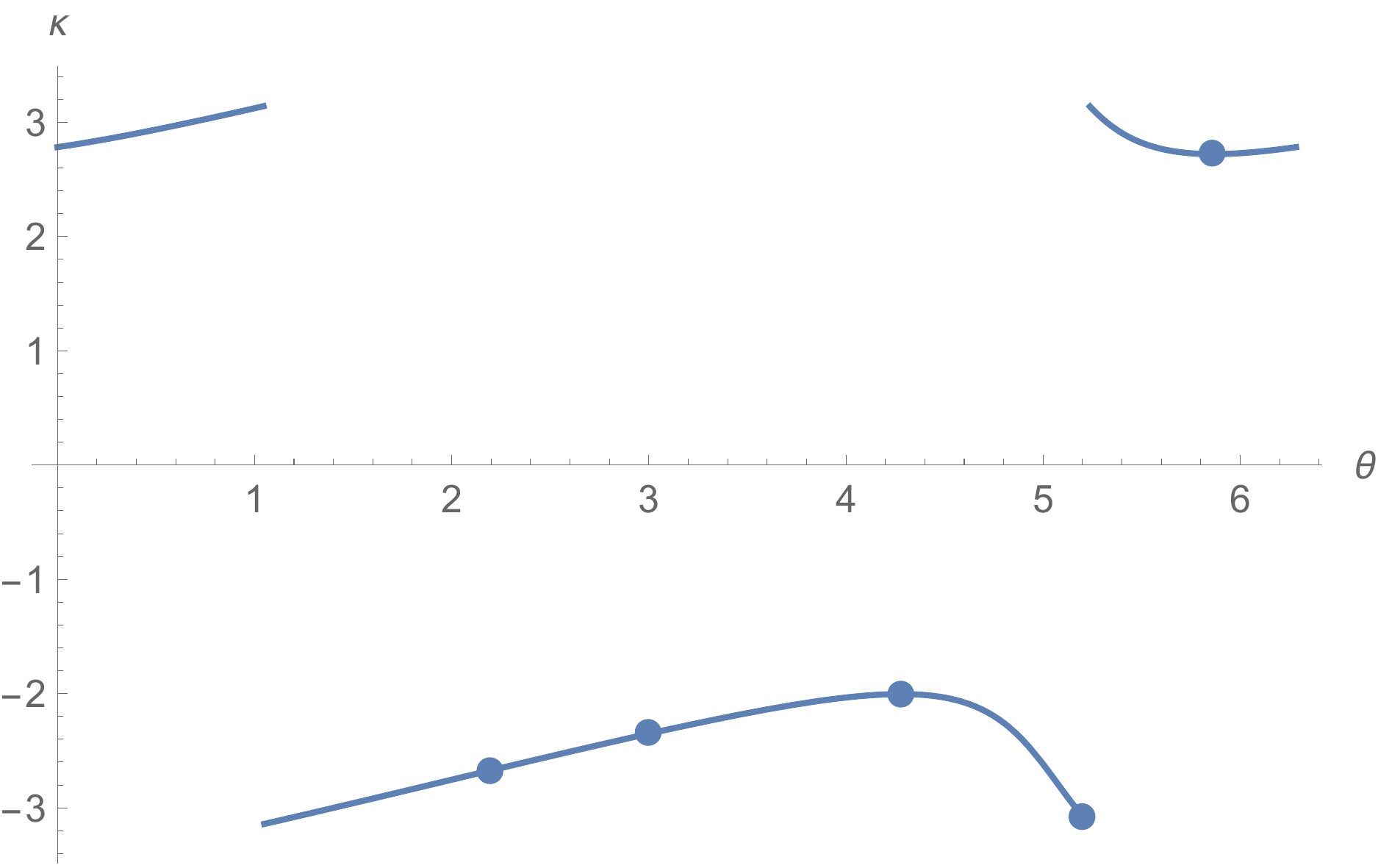}
      \caption{Discontinuous $\kappa_2$; $\kappa'=0$ points are tangent}
      \end{subfigure}
  \end{center}%
  \raisebox{230pt}[00pt][0pt]{
    \makebox[-5pt][l]{
		\hspace*{15pt}$(x,y)$}}%
  \raisebox{330pt}[00pt][0pt]{
    \makebox[-5pt][l]{
      \hspace*{15pt}$\kappa(\theta_1) = \theta_1 + 2n\pi$}}%
  \raisebox{235pt}[00pt][0pt]{
    \makebox[0pt][l]{
      \hspace*{100pt}
      $\kappa(\theta_2) = \theta_2 + 2(m+1)\pi$
  }}
  \caption{If we parameterize positions on a circular path using the 
    vehicle orientation	$\theta$ associated with the tangent, then 
    $\kappa(\theta)$
	(shaded) is the angle of the line connecting the
    point on the circle to a point $(x,y)$ outside the circle.}
  \label{fig:tangents}
\end{figure}

First we needed to encode in our proving environment
a basic understanding of the way circular turns may be combined with
straight paths that exit the turns on a
tangent. There are two tangent lines to a circle, anchored at orientations $\theta_1$
and $\theta_2$, which arrive at any particular point $(x,y)$ outside
the circle (see Fig. \ref{fig:tangents}).
One of the tangents is not useful because for counter-clockwise turns,
it always results in a path with a discontinuous derivative. This is
geometrically obvious to a human by inspection, but somewhat
challenging to formalize in Coq.

\begin{figure}[htbp]
  \begin{center}
      \includegraphics[width=.42\textwidth]{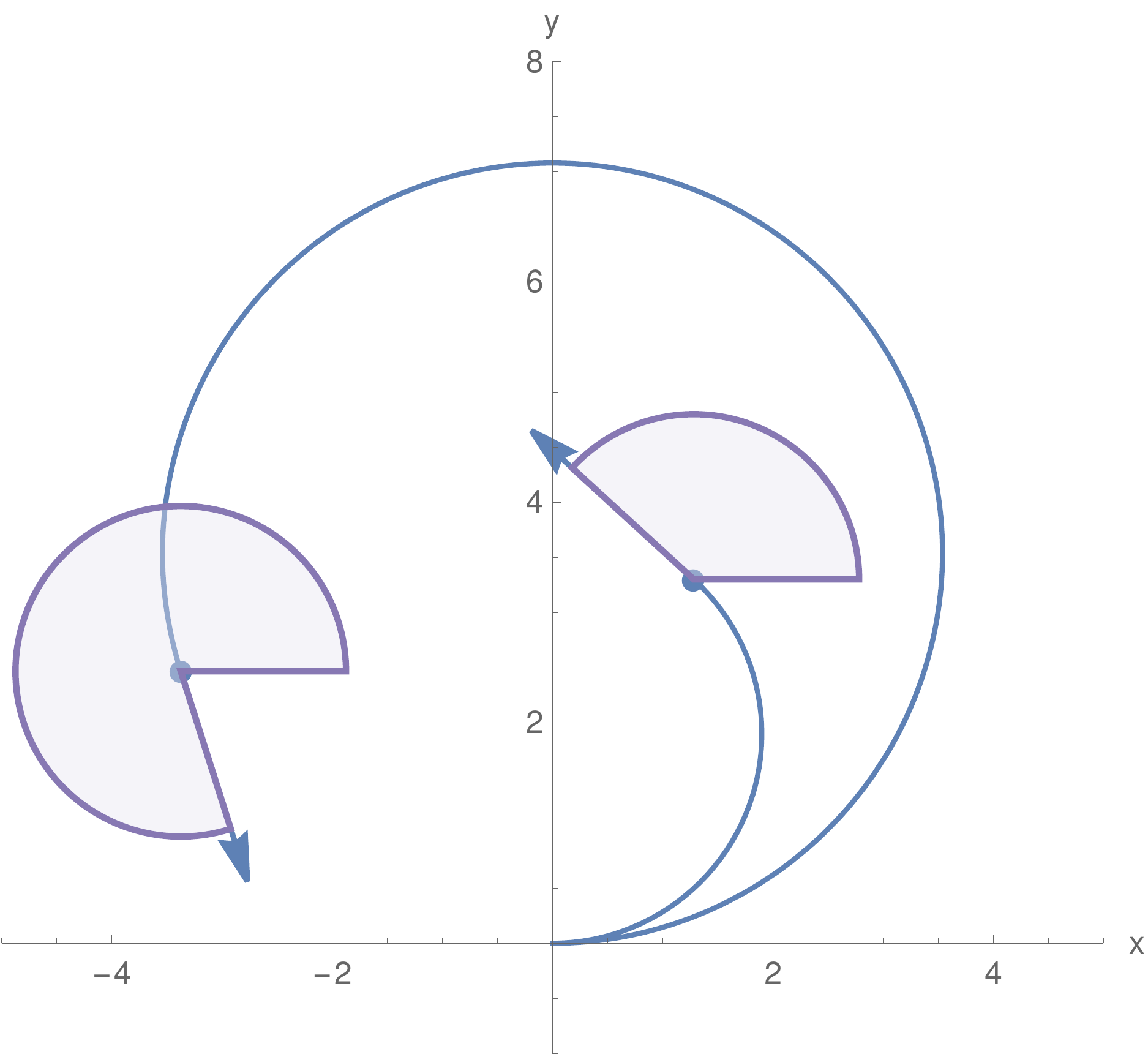}%
  \raisebox{80pt}[00pt][0pt]{
    \makebox[-5pt][l]{
		\hspace*{-120pt}$(x_1,y_1)$}}%
  \raisebox{95pt}[00pt][0pt]{
    \makebox[-5pt][l]{
      \hspace*{-95pt}$\theta_m(x_1,y_1)$}}%
    \raisebox{62pt}[00pt][0pt]{
    \makebox[-5pt][l]{
		\hspace*{-170pt}$(x_2,y_2)$}}%
  \raisebox{75pt}[00pt][0pt]{
    \makebox[-5pt][l]{
      \hspace*{-190pt}$\theta_m(x_2,y_2)$}}%
        \end{center}%
  \caption{Two example points illustrating the geometric intuition for $r_m$ and $\theta_m$ using circular arcs that connect them with the origin. Each arc has a radius of $r_m$, and a tangent of $\theta_m$ at its end, where each parameter is calculated with the coordinates of the final point.}
  \label{fig:thetam}
\end{figure}

Using a chord lemma from geometry we can infer that a vehicle
approaching $(x,y)$ from a circular turn will do so at an angle of
$\theta_m = 2\atan_2(y,x)$ and that the radius required to reach it
will be $r_m=(x^2+y^2)/(2y)$, see Fig. \ref{fig:thetam}. We find that
decreasing the radius of the turn decreases the angle, while
increasing it makes $(x,y)$ unreachable via a tangent line. Thus 
$\theta_m$ defines an upper bound on the approach angle.
From inspection, we can see that the angle of the second tangent exceeds 
this boundary.

In order to formalize this geometric intuition, we define a function
which given the orientation $\theta$ of the vehicle on a turning path
of radius $r$ would return the angle from the vehicle to the point
$(x, y)$,
\begin{equation}
  \kappa(\theta) = \atan\left(\frac{y - r(1 - \cos(\theta))}
        {x - r\sin(\theta)}\right).
\end{equation}
Both by construction, and by the periodicity of $\sin$ and $\cos$, we
note that $\kappa$ is periodic with period $2\pi$.  For the remainder
of this section we will restrict the domain of $\kappa$ to $(0, 2\pi)$
for $r>0$, and $(-2\pi,0)$ for $r<0$.  As illustrated in
Fig.~\ref{fig:tangents}b, the function $\kappa$ is not continuous for all values
of the destination point $(x,y)$.

We define a series of functions based on a
\twoargument{} arctangent and different branch cuts, which have distinct, 
overlapping, and complementary domains upon which $(x,y)$ yields a
continuous function.
\begin{align}
    \kappa_2(\theta) ={}&
    \atan_2\big(y - r(1 - \cos(\theta)), x - r\sin(\theta)\big)  \\
    \begin{split}
        \kappa_3(\theta) ={}&
        \atan_2\big(-(y - r(1 - \cos(\theta))), \\
            &\qquad\qquad\qquad -(x - r\sin(\theta))\big) + \pi
    \end{split}\\
    \begin{split}
        \kappa_4(\theta) ={}&
        \atan_2\big(-(x - r\sin(\theta)), \\
            &\qquad\qquad\qquad y - r(1 - \cos(\theta))\big) + \pi/2
    \end{split}
\end{align}
Henceforth, when we refer to properties of $\kappa$, we are choosing
a variant with the branch cut oriented so that there is no
discontinuity for the given destination point $(x,y)$.

When $\kappa$ is continuous, we show that the unique maximum and
minimum values $\kappa(\theta_1)$ and $\kappa(\theta_2)$ correspond to
the angles of the correct and incorrect tangent lines respectively
(for $r > 0$, if $r < 0$ the maxima and minima are reversed). We prove
that $\kappa(\theta) = \frac{\theta_m}{2}$ implies that $\theta = 0$
or $\theta = \theta_m$.
Since $\kappa$ takes on the value $\frac{\theta_m}{2}$, it must be that
$\kappa(\theta_1) \ge \frac{\theta_m}{2} \ge \kappa(\theta_2)$.

Our choice of domain ensures that $0$ is not between $\theta_1$ and
$\theta_2$, so we can use the Intermediate Value Theorem to show that
$\theta_m$ is in-between $\theta_1$ and $\theta_2$ in the
domain. Because $\theta_m$ is a limiting value of the approach angle,
we can eliminate $\theta_2$, which is always outside of the allowable
range, leaving $\theta_1$ as the angle of approach that ensures path
continuity.

We calculate extremal values of $\kappa$, $\theta_1$ and $\theta_2$, by
setting the derivative of $\kappa$ to zero, and solving for the
argument.
Fortunately, each variant of the $\kappa$ function for 
which the destination point $(x,y)$ yields a continuous function has the same derivative
\begin{equation}
	\kappa'(\theta) = \frac{r((2r-y)\tan^2(\theta/2) - 2x\tan(\theta/2) + y)}{D(\theta)},
  \label{eqn:kappaderv}
\end{equation}
for $\theta\notin\{0, \pi\}$.  The sign of the
denominator 
\begin{equation}
	\begin{aligned}
	D(\theta) &= 2 (1-\cos(\theta))/\sin^2(\theta) \\
	&\quad\cdot\left((y - r(1-\cos(\theta)))^2+(x - r\sin(\theta))^2\right)
	\end{aligned}
\end{equation}
is always positive, and so the sign of $\kappa'$ is directly
related to the sign of the quadratic function in the numerator; the
task of calculating the maximum and minimum is reduced to the
problem of solving a quadratic in $\tan(\theta/2)$.
The solution associated with the maximum value of $\kappa$ is given in
Eq.~\eqref{eqn:Theta}.

Reasoning about the continuity of the $\kappa$ variants, handling
their derivatives as the angle crosses the branch cut, and ordering of
roots and angles to establish what ``in-between'' means in an angular
domain that is a clock system is contained within the file
\emph{strt.v} and its corresponding documentation.

\subsection{Turn-to-Bearing Path Properties}
\label{sec:ttb_path_properties}

Parameters for turn-to-bearing trajectories must be selected in a
way that the radius and angle of departure from the turn lead from the
starting point to the ending point, and so that the distance is
consistent with the path. In this
section, we state basic results about paths, and select a few proofs
about which we provide some details in order to give a flavor of the
reasoning in the library.

\begin{figure}[htb]
\centering
  \begin{subfigure}[t]{0.45\textwidth}
  \centering
    \includegraphics[scale=.45,trim={50 50 50 50},clip]{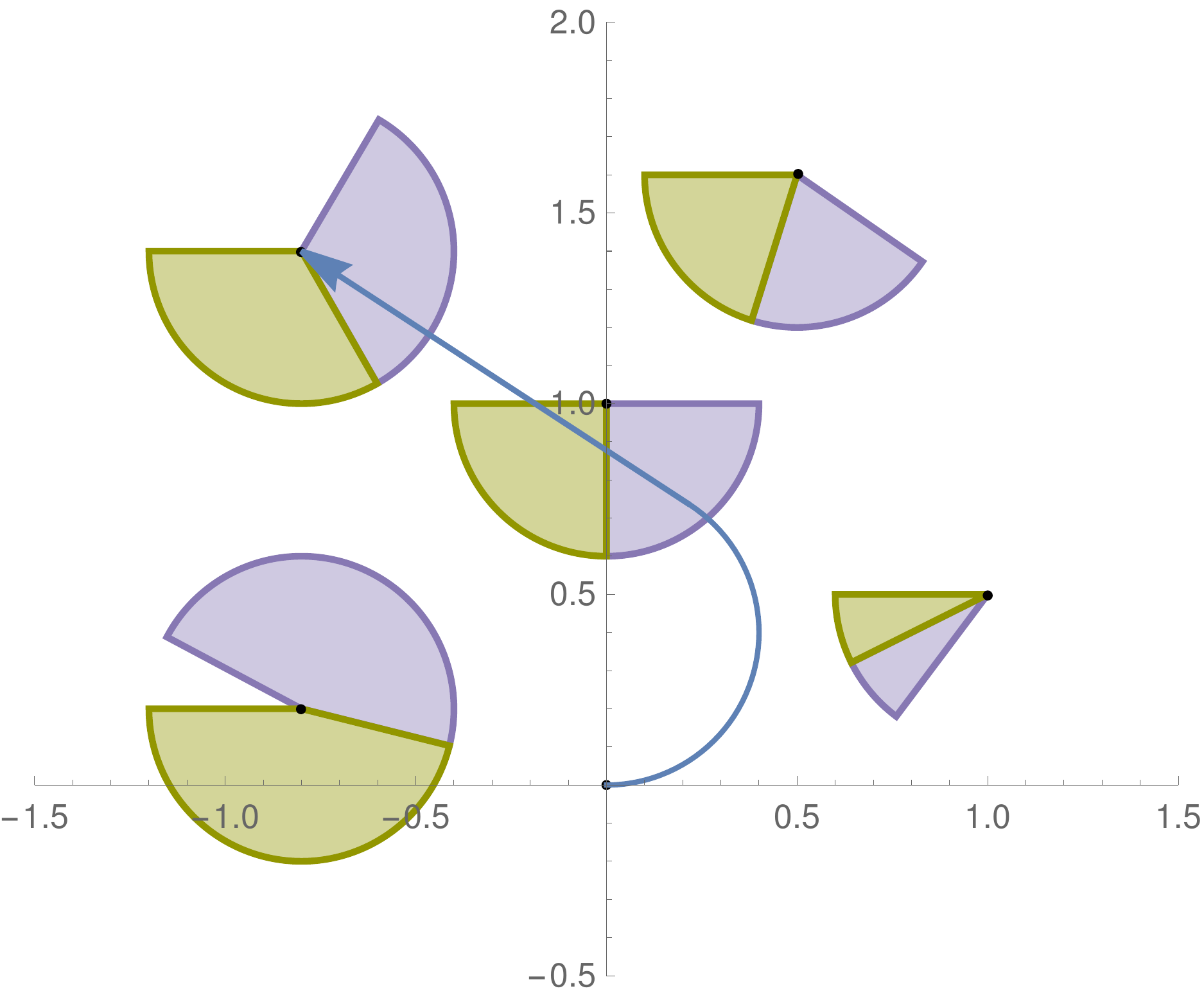}
    \caption{Approaches for different points.}
    \raisebox{51pt}[00pt][0pt]{
      \makebox[0pt][l]{
        \hspace*{-15pt}
        \includegraphics[angle=-147,scale=.013]{figs/airplane_topview_CC.png}}}
  \end{subfigure} \\
  
  \begin{subfigure}[t]{0.45\textwidth}
  \centering
    \includegraphics[scale=.43,trim={30 0 0 0},clip]{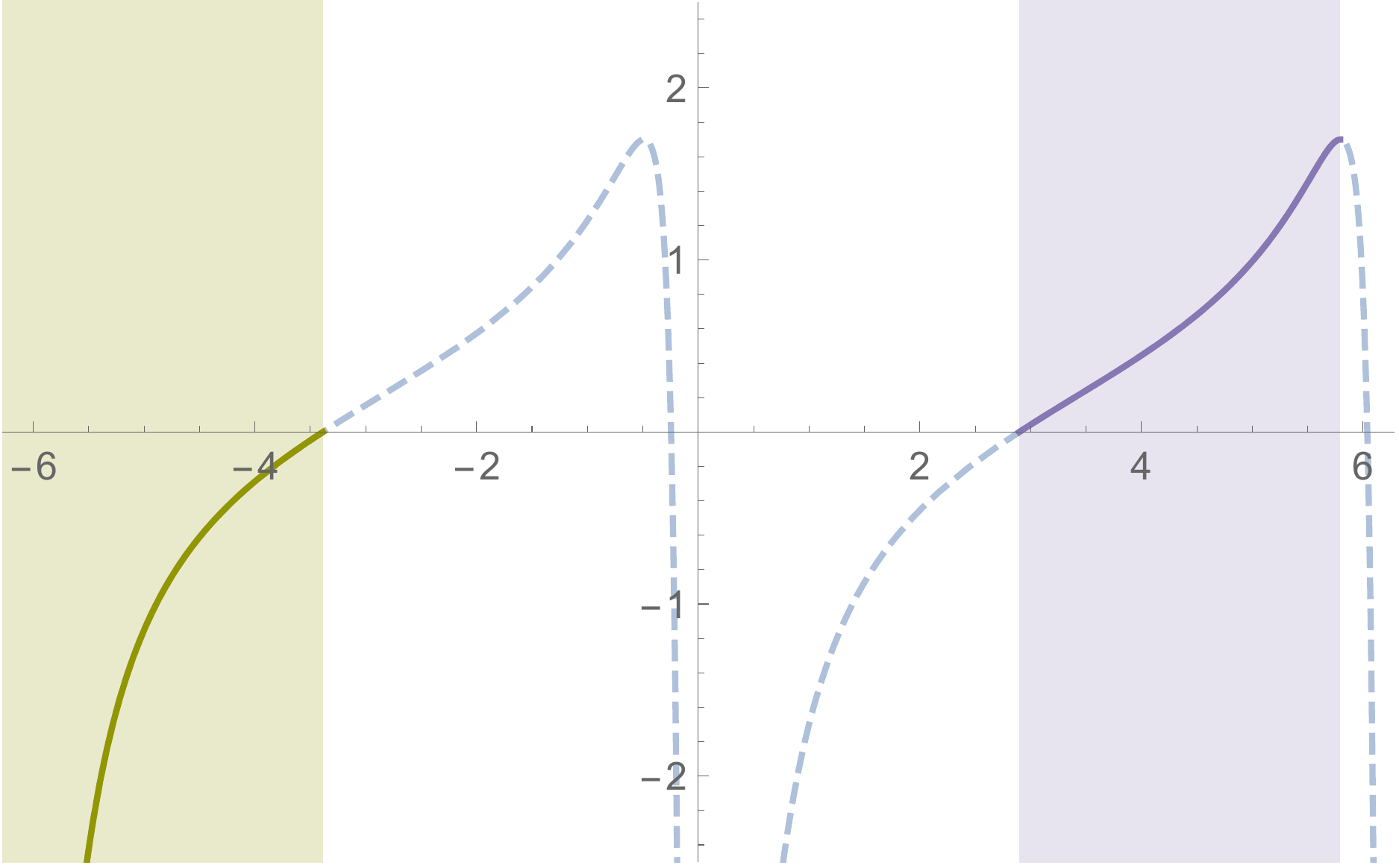}
    \caption{$r$ vs. $\theta$ for a single point $(-0.8,0.2)$.}
  \end{subfigure}

  \caption{Relationship between allowable angle of approach and
    required radius to achieve that angle. Choosing angular ranges of
    approach also entails a turn direction; left turns are marked with
    violet and right turns marked with green.}
  \label{fig:limited-approach}
\end{figure}

We can construct a turn-to-bearing trajectory by first choosing an angle of
approach $\theta$ for a point $(x,y)$, and then computing the turn radius
required to arrive there with that orientation. The angle of approach
is constrained because of the initial position and angle of the
aircraft and the required kinematics.
Fig. \ref{fig:limited-approach}a shows the initial position of an aircraft and for a series of example reachable points, plots shows the angular extent of feasible approaches using pie slice-shaped circular segments; an example path is shown to illustrate one possibility to reach one of the points. Fig. \ref{fig:limited-approach}b graphically plots, for the lower left point, Eq. \ref{eq:rcalc}, i.e. what $r$ must be for each allowable choice of $\theta$ at that point.

\begin{theorem}[Turn-to-bearing dependent radius]
  A vehicle following a turn-to-bearing trajectory
  can approach point $(x,y)$ with a chosen angle $\theta$ when
  \begin{equation}
    \begin{split}
  (0 < \theta_m \land (\theta_m/2 < \theta \le \theta_m \lor
    - 2\pi < \theta < \theta_m/2 - 2\pi)) \lor \\
    (\theta_m < 0 \land (\theta_m \le \theta < \theta_m/2 \lor
    \theta_m/2 + 2\pi < \theta < 2\pi))
    \end{split}
  \end{equation}
using radius
\begin{equation}
  R(x,y,\theta) = \frac{x \sin(\theta) - y \cos(\theta)}
                             {1 - \cos(\theta)}.
\label{eq:rcalc}
\end{equation}
\end{theorem}

Similarly, we can also construct a turn-to-bearing trajectory by first choosing
a turn radius $r$, and then computing the angle of approach that the radius
entails when we arrive at $(x,y)$. The choice of radius is constrained if
the target point is on the same side as the direction of the turn,
because the turn must be rapid enough to orient the aircraft in the
direction of the target point before it has passed it.

\begin{theorem}[Turn-to-bearing dependent approach angle]
  A vehicle following a turn-to-bearing trajectory
  can approach point $(x,y)$ using a turn with chosen radius $r$ when
  \begin{multline*}
    \left(0< y \land r \le \frac{x^2 + y^2}{2 y}\right) \lor
    \left(y=0 \land x < 0\right) \lor\\
    \left(y < 0 \land \frac{x^2 + y^2}{2 y}\le r\right)
  \end{multline*}
and the angle of approach is
\begin{multline}
  \Theta(x,y,r) =\\
  \begin{cases}
	  2\atan\left(\frac{x - \sqrt{x^2 - (2 r - y) y}}{(2 r - y)}\right) + P &  2 r-y \neq 0 \\
    2\atan\left(\frac{y}{2 x}\right) & 2 r - y = 0 \land x > 0 \\
    \pi \sign(r) & 2 r - y = 0 \land x \leq 0
    \end{cases}
    \label{eqn:Theta}
\end{multline}
where $P=P(x,y,r)$ is a phase correction given by
\begin{multline*}\label{eqn:pce}
  P(x,y,r) =\\
  \begin{cases}
    0 & (0 < r \land ((0 < x\land 0 < y) \lor x \leq 0 \land 2r < y)) \\
      & \lor (r < 0 \land ((x < 0 \land y < 0) \lor y < 2r)) \\
    2\pi & 0 < r \land (0 \leq x \land y < 0 \lor x < 0 \land y < 2r)) \\
    -2\pi & r < 0 \land (0 \leq x \land 0 < y \lor x < 0 \land 2r < y).
    \end{cases}
\end{multline*}
\end{theorem} 

\begin{figure}
  \begin{center}
      \begin{subfigure}[t]{.49\textwidth}\centering
      \includegraphics[scale=.4]{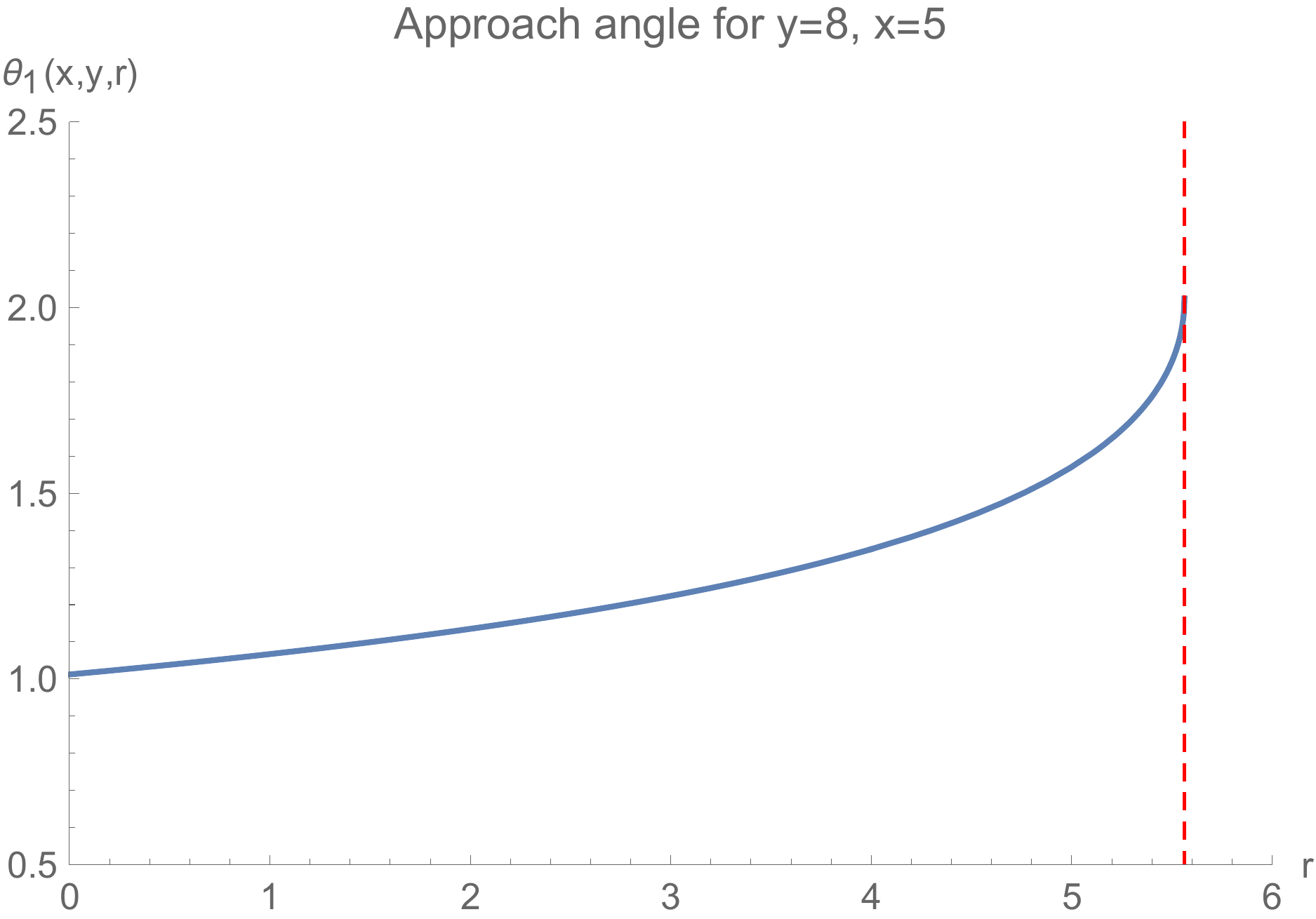}
    \caption{Varying $r$ for a single $(x,y)$.
      }
    \end{subfigure}\hfil
    \begin{subfigure}[t]{.49\textwidth}\centering
      \includegraphics[scale=.4]{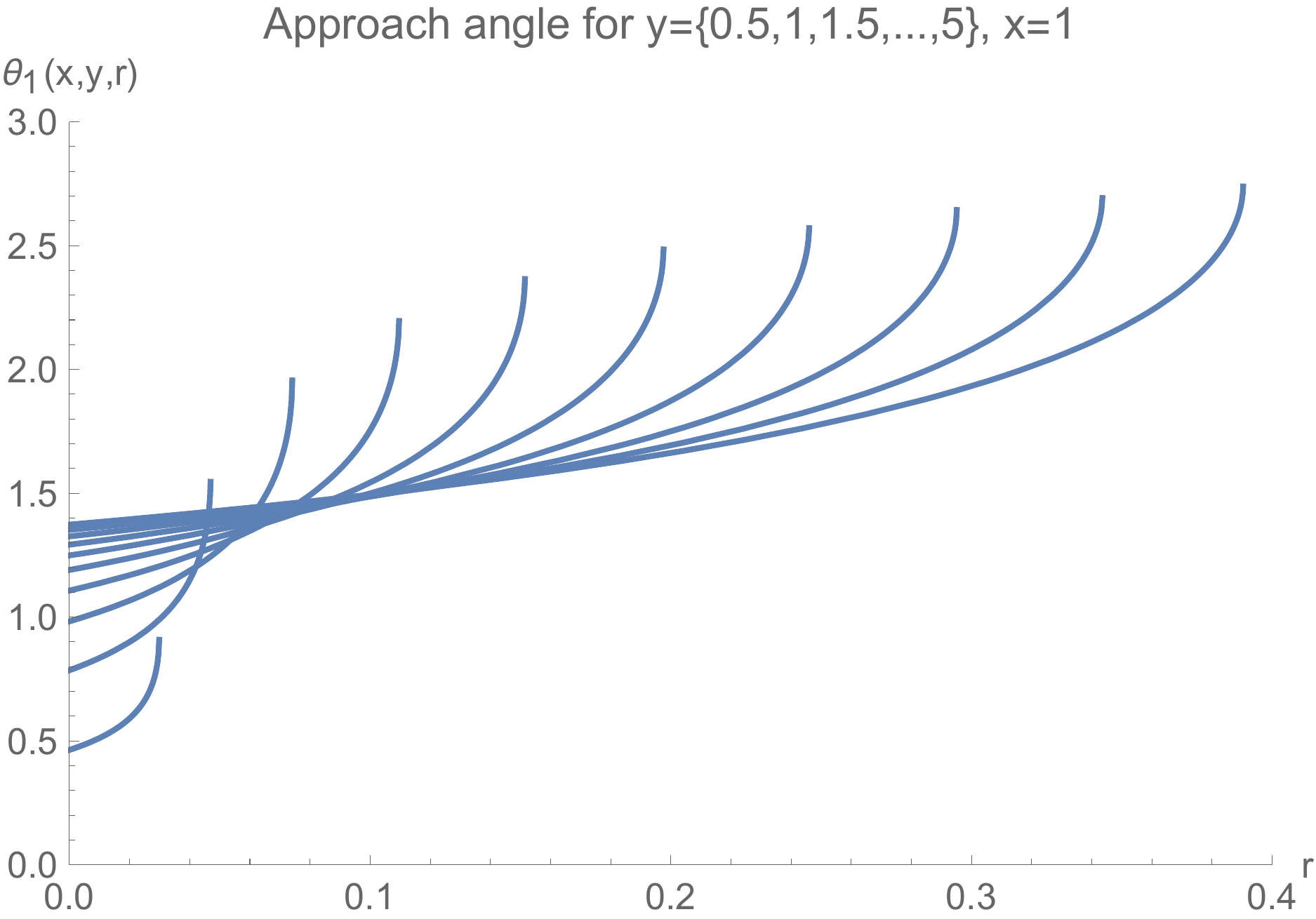}
      \caption{Varying $r$ for a sequence of points.}
      \end{subfigure}
  \end{center}
\caption{Plot of the first piece of $\Theta(x,y,r)$ from Eq.~\eqref{eqn:Theta}}
  \label{fig:theta1-piece}
\end{figure}

It is not surprising that for fixed $(x,y)$, the first piece of
$\Theta(x,y,r)$ is not differentiable or even always defined at
$r=r_m$. What is surprising is that even if we define the endpoint to
ensure the value of the function is finite, its rate of change is
unbounded at the end of the interval. 
We initially expected that we could simply extend our $\theta_1$ curve to create a function with a continuous derivative, but in the end had to settle for creating a piecewise continuous function using the limiting value at the cutoff point, which turned out to be enough for our purposes.
This is illustrated in
Fig. \ref{fig:theta1-piece}, and made formalizing the relationship between
the length of circular arc path segments and the rest of the
turn-to-bearing kinematics a longer process than we had expected. 

This geometry appears in a variety of contexts, including
\cite[p. 15]{platzer2017}, which has another
expression that may be used to solve for the angle. We leave it to the
interested reader to show the equivalence between the result we have
proved, and alternate formulations. We also found a remarkable 
simplification for the tangent path length:

\begin{theorem}[Straight path segment expression]
For a turn-to-bearing trajectory given by $(r,\Theta(x,y,r))$, that
starts at the origin with $\theta_0 = 0$ and passes through $(x,y)$, the square of the
distance traveled on a straight line before we arrive at $(x,y)$ is
given by
\begin{multline}
(x - r\sin (\Theta(x, y, r)))^2 + (y - r(1 - \cos(\Theta(x, y, r))))^2 \\
= x^2 - (2 r - y) y.
\end{multline}
\end{theorem}

%% file: horiz-analysis_of_horizontal_motion.tex
\label{pointwise-collision}

This section describes the application of our turn-to-bearing Coq library to formalize and formally verify an exact, non-trivial timing property of these trajectories.

\begin{figure}[htbp]
\centering
  \includegraphics[scale=.35,trim={12 0 0 0}, clip]{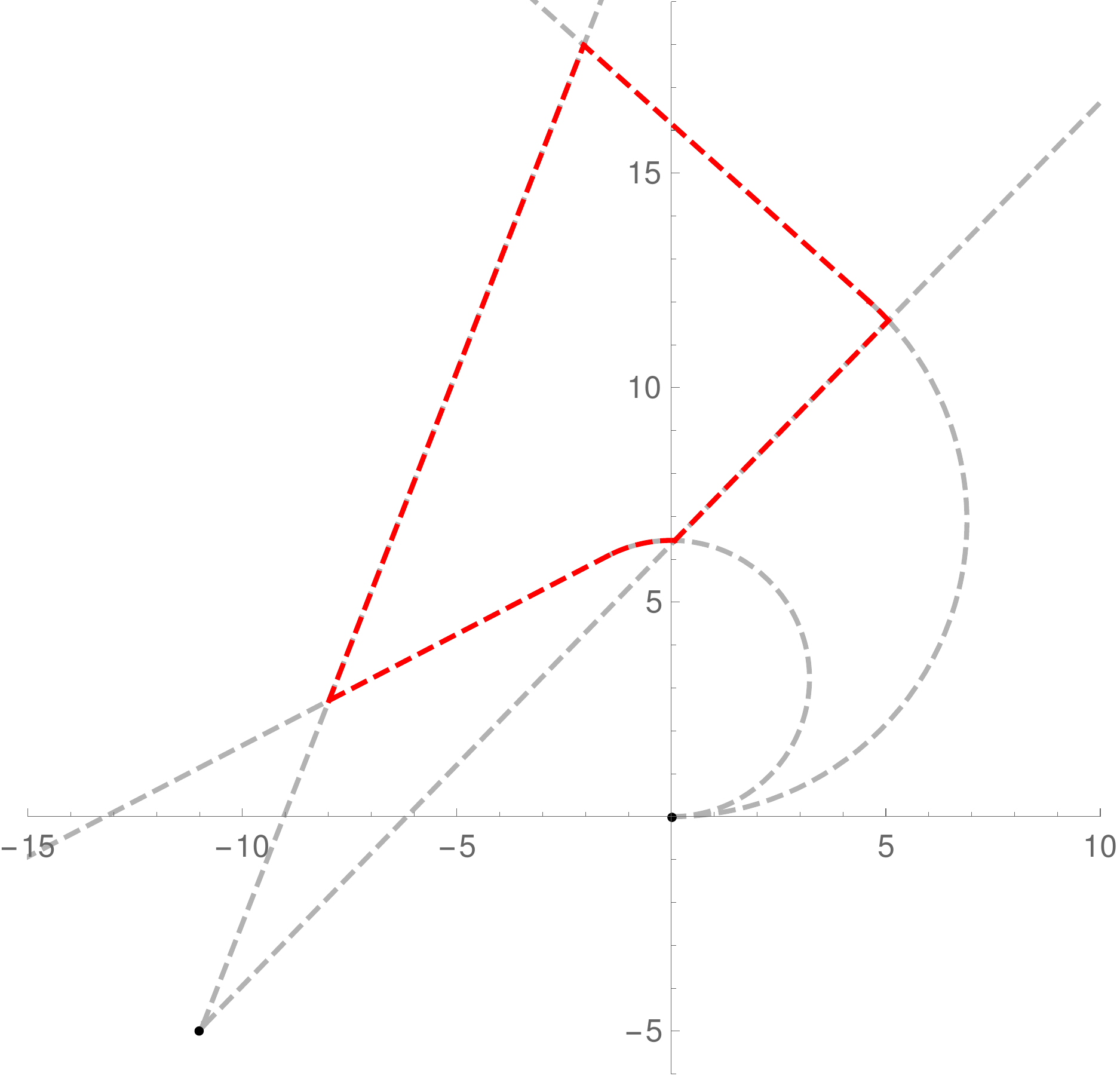}
  \raisebox{15pt}[0pt][0pt]{
    \makebox[0pt][l]{\hspace*{-170pt}
      \includegraphics[angle=-94,scale=.012]{figs/airplane_topview_CC.png}}}
  \raisebox{52pt}[0pt][0pt]{
    \makebox[0pt][l]{\hspace*{-100pt}
      \includegraphics[angle=-147,scale=.012]{figs/airplane_topview_CC.png}}}
  \caption{A two-aircraft encounter. Dashed lines show the edges of the reachable area for each
aircraft over all time, i.e. all possible future positions throughout the entire
encounter. The intersection of reachable areas---outlined in red---is an
overapproximation of the conflict area, containing all possible future
collisions.}\label{o0s}
\end{figure}

Having formalized turn-to-bearing paths, we need to reason about when (timing) and where (geometry) collisions might occur. The geometry of the reachable envelope for a turn is bounded by edges that are combinations of circular arcs and straight lines; the intersection of these areas can be computed in a straightforward manner. In other words, it is straightforward to overapproximate the conflict area as shown in Fig. \ref{o0s}.

In general, a collision can occur if there exists a point such that both aircraft can reach that point at the same time. This section considers the theorems and equations necessary to compute the earliest and latest possible times that aircraft can be at a \emph{given} point. Intuitively, the possible locations of an aircraft are contained within an area that moves over time, a propagating wave within the reachable envelope with a leading and lagging edge. This wave of position possibilities can be computed via piecewise equations. Although it is intractable to exhaustively search over all points for the exact collision times, these theorems will permit us to calculate a sound overapproximation for collisions in later sections.

\subsection{Pointwise Collision Timing}
\label{ss:ptwisecolltime}

\begin{figure*}[htb]
    \centering
    \begin{subfigure}[t]{0.32\textwidth}
        \centering
        \includegraphics[width=\textwidth]{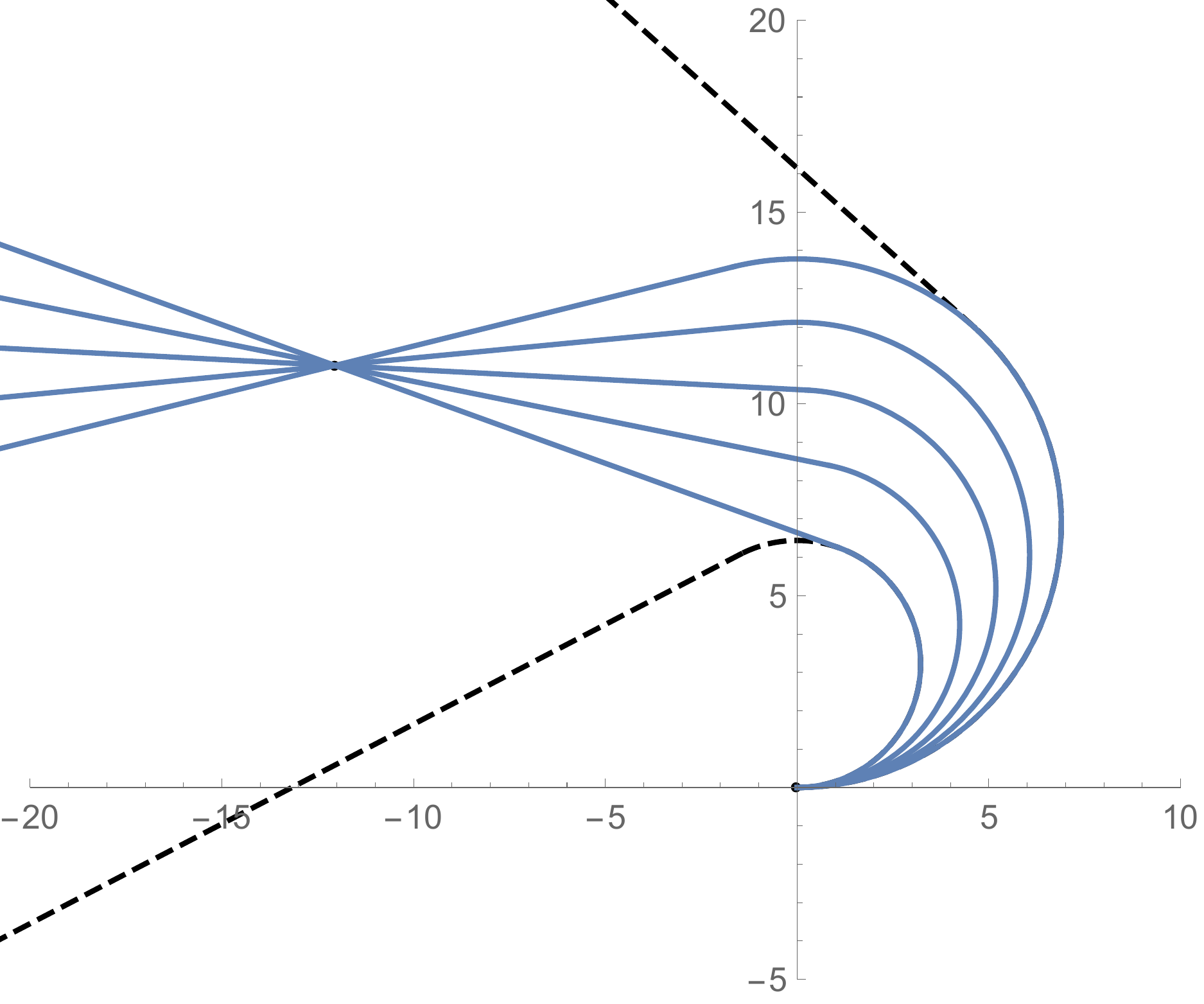}
        \caption{}
    \end{subfigure}
    \begin{subfigure}[t]{0.32\textwidth}
        \centering
        \includegraphics[width=\textwidth]{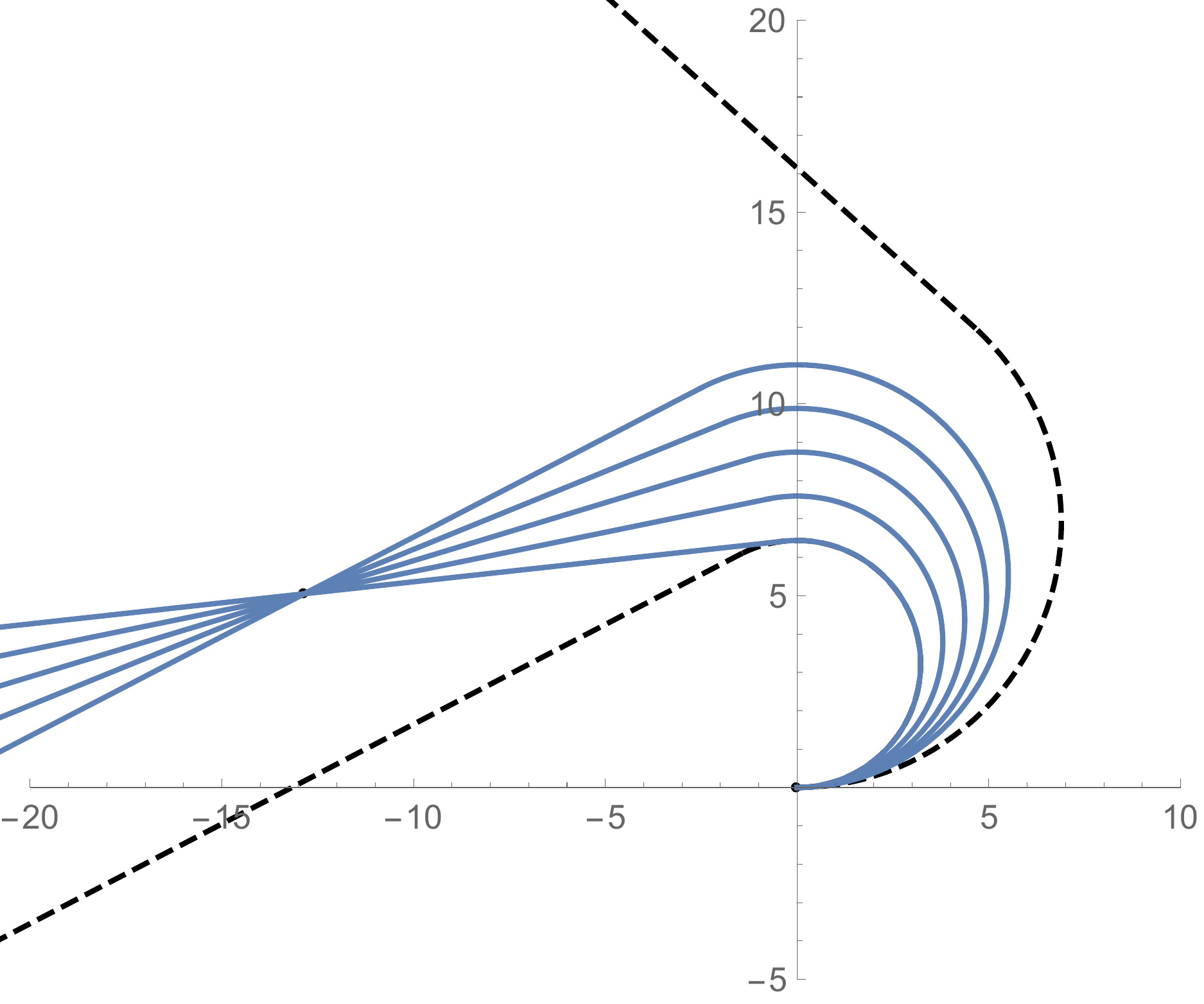}
        \caption{}
    \end{subfigure}
\caption{Paths from the set of possible
turn-to-bearing trajectories that reach two example points in
space. The edges of the reachable envelope for a non-deterministic left turn are shown as a set of dashed lines;
any point in the reachable envelope is reachable via these kinematics.}
\label{fig:reaching}
\end{figure*}

We define the reachable envelope
\begin{equation}
	E=\{p\mid\exists\ (\theta_c,r, s, u), \ \chi(r, \theta_c, s)
  \land u>0\land J(u)=p\}
\end{equation}
for a
vehicle to be the set of points that are reachable over the range of
possible future trajectories.  For any point in the
reachable envelope $p\in E$, there is a set of trajectories
$\mathcal{T}(p)=\{J(\cdot) \mid
	\exists\ (\theta_c,r, s, u), \ \chi(r, \theta_c, s)
  \land u>0\land
  J(u)=p\}$
that can
reach that point.
  Each trajectory $J\in\mathcal{T}(p)$ corresponds
with a
different choice of radius and final bearing (which determine the
path), and future ground speed $s(t)$.  Figs.~\ref{fig:reaching}a and b 
illustrate two different points in the reachable envelope of the
ownship from Fig. \ref{fig:ottb} and possible paths taken from the 
family of trajectories that could reach each point.

There is a corresponding set of arrival
times $I(p) = \left\{ t_a \mid J\in \mathcal{T}(p)\land
J(t_a)=p\right\}$ at which a vehicle can arrive at
$p$. The earliest and latest arrival times
for a single vehicle at a point $p$ are then given by 
\begin{align}
    t_e(p) &= \inf I(p) \label{eq:te_p}\\
    t_l(p) &= \sup I(p). \label{eq:tx_p}
\end{align}

To determine whether a collision between two aircraft is possible, 
we must look at the earliest and latest arrival times for each aircraft 
at each point in its reachable envelope. 
To organize this analysis, we first define four logical predicates
that express whether the earliest and latest arrival time at point $p$
in the reachable area occur when the other vehicle may also be located
at that point. Each time variable $t$ in the subsequent equations has
a subscript indicating whether the time is the earliest possible ($_e$) or
latest possible ($_l$) time of arrival, and a superscript indicating
which aircraft timing is referenced, $^i$ for intruder or $^o$ for
ownship.
\begin{align}
W^i_e(p) & = t^o_e(p) \leq t^i_e(p) \leq t^o_l(p) \\
W^i_l(p) & = t^o_e(p) \leq t^i_l(p) \leq t^o_l(p) \\
W^o_e(p) & = t^i_e(p) \leq t^o_e(p) \leq t^i_l(p) \\
W^o_l(p) & = t^i_e(p) \leq t^o_l(p) \leq t^i_l(p)
\end{align}
We combine these to define two predicates to evaluate safety, one
using the earliest arrival time, and the other using the latest
arrival time,
\begin{align}
  W_e(p) & = W^i_e(p) \lor W^o_e(p) \\
  W_l(p) & = W^i_l(p) \lor W^o_l(p)
\end{align}
For two aircraft we define a conflict area $C=E^o\cap E^i$
to reflect the geometry of the intersection of future paths without
timing considerations. We prove:
\begin{theorem}[Leading Lagging Equivalence]
  For all points $p\in C$, the predicates $W_e(p) = W_l(p)$ are equal, 
  so we can drop the subscript.
\end{theorem}

\begin{theorem}[Pointwise Safety]
  $W(p)$ correctly establishes safety at point $p$: when it is true,
  there exist circumstances that lead to collision at $p$, and when it is not
  there are no circumstances that lead to collision at $p$.
\end{theorem}

\begin{theorem}[Collision Timing]\label{thm:collision_timing}
  For each point $p\in C\land W(p)$, a
  collision may only occur in the time interval
  $[\max\left(t^i_e(p),t^o_e(p)\right),\min\left(t^i_l(p),t^o_l(p)\right)]$,
  and under the assumptions, no collision may occur outside this time interval.
\end{theorem}

We can directly relate timing of a trajectory between two points to
the range of path lengths for different possible paths connecting the
points. The earliest arrival time to reach a point $p$, $t_e(p)$ is
achieved by the trajectory following the shortest path and the highest
ground speed, i.e.  $\inf I(p) = \frac{d_{\min}(p)}{\sb}$, where
$d_{\min}(p)$ is the length of the shortest path from the starting
point to $p$. The latest arrival time $t_l(p)$ is achieved by the trajectory
following the longest path with the slowest ground speed, i.e.
$\sup I(p) = \frac{d_{\max}(p)}{\sa}$, where $d_{\max}(p)$ is the
length of longest path from the starting point to $p$. In this way, we
convert the problem of computing collision timing into a problem
computing the range of possible path lengths between two points.

\subsection{Path Length Properties}

We can define a function that computes the length of the path for a
deterministic, left-turning turn-to-bearing trajectory starting from
the origin with orientation $\theta_0=0$, passing through $(x,y)$ with
orientation $\theta$, using a turn of radius $r$:
\begin{equation}
  L(x,y,\theta,r) = r\theta + \left\lVert (x,y) -
                    r (\sin\theta,1 -\cos\theta) \right\rVert.
\end{equation}
As discussed in Sec~\ref{sec:ttb_path_properties}, turn-to-bearing kinematics
constrain the parameters for $L$,
i.e., its arguments cannot all be chosen independently. Assume we fix
the point we wish to reach, $(x,y)$. We can independently choose the angle
of approach $\theta$ to the final point, and that determines the
turn radius of the maneuver. Alternatively, we can choose the radius
of the turn, and compute the angle of approach to the point.

A central insight here is that for paths with the same starting and
ending points, the path with a larger angle of approach will have a
larger radius; and the path with a larger radius will be longer.  More
precisely:
\begin{theorem}[Approach angle orders \turntobearing{} path radii]
Given two turn-to-bearing paths, $(r_1,\theta_1)$ and $(r_2,\theta_2)$
that pass through the same point $(x,y)$, if $\theta_1>\theta_2>0$, then the
radius of the first path $r_1$ is longer than the radius of the second
path $r_2$, i.e. $r_1 > r_2$:
\begin{equation}
(\theta_1>\theta_2>0) \rightarrow R(x,y,\theta_1) > R(x,y,\theta_2)
\end{equation}
\label{thm:radius-ordering}
\end{theorem}

\begin{theorem}[Radius orders turn-to-bearing path lengths]
Given two turn-to-bearing paths, $(r_1,\theta_1)$ and $(r_2,\theta_2)$
that pass through the same point $(x,y)$, if $r_1>r_2>0$,
then the first path length $L_1$ is greater than the second path
length $L_2$, i.e. $L_1 > L_2$:
\begin{equation}
\begin{split}
    & 
    (r_1>r_2>0) \rightarrow \\
    & \quad
L(x,y,\Theta(x,y,r_1),r_1) > L(x,y,\Theta(x,y,r_2),r_2)
\end{split}
\end{equation}
\label{thm:path-length-ordering}
\end{theorem}

\subsubsection{Maximum and minimum path lengths}

At each point in the reachable area, we can use the ordering of path
lengths implied by Thms. \ref{thm:radius-ordering} and
\ref{thm:path-length-ordering} to find the minimum and maximum length
path possible for uncertain turn-to-bearing motion constrained by
non-deterministic bounds.

\begin{theorem}[Minimum bearing-constrained path length]
  For turn-to-bearing kinematics, given interval constraints on final
  bearing $[\tha, \thb]$ and turn radius
  $[\ra,\rb]$ where $0 < \ra$ and $0 < \tha$,
  and a reachable point $(x,y)$, the minimum path length is given by Eq.~\eqref{eq:min1}.
 \label{thm:minlen}
\end{theorem}

\begin{theorem}[Maximum bearing-constrained path length]
  For turn-to-bearing kinematics, given interval constraints on final bearing
  $[\tha, \thb]$ and turn radius
  $[\ra,\rb]$ where $0 < \ra$ and $0 <
  \tha$, and a reachable point $(x,y)$, the maximum
  path length is given by Eq.~\eqref{eq:max1}.
\label{thm:maxlen}
\end{theorem}

\begin{table*}[hbtp] 
\begin{align}
  d_{\text{min}}(x,y) &=
  \begin{cases}
  L(x,y,\Theta(x,y,\ra),\ra) &
  \tha\le\Theta(x,y,\ra)\le\thb\\ 
  L(x,y,\tha,R(x,y,\tha)) & 
  (x^2+y^2>2\ra y\land
  (0\le y(\ra\le r_m\lor y=0)
  \land\tha<\theta_m)\lor \\
  & \ \ \ \ (
  y<0\land\theta_m<0)
  )\land\Theta(x,y,\ra)<\tha \\
  L(x,y,\theta_m,r_m) &
  \ra\le
  r_m\le\rb\land\theta_m\le\max(\tha,\Theta(x,y,\ra))
  \end{cases} \label{eq:min1}\\
  d_{\text{max}}(x,y) &=
  \begin{cases}
    L(x,y,\Theta(x,y,r_\beta), \rb) &
    x^2+y^2>2\rb y\land
                            \Theta(x,y,\rb)\le\thb\\ 
    L(x,y,\thb,R(x,y,\thb)) & 
    (x^2+y^2 >2\rb y\land\thb<\Theta(x,y,\rb))\lor
    (\ra\le r_m\le\rb\land\thb<\theta_m)%
\\
    L(x,y,\theta_m,r_m) &
    \ra\le
    r_m\le\rb\land\theta_m\le\thb
  \end{cases} \label{eq:max1}
\end{align}
\end{table*}

\subsubsection{Right and uncertain turns}

So far we have looked only at left turns, where the circle that
defines our turn radius is positioned to the left of the vehicle, and
the change in bearing is a relative angle in radians, positive
according to the usual counter-clockwise convention. For
non-deterministic left turns, $0 < \ra\leq \rb$ and $0 \leq
\tha \leq \thb$.

We can handle other types of turns via symmetry. For right turns, we
choose the convention of identifying turning trajectories using radii
with negative numbers, and giving relative bearing with negative
numbers as well. We describe non-deterministic right turns using
parameters such that $\ra\leq \rb < 0$ and $\tha \leq
\thb < 0$. For this convention, the path length for right
turns is given by:
\begin{equation}
L^{\text{right}}(x,y,\theta,r)=L(x,-y,-\theta,-r).
\end{equation}
The function that determines the maximum and minimum distance for
right turns is related to that for left turns in the following way:
\begin{multline}
d^{\text{right}}(x,y,\tha,\thb,\ra,\rb) = \\
d(x,-y,-\thb,-\tha,-\rb,-\ra)
\end{multline}
for both minimum and maximum distance.

We can compute the distances associated with non-deterministic forward
motion that might include either a left or a right turn, by requiring
$\rb< 0 < \ra$ and $\tha \leq 0 \leq \thb$. The
distance function then relates to the left and right distance
functions:
\begin{multline}
d^{\text{either}}(x,y,\tha,\thb,\ra,\rb) = \\
\begin{cases}
d(x,y,0,\thb,\ra,\infty) & y > 0 \\
d^{\text{right}}(x,y,\tha,0,-\infty,\rb) & y < 0 \\
x & y = 0
\end{cases}
\end{multline}

\subsection{Exact Timing Wavefront}
\label{sec:wavefront}
\begin{figure*}[htbp]
    \centering
    \begin{subfigure}[t]{0.32\textwidth}
        \centering
        \includegraphics[width=\textwidth]{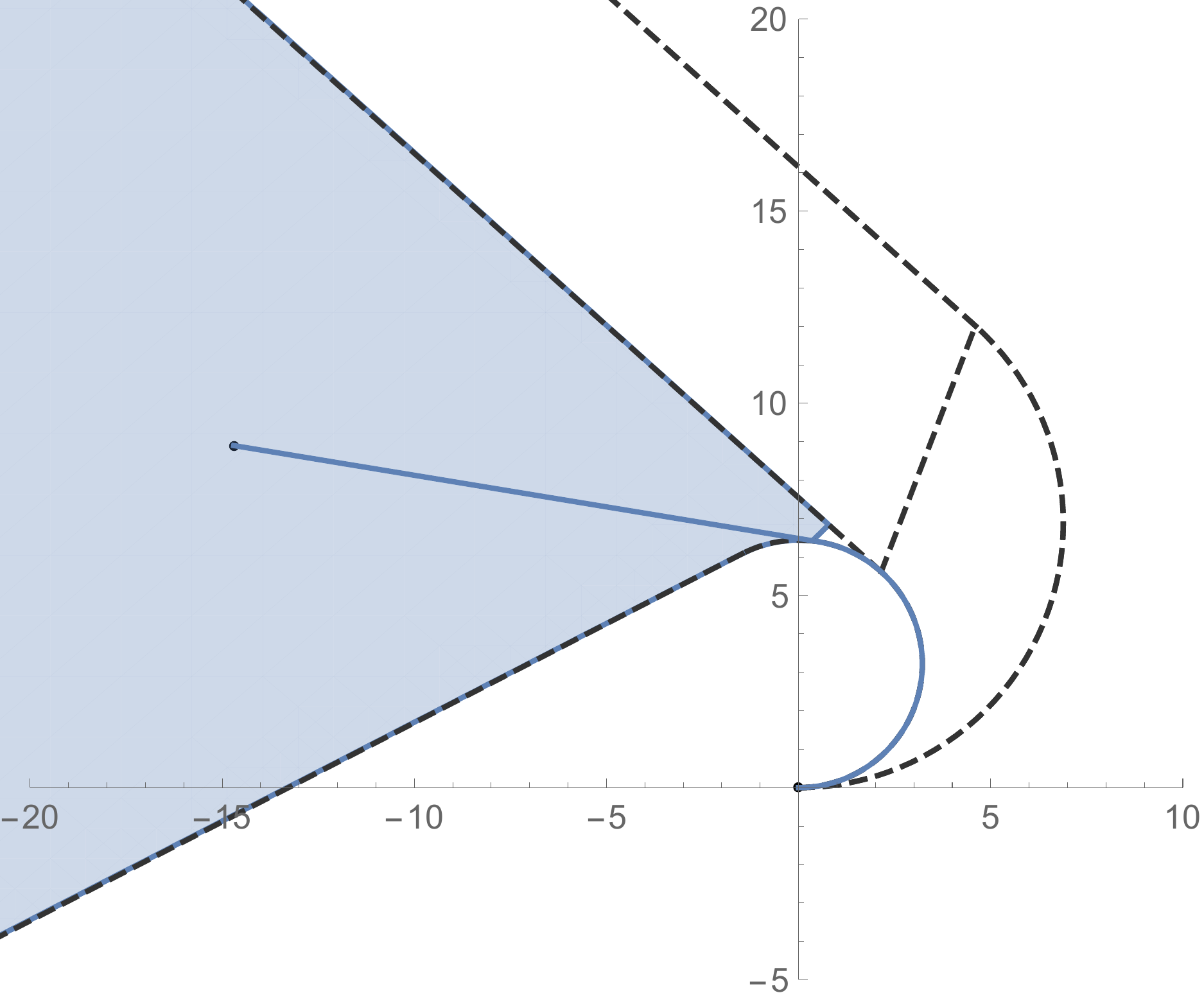}
        \caption{$r=\ra$}
        \label{mindist-circ}
    \end{subfigure}
    \begin{subfigure}[t]{0.32\textwidth}
        \centering
        \includegraphics[width=\textwidth]{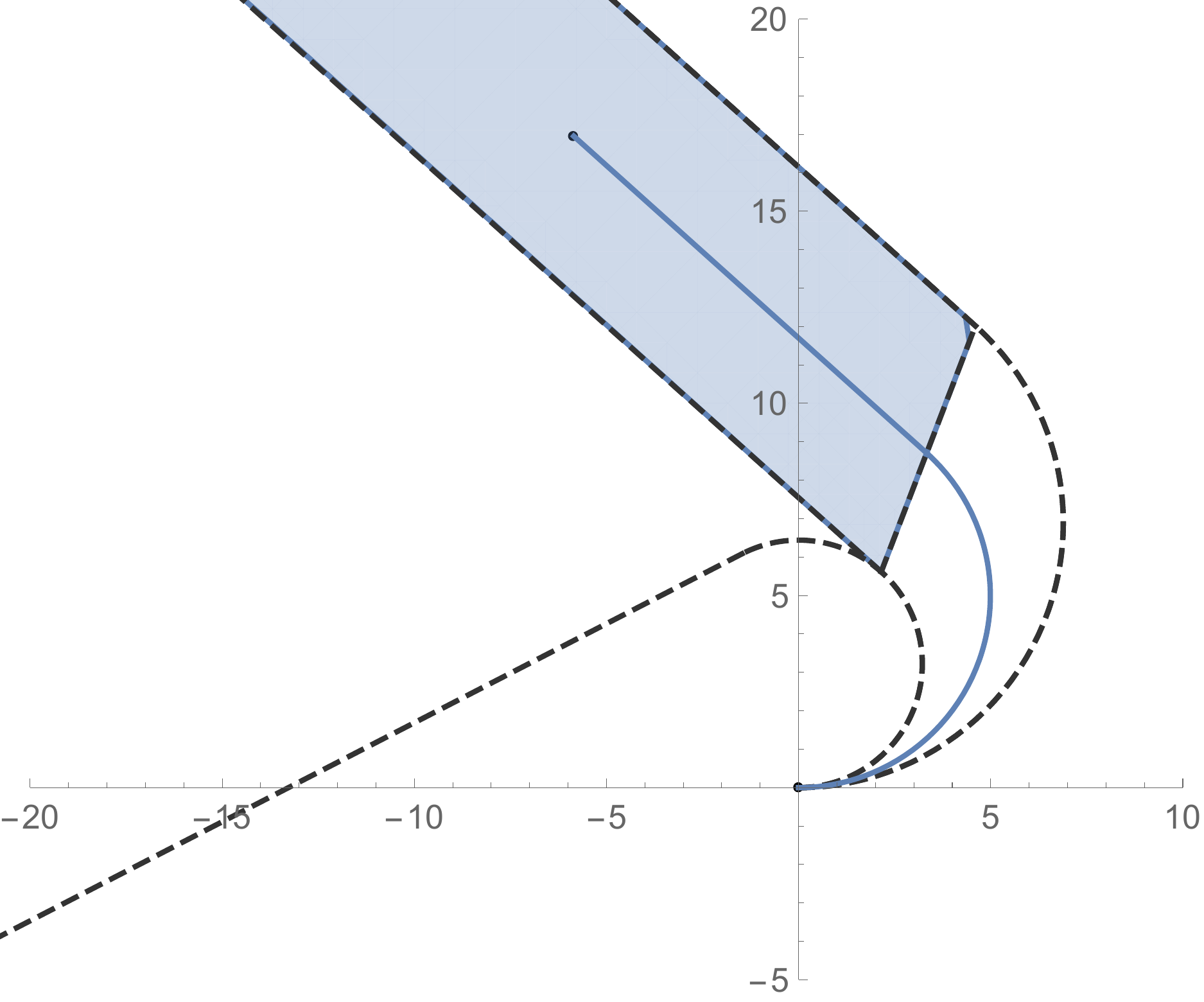}
        \caption{$\theta_c=\tha$}
        \label{mindist-lin}
    \end{subfigure}
    \begin{subfigure}[t]{0.32\textwidth}
        \centering
        \includegraphics[width=\textwidth]{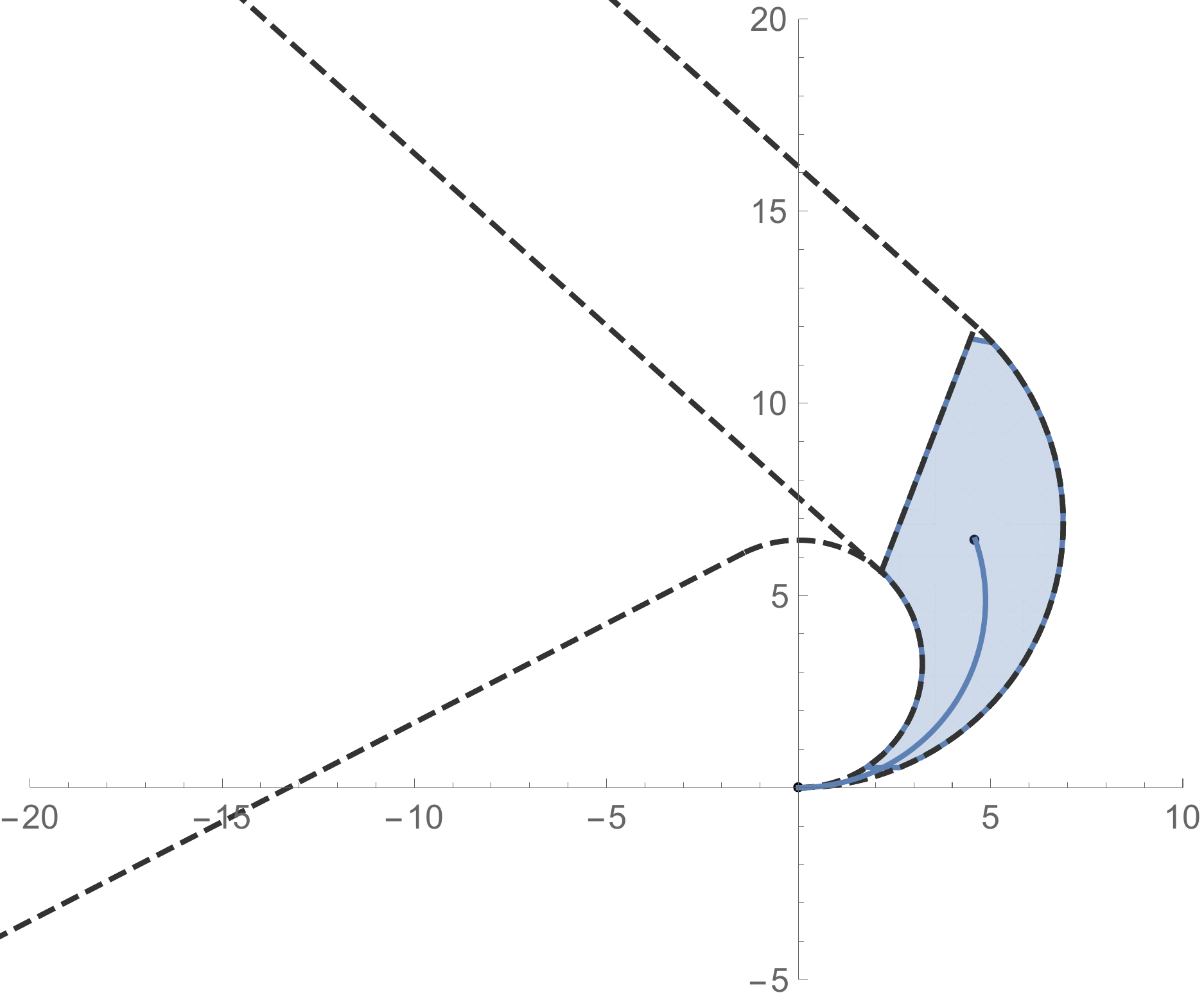}
        \caption{$r=r_m \land \theta_m\le\theta_c$}
    \label{mindist-turn}
    \end{subfigure}
\caption{Strategies described by Eq.~\eqref{eq:min1} to find the minimum
  distance turn-to-bearing trajectory from the origin to a particular point for each of
  three possible regions. Example trajectories illustrate the strategy
  for a single point indicated in each region. }
    \label{configA-min}
\end{figure*}

\begin{figure*}[htbp]
    \centering
    \begin{subfigure}[t]{0.32\textwidth}
        \centering
        \includegraphics[width=\textwidth]{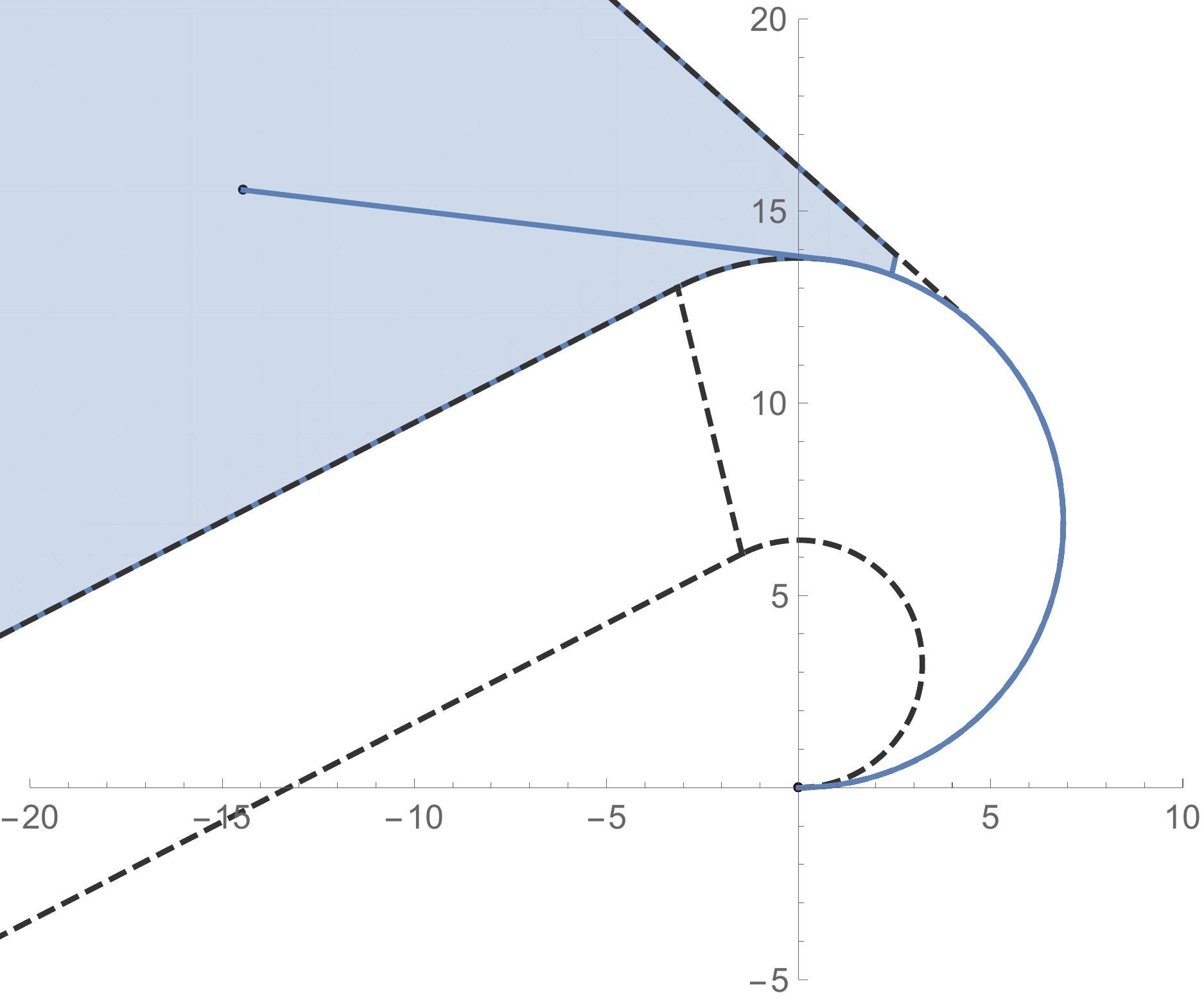}
        \caption{$r=\rb$}
        \label{maxdist-circ}
    \end{subfigure}
    \begin{subfigure}[t]{0.32\textwidth}
        \centering
        \includegraphics[width=\textwidth]{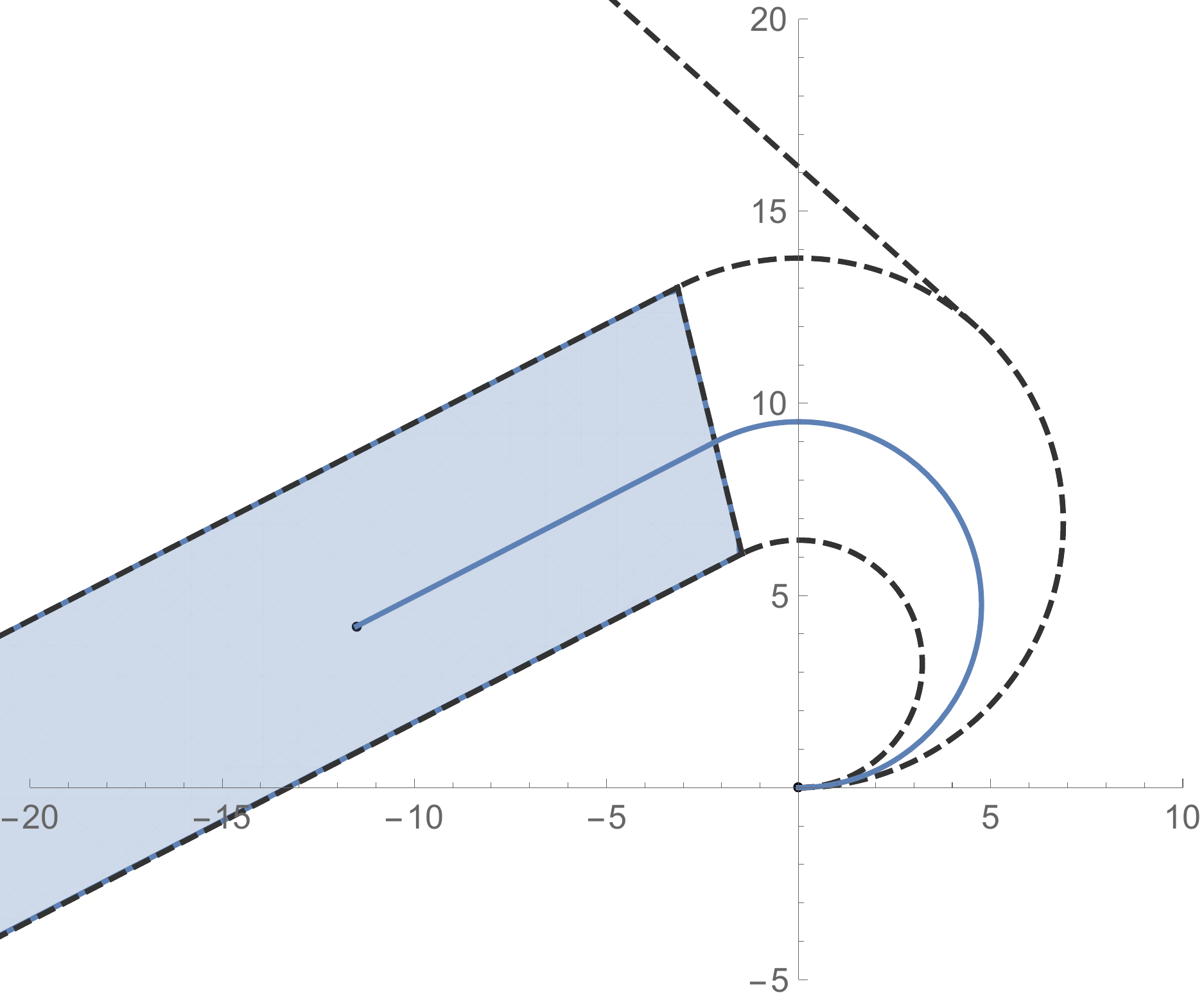}
        \caption{$\theta_c=\thb$}
        \label{maxdist-lin}
    \end{subfigure}
    \begin{subfigure}[t]{0.32\textwidth}
        \centering
        \includegraphics[width=\textwidth]{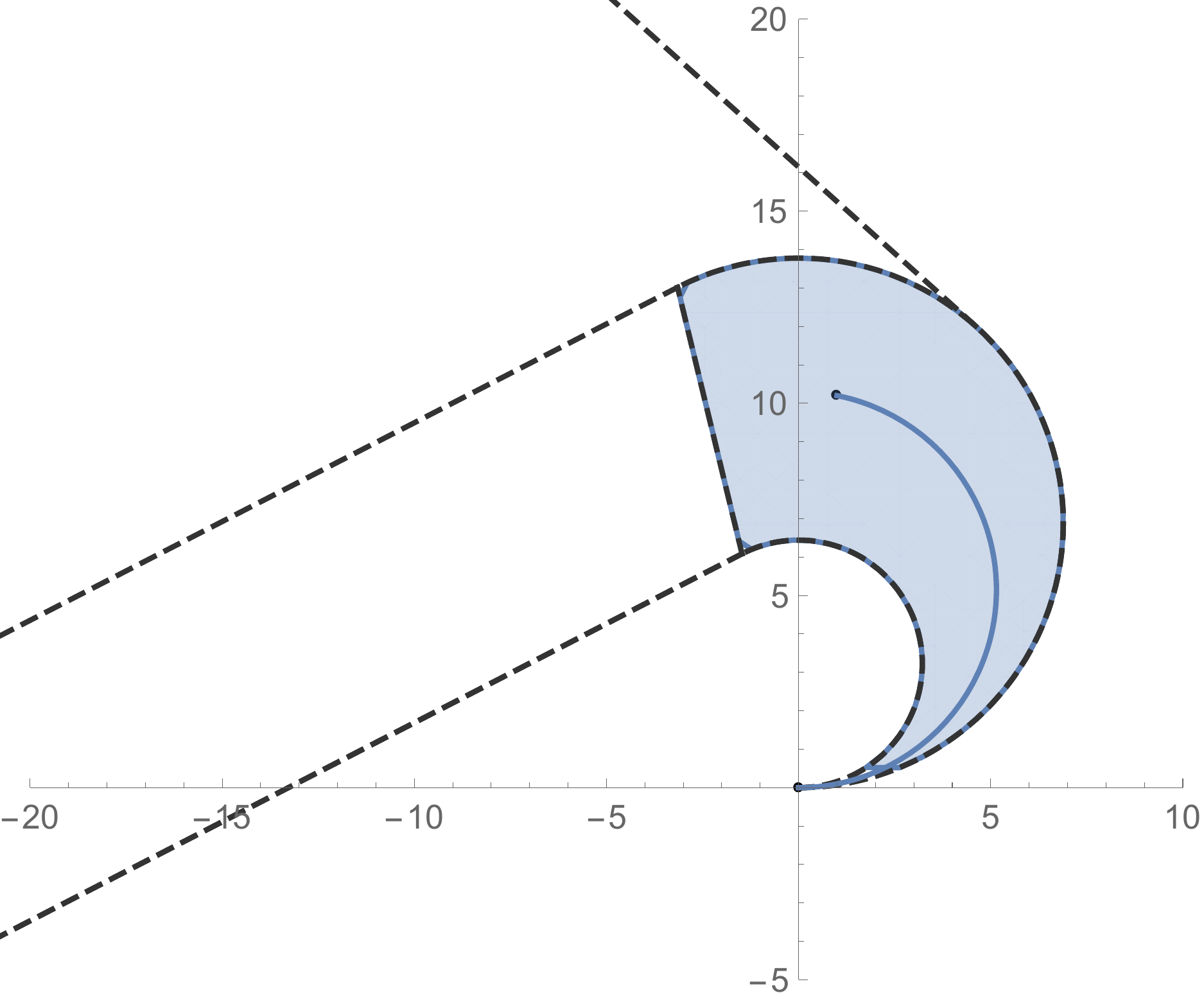}
        \caption{$r=r_m \land \theta_m\le\theta_c$}
        \label{maxdist-turn}
    \end{subfigure}
\caption{Strategies described by Eq.~\eqref{eq:max1} to find the maximum
  length turn-to-bearing trajectory from the origin to a particular point when the point is in each of
  three possible regions. Example trajectories illustrate the strategy
  for a single point indicated in each region.}
    \label{configA-max}
\end{figure*}

The observations in Thms. \ref{thm:minlen}--\ref{thm:maxlen} allow us
to subdivide the reachable envelope into different areas, using a
piecewise function to describe the timing. Figs.~\ref{configA-min}
and \ref{configA-max} illustrate, for a single vehicle and a
particular choice of parameters, the different strategies that
maximize and minimize path length, and the areas associated with each
strategy.  The bounding areas that enclose uniform strategies are
shown with dashed lines that illustrate the limits where each
strategy is appropriate for finding minimum and maximum length. For
turns with different parameters, these shapes change
accordingly.

This means that if we want to find the shortest and longest paths to a
point, we first consider paths with the smallest and greatest radii,
$\ra$ and $\rb$. Figs. \ref{configA-min}a and \ref{configA-max}a
illustrate individual trajectories that have minimum and maximum
length for our example maneuver, constructed by using the minimum and
maximum radii allowed.  For some points, the most extreme turns could
not produce trajectories that arrive at $p$, because the final
bearings required by such trajectories are outside the parameters 
set for the motion, or because the points are inside the turning
circle. Figs. \ref{configA-min}b and \ref{configA-max}b illustrate
individual trajectories that have minimum and maximum length for our
example maneuver in this case. These are constructed by choosing radii
that lead to most extreme values of bearing, so that the trajectory
both reaches $p$, and does so with an orientation that is allowed by
the parameters of our turn.  Finally, there are points in the
reachable envelope that are reachable as part of the initial turn. For
these points, this initial turn is the maximum-length path. If the
bearing at point $p$ is outside the allowable range, then this is also
the minimum-length path. Figs.~\ref{configA-min}c and
\ref{configA-max}c illustrate individual trajectories that have
minimum and maximum length for our example maneuver, which must be
constructed as circular arcs.

\begin{figure*}[htbp!]
    \centering
    \begin{subfigure}[t]{0.45\textwidth}
        \centering
        \includegraphics[scale=.33]{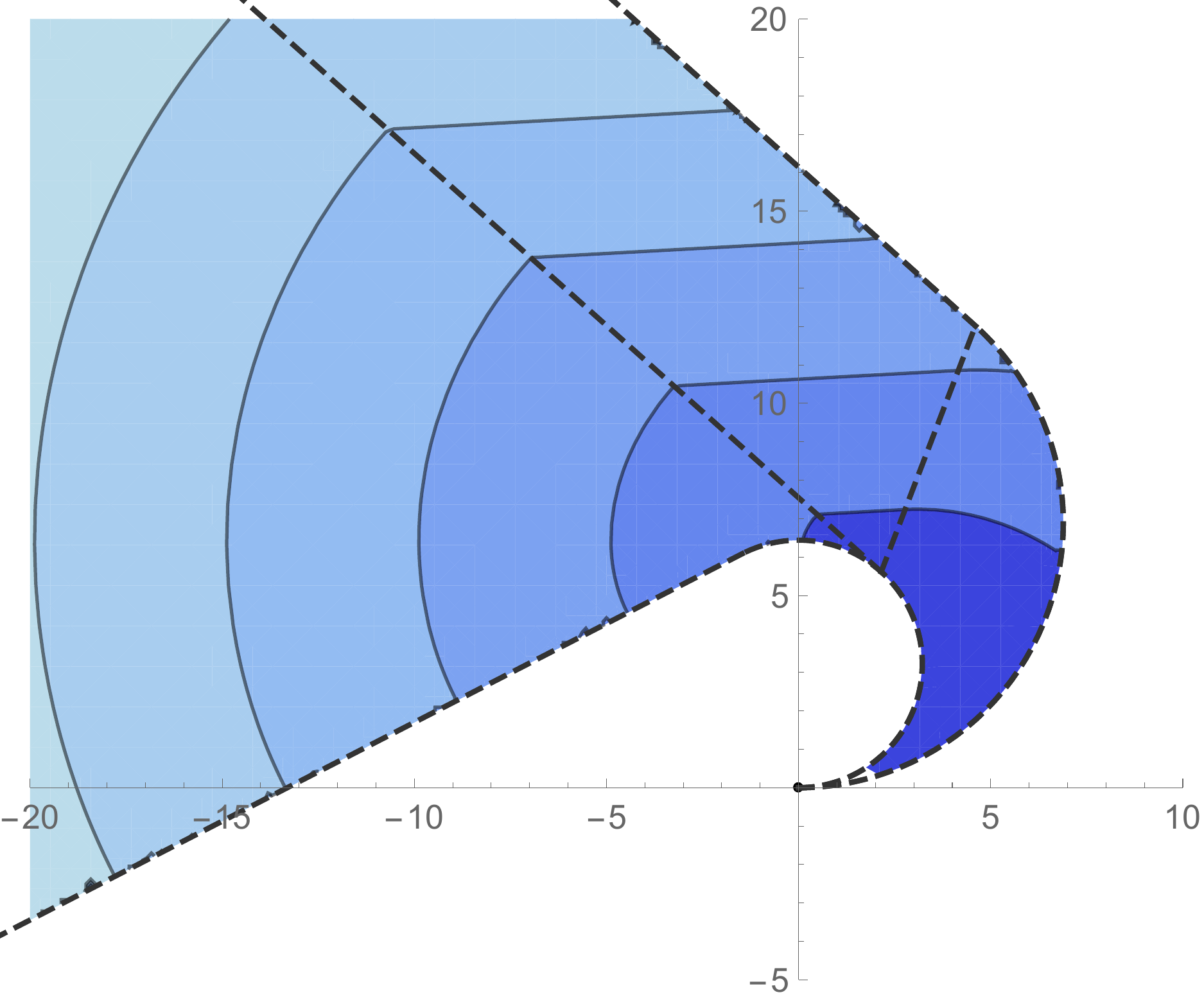}%
        \raisebox{40pt}[0pt][0pt]{
          \makebox[0pt][l]{
            \hspace*{-80pt}
            \includegraphics[angle=-147,scale=.013]{figs/airplane_topview_CC.png}}}
        \caption{Minimum time to arrive, $t_e(p)$.}
    \end{subfigure}
    \begin{subfigure}[t]{0.45\textwidth}
        \centering
        \includegraphics[scale=.33]{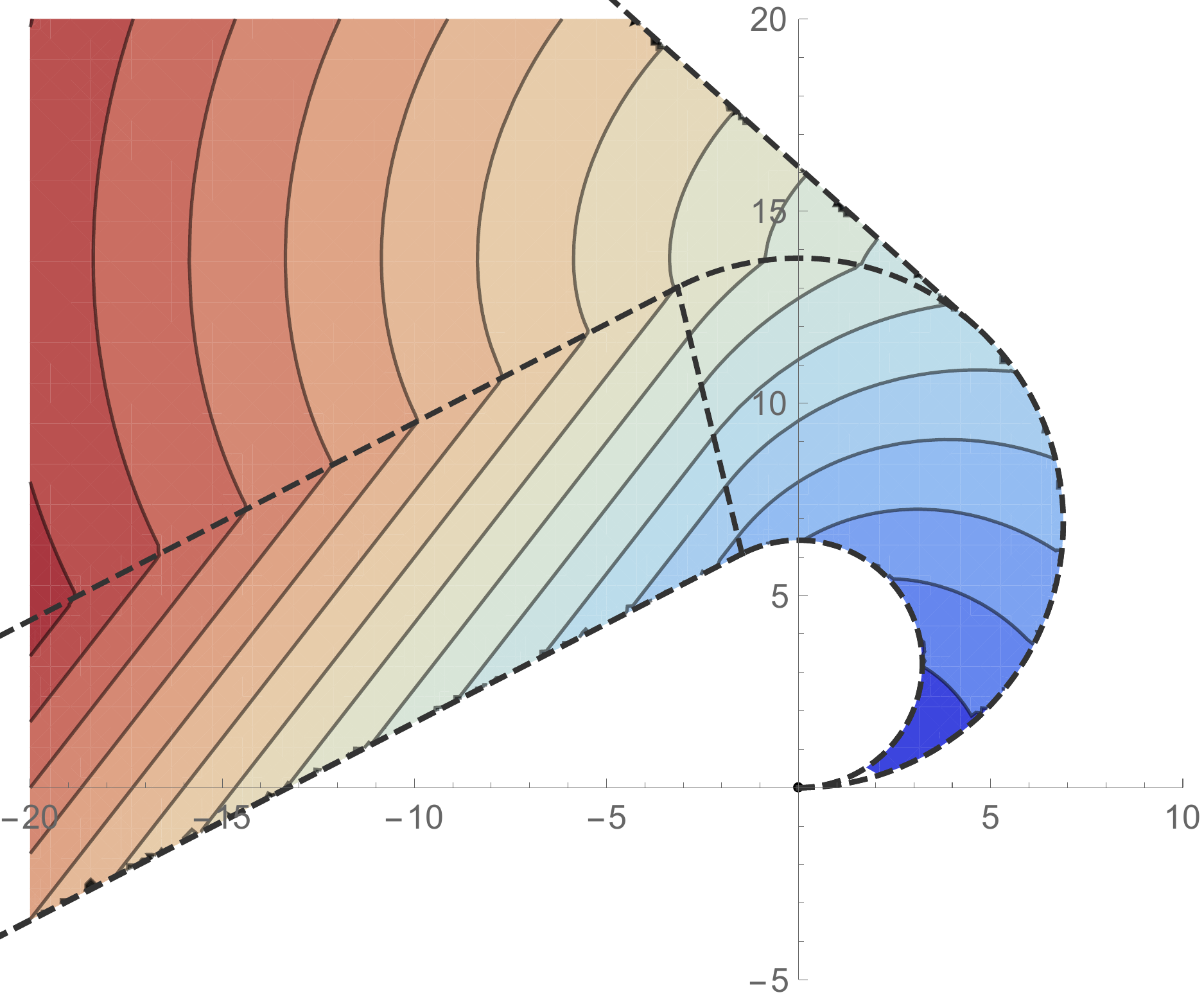}%
        \raisebox{40pt}[0pt][0pt]{
          \makebox[0pt][l]{
            \hspace*{-80pt}
            \includegraphics[angle=-147,scale=.013]{figs/airplane_topview_CC.png}}}
        \caption{Maximum time to arrive, $t_l(p)$.}
    \end{subfigure}
    \begin{subfigure}[t]{0.07\textwidth}
        \centering
        \includegraphics[scale=.4,trim={0 50 0 0},clip]{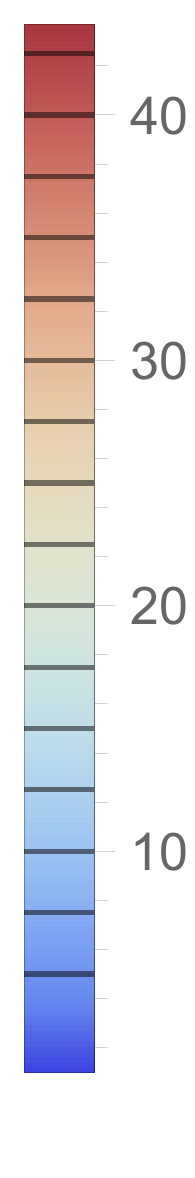}
    \end{subfigure}
    \caption{Contour plot describing timing for a left turn with
      parameters used in  Fig. \ref{fig:ottb}. Each $(x,y)$ position
      in the Cartesian plane is associated with a time to arrive at
      that position starting from $(0,0)$ with orientation
      $\theta_0=0$, following turn-to-bearing kinematics. }
\label{distance-traveled}
\end{figure*}

In Fig.~\ref{distance-traveled} we use the parameters of our example
maneuver in Fig.~\ref{fig:ottb}, combining all of the results from
Thms.~\ref{thm:minlen}--\ref{thm:maxlen} together into a single
contour plot of the earliest and latest times to reach each point in
the reachable envelope.  The contour lines plotted in
Figs. \ref{distance-traveled}a and b
can be thought of as the outer and
inner boundaries (respectively) of the irregular annulus at the
instant corresponding to the value of the contour. As time progresses,
this annulus expands, so we can treat this like a propagating wave,
an area that moves over time and encompasses all the possible positions that the vehicle may be in at each future moment.

The
contours of equal timing for the minimum and maximum arrival times
represent the shape of the leading and trailing edge of this wave,
respectively.  In addition to modeling and analyzing ranges of
possibilities for turn-to-bearing kinematics, we will find that adding
non-determinism also allows us to evaluate timing safety for small
perturbations of turn-to-bearing---types of motion whose combination
of trajectory and speed is sufficiently close, but not exactly the
same.

The theorems and equations within this section give exact, formally 
verified expressions that describe how to compute collision timing 
parameters $t_e(p)$ and $t_l(p)$ for two turning vehicles. Backed up by Thm.~\ref{thm:collision_timing}, these parameters describe the earliest 
and latest times that the vehicles may collide at a \emph{given} 
point $p$ that is in the conflict area $C$.

The next three sections that follow develop the mechanics for quantifying these calculations over \emph{all} the points $p$ in $C$.
Quantifying the timing computation over the points in $C$ is an important, practical step because it provides a parameter that can be used to fully characterize encounter timing.

%% file: happrox-introduction.tex
\label{sec:happrox-intro}

In this section, we develop an overall approach to quantifying the timing computations over the conflict area. We will describe the objective and intuition behind the quantification step, discuss the problems we encounter in more detail, and then create a framework for solving it. The following two sections fill in the remaining details of how we executed the approach.

To form some intuition about our algorithm,
recall the example two-aircraft encounter (Fig.~\ref{o0s}) where both aircraft
follow uncertain turn-to-bearing trajectories.  Dashed lines show the edges of
the reachable envelopes $E^o$ and $E^i$ that contain the positions of each
aircraft during the encounter, and illustrate how the conflict area $C$,
outlined in red, is found by taking the intersection of the envelopes. 
\begin{figure*}[htbp]
\centering
\hfil
\begin{subfigure}[t]{0.3\textwidth}
\centering
      \includegraphics[width=\textwidth,
      trim={12 0 0 0}, clip]{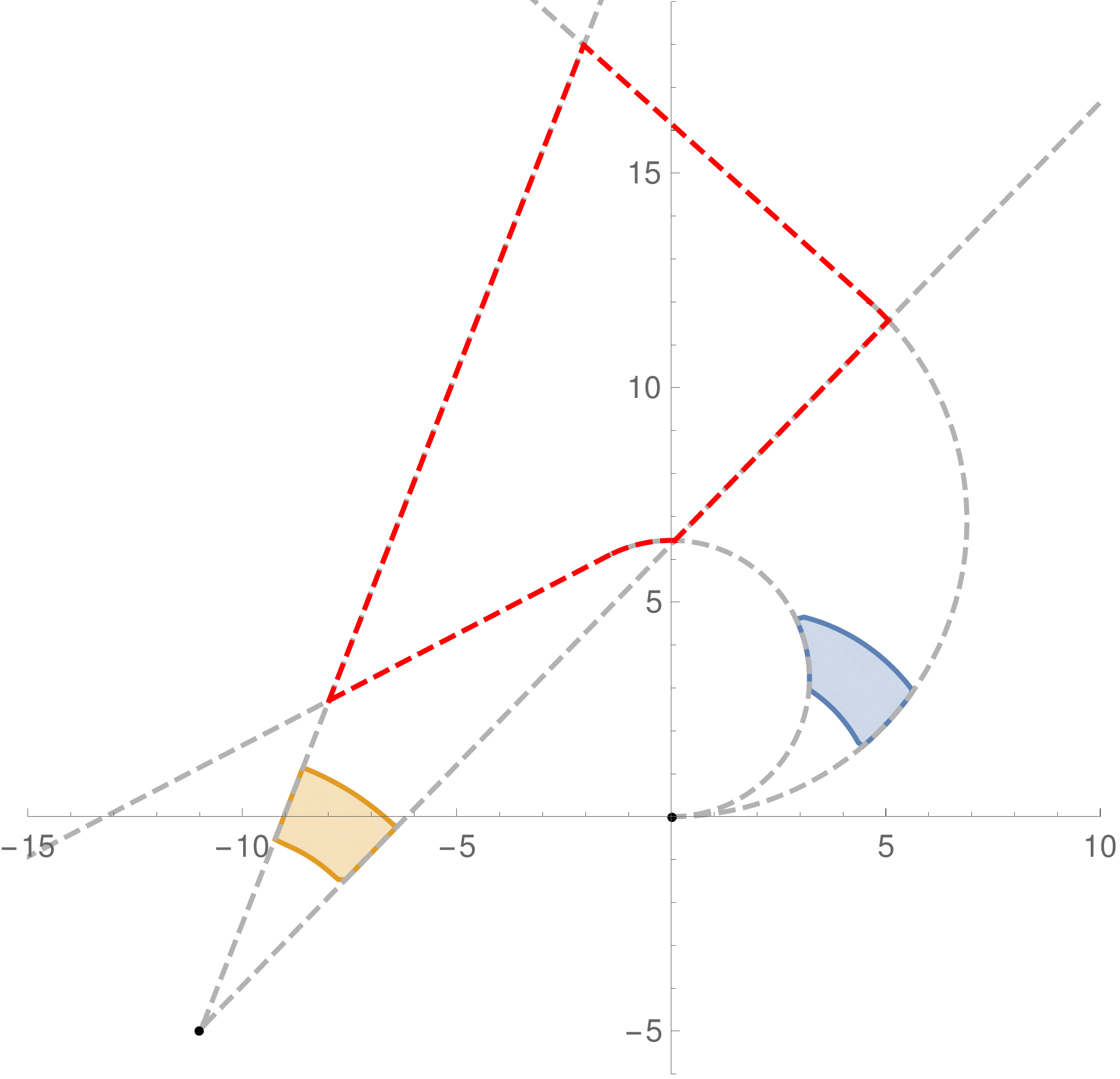}
\caption{$t=8$}  \label{o12s}
\end{subfigure} \hfil
\begin{subfigure}[t]{0.3\textwidth}
  \centering
      \includegraphics[width=\textwidth,
      trim={12 0 0 0}, clip]{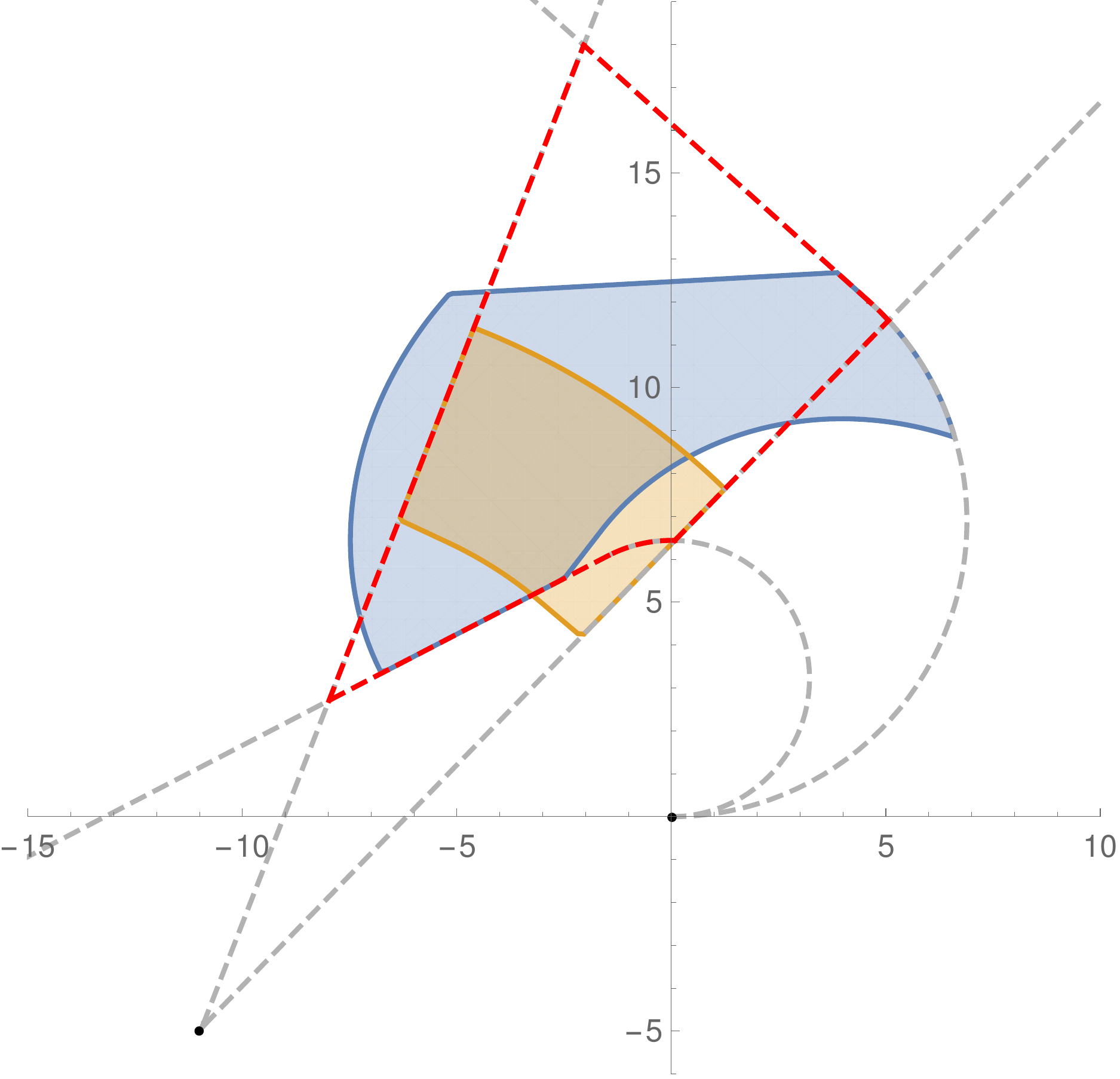}
  \caption{$t=16$}
  \label{o18s}
\end{subfigure}
\hfil
\begin{subfigure}[t]{0.3\textwidth}
\centering
      \includegraphics[width=\textwidth,
      trim={12 0 0 0}, clip]{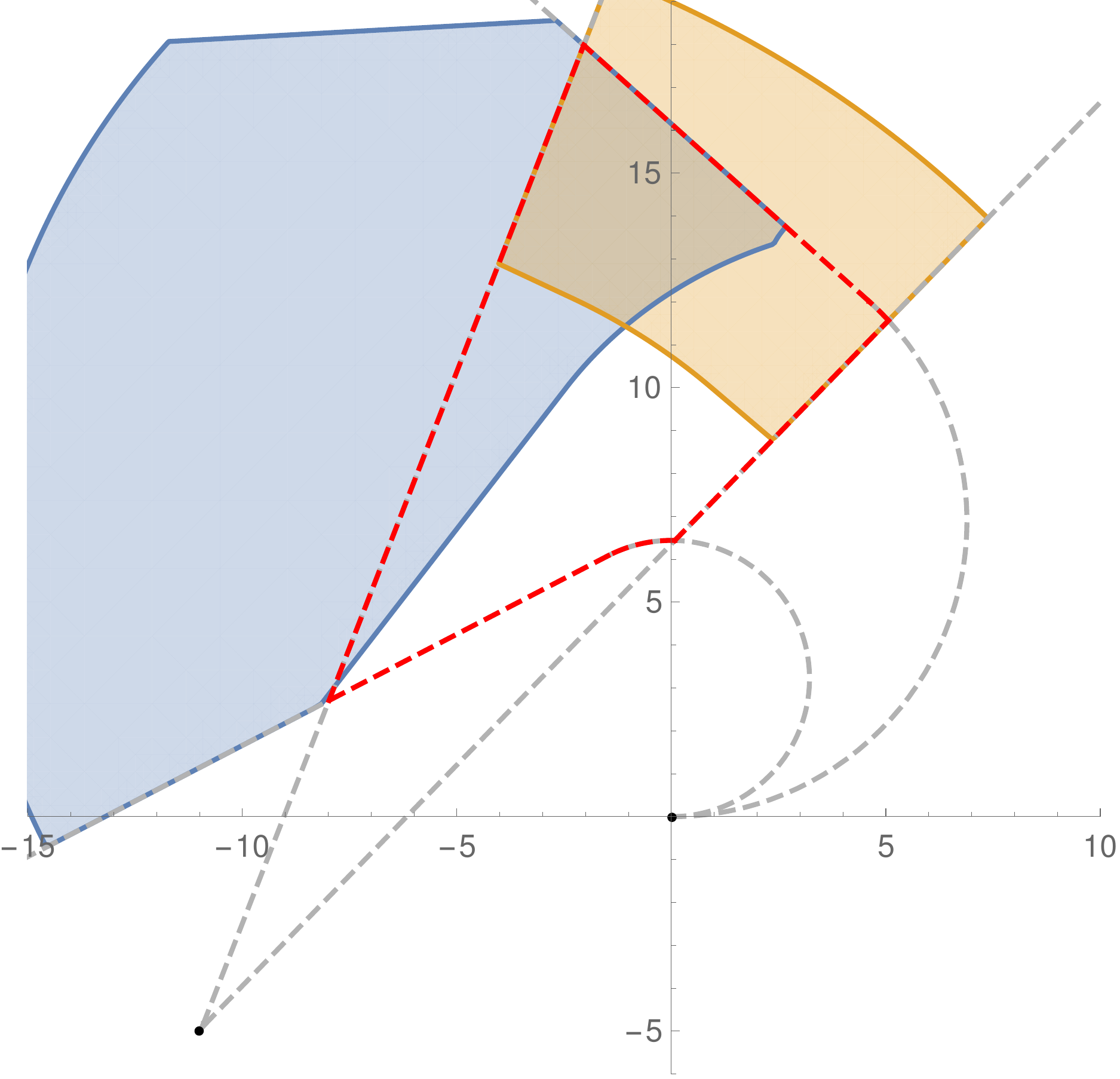}
\caption{$t=24$} \label{o24s}
\end{subfigure}
\caption{Timing for a single scenario where two vehicles will make
simultaneous, independent, non-deterministic turns with different
turn-to-bearing assumptions. The reachable areas for the vehicles are shown at
the moments in time indicated. The instantaneous reachable area is a
propagating wave that moves outwards over time, encompassing the all possible
positions that the vehicle may be located at each moment.}
\label{o}
\end{figure*}
We observe that the
earliest and latest times can be computed by appropriately tracking
the propagating timing wavefronts. To see how this works, refer to Fig.~\ref{o},
which shows a series of snapshots of the future of the encounter from
Fig.~\ref{o0s} (at times $t=8,16,$ and $24$) with shaded areas showing the set
of possible positions in which each aircraft can be found at that moment.
These moving areas are precisely the waves of possible positions described in
Sec.~\ref{sec:wavefront}. Their front and back edges are defined by the level
sets of the timing computation, i.e. the contours of
Fig.~\ref{distance-traveled}. Our objective is to solve for the time interval
during which these propagating regions overlap.

Writing down the expression for timing parameters that span the entire
encounter is straightforward. Recall from Thm.~\ref{thm:collision_timing} that
the earliest and latest collision times possible in an area $C$ are given by
\begin{align}
  \label{eq:te1}
t_e & =  \inf_{p\in C \land W(p)} \max\left(t^i_e(p),t^o_e(p)\right) \\
t_l & = \sup_{p\in C \land W(p)} \min\left(t^i_l(p),t^o_l(p)\right)
\label{eq:tx1}
\end{align}

From the relationship between the timing of a trajectory at a point and the
range of possible path lengths to arrive at that point, it suffices to analyze
the path lengths (quantified over the conflict region), instead of considering
timing directly (Sec.~\ref{ss:ptwisecolltime}). 

However, it is not obvious how to
solve for
$t_e$ and $t_l$ over every point in a region by using the minimum and maximum
path lengths because there are a number of problems. 
\begin{figure*}[htbp]
    \centering
    \begin{subfigure}[t]{0.32\textwidth}
        \centering
  \includegraphics[width=.98\textwidth]{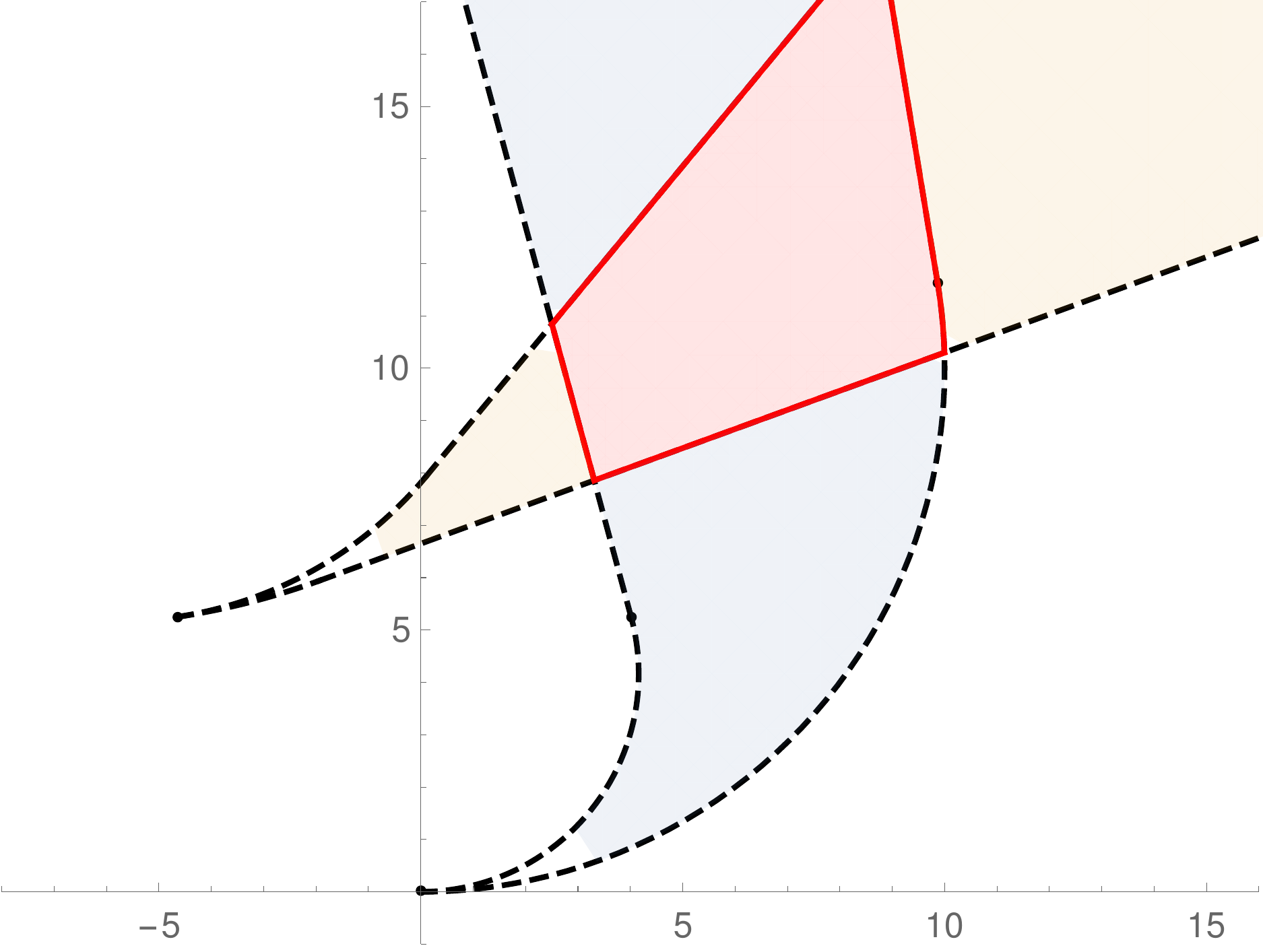}%
\raisebox{13pt}[00pt][0pt]{
  \makebox[0pt][l]{\hspace*{-114pt}\large \includegraphics[angle=-147,scale=.01]{figs/airplane_topview_CC.png}}}%
\raisebox{47pt}[00pt][0pt]{
  \makebox[0pt][l]{\hspace*{-148pt}\large \includegraphics[angle=-140,scale=.01]{figs/airplane_topview_CC.png}}}%
\caption{}
    \end{subfigure}
    \begin{subfigure}[t]{0.32\textwidth}
        \centering
  \includegraphics[width=\textwidth]{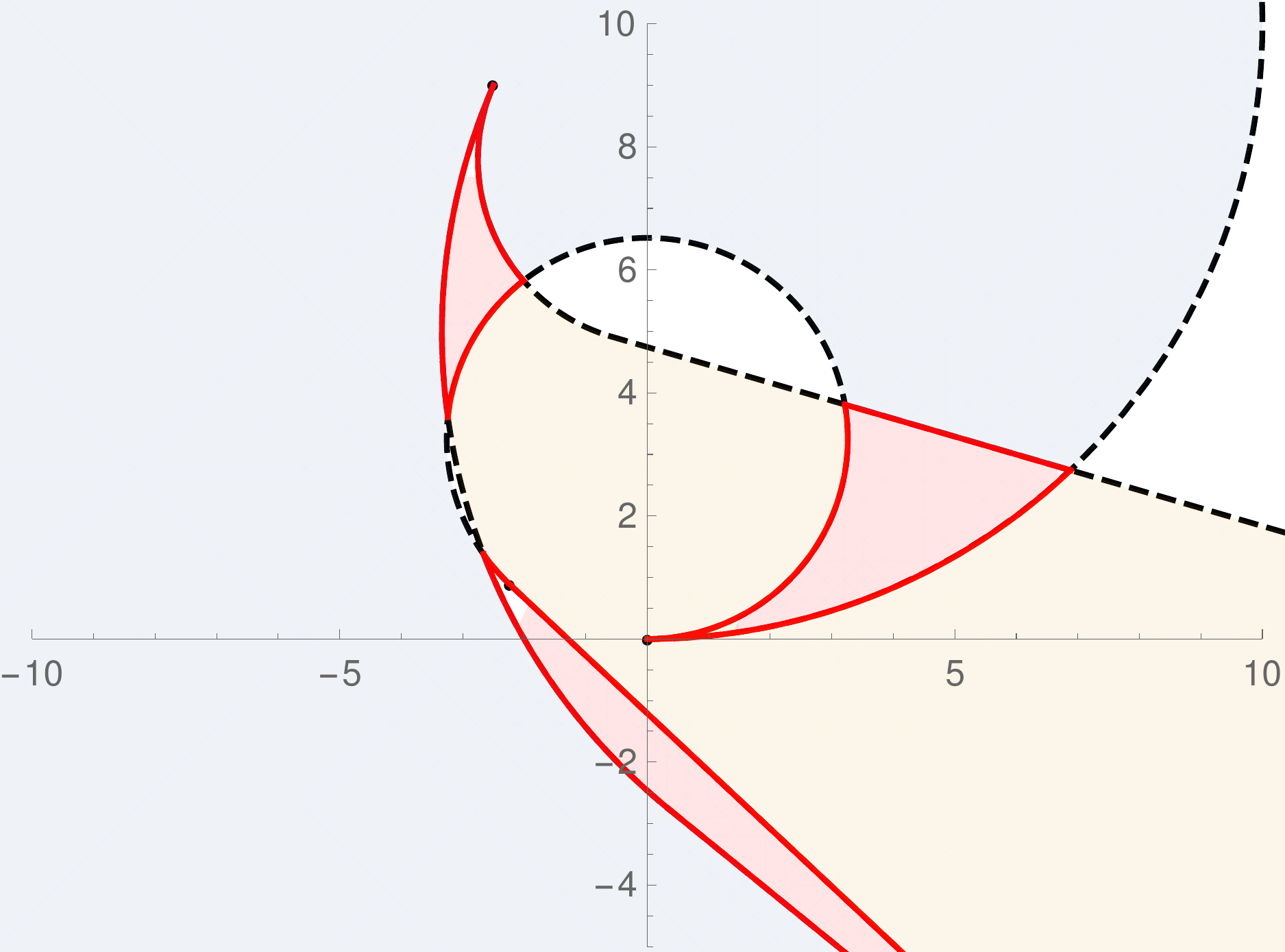}%
\raisebox{46pt}[00pt][0pt]{
  \makebox[0pt][l]{\hspace*{-90pt}\large \includegraphics[angle=-147,scale=.012]{figs/airplane_topview_CC.png}}}%
\raisebox{102pt}[00pt][0pt]{
  \makebox[0pt][l]{\hspace*{-108pt}\large \includegraphics[angle=97,scale=.012]{figs/airplane_topview_CC.png}}}%
        \caption{}
    \end{subfigure}
    \begin{subfigure}[t]{0.32\textwidth}
        \centering
      \includegraphics[width=\textwidth]{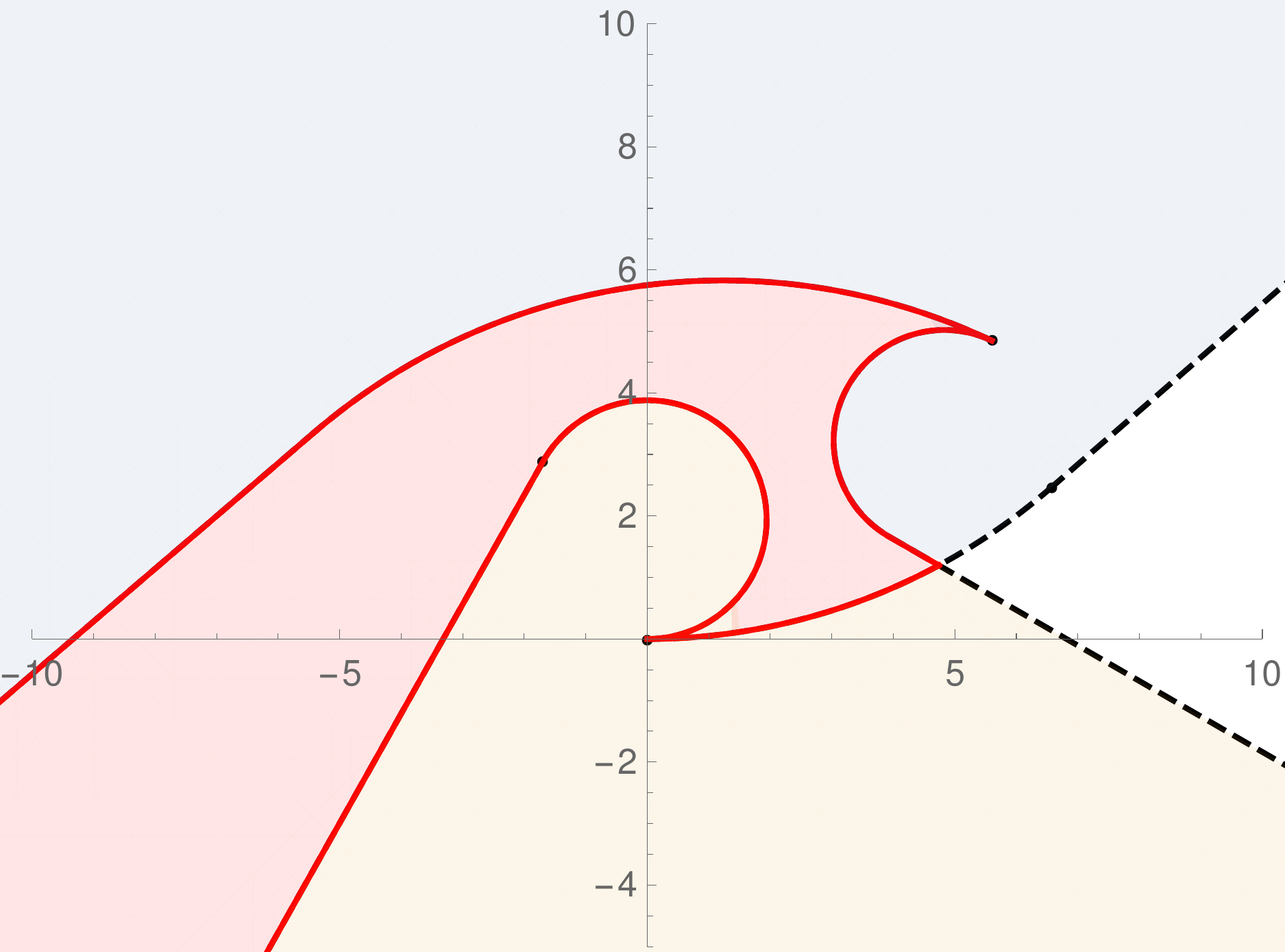}%
\raisebox{46pt}[00pt][0pt]{
  \makebox[0pt][l]{\hspace*{-90pt}\large \includegraphics[angle=-147,scale=.012]{figs/airplane_topview_CC.png}}}%
\raisebox{69pt}[00pt][0pt]{
  \makebox[0pt][l]{\hspace*{-46pt}\large \includegraphics[angle=8,scale=.012]{figs/airplane_topview_CC.png}}}%
        \caption{}
    \end{subfigure}
\caption{Geometry of three different scenarios where two vehicles will make
    simultaneous, independent, non-deterministic turns with different
    turn-to-bearing assumptions. The blue and yellow shading shows the
    reachable areas for the vehicles quantified over all time; it is equivalent
    to the union of instantaneous reachable areas shown in Fig. \ref{o} for
    each future moment. The conflict area, where collisions might occur, is the
    intersection of the reachable areas, shaded in red.}
\label{fig:conflict}
\end{figure*}
The first problem we encounter is that our path length expressions in Eqs. \eqref{eq:min1} and \eqref{eq:max1} are piecewise, so we need some strategy for treating each piece separately. The second problem is that the domain over which the path length expressions apply is not convex and thus makes optimization more challenging. Fig.~\ref{fig:conflict} illustrates this with several examples of conflict areas $C$ with different initial conditions.
The conflict areas over which we need to quantify our timing calculation are neither convex nor necessarily simply connected. The framework we develop in this section will address the first two problems directly.
The third problem is that the parameters $x$ and $y$ over which we are quantifying are in multiple places in the path length expressions, sometimes sprinkled within and between nested layers of transcendental functions. And it is not obvious how to enforce appropriate constraints on the domain of $x$ and $y$. The straightforward approach of sampling of the conflict region and checking appropriate constraints
would be both unsound (i.e. might indicate safety at times when collisions are
possible) and computationally expensive.
We discuss this problem further at the end of the section, and present a solution in Sections 6 and 7 that is sound and computationally efficient.

To address the first two problems above, we divide the reachable envelope (domain) into different areas, according to the expressions that compute the minimum and maximum length paths, and then subdivide these pieces further to account for the conflict area and the motion possibilities of the other vehicle. The result is a covering of polygons where each is convex, and contains only a single uniform path length expression corresponding to earliest and latest timing for each vehicle at each point within the domain.

%% file: happrox-approach.tex
\begin{figure}[htbp]
  \begin{center}
      \includegraphics[scale=.45]{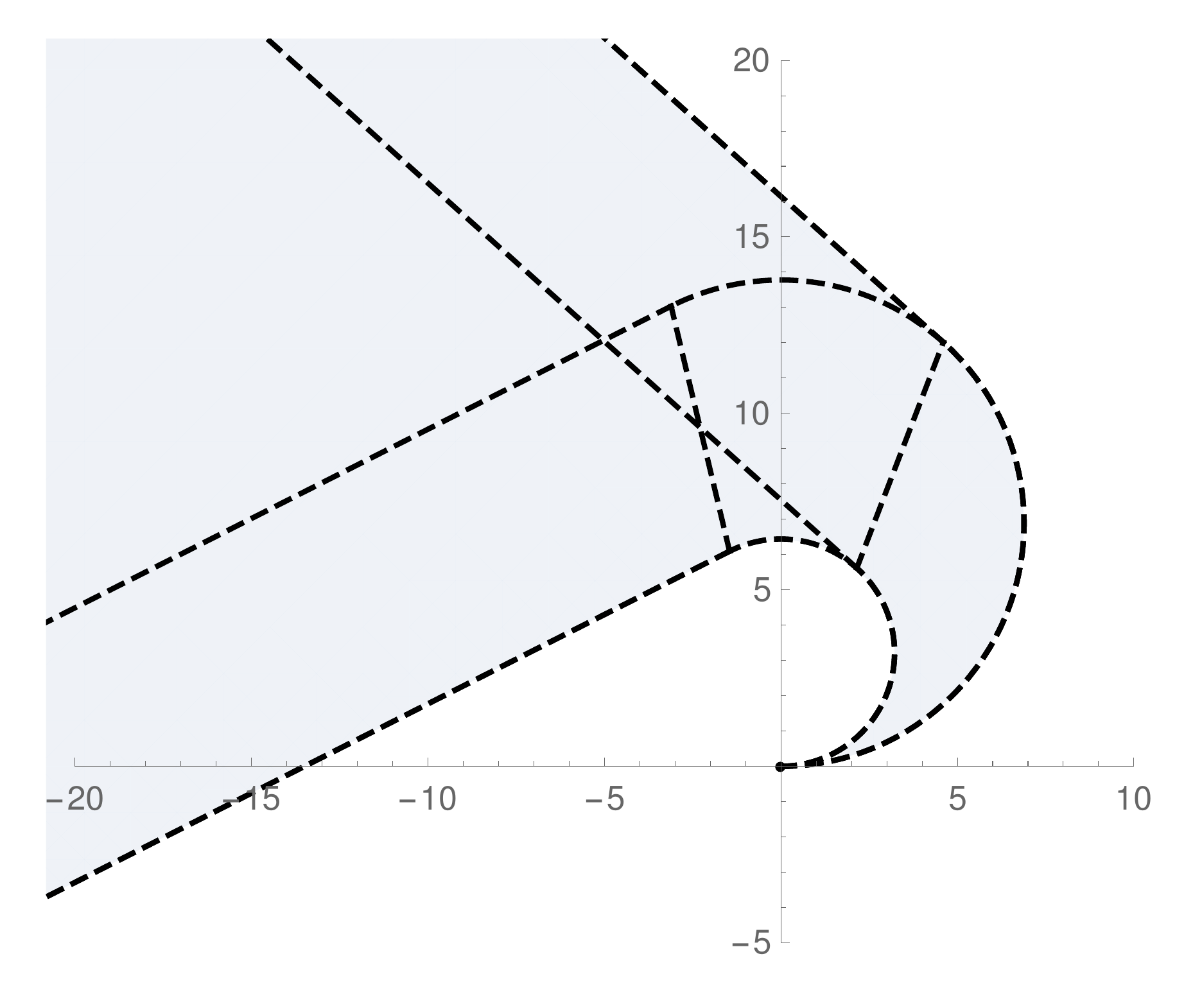}%
  \raisebox{58pt}[00pt][0pt]{
  \makebox[0pt][l]{
  \hspace*{-105pt}
  \includegraphics[angle=-147,scale=.015]{figs/airplane_topview_CC.png}}}%
  \raisebox{90pt}[00pt][0pt]{
  \makebox[0pt][l]{\hspace*{-170pt}\large G}}%
    \raisebox{140pt}[00pt][0pt]{
  \makebox[0pt][l]{\hspace*{-200pt}\large F}}%
    \raisebox{160pt}[00pt][0pt]{
  \makebox[0pt][l]{\hspace*{-140pt}\large E}}%
    \raisebox{129pt}[00pt][0pt]{
  \makebox[0pt][l]{\hspace*{-130pt}\large D}}%
    \raisebox{98pt}[00pt][0pt]{
  \makebox[0pt][l]{\hspace*{-115pt}\large C}}%
      \raisebox{120pt}[00pt][0pt]{
  \makebox[0pt][l]{\hspace*{-105pt}\large B}}%
      \raisebox{90pt}[00pt][0pt]{
  \makebox[0pt][l]{\hspace*{-80pt}\large A}}%
  \end{center}
\caption{A plot of four trajectories, with initial position $p_0=(0,0)$ at the
    origin, and initial orientation $\theta_0=0$ along the $x$-axis. These
    correspond to the
  limiting radii and bearings allowed for a non-deterministic left
  turn with $[\theta_\alpha,\theta_\beta]=[2.41,3.62]$, and
  $[r_{\text{min}},r_{\text{max}}]=[3.22,6.89]$. All radii and
  bearings in between these limits are also possible.
  }  \label{fig:configA}
\end{figure}

First, we create a sound overapproximation for the geometry of the conflict
area using a set of convex polygonal sets in which the four different
expressions for the timing computation (i.e. front and back edges of the waves
for both vehicles) each belong to only one piece of the timing equations, 
Eqs.~\eqref{eq:min1} and \eqref{eq:max1}. 
For each vehicle with reachable envelope $E$, we create partitions using the
domains $\{D_j\}$ of the timing equations matching the parameters of its
kinematics, i.e. $F_j=\{ D_j\cap E, E\setminus D_j\}$. Then we create a refined
partition $R$ of the $F_j$ domains that satisfies $\forall j, R\le F_j$. Here
the $\le$ operator applied to partitions indicates that the left hand side
partition is a refinement of the right hand side. 
Fig.~\ref{fig:configA} shows what $R$ looks like for the maneuver in Fig.~\ref{fig:ottb}. 

We then find a partition $P$, satisfying $P\le (R^i\cap E^o)$ and $P\le
(R^o\cap E^i)$, that refines the partitions produced by the intersection of the
reachable area partitions with the conflict area $C$.
To create a sound polygonal overapproximation for this final partition $P$, we
can follow the procedure above using polygonal approximations that describe
each of the domains in the timing equation. In doing this, we relax the
requirement that the sets subdividing the reachable area be partitions,
allowing polygonal sets used for the initial part of the turn to overlap to
ensure convexity. We describe these approximations and the adjustments they
require in the next subsection.

For each polygon, we create a sound approximation of the edges of the two position waves within it using expressions that are circular or linear, so that we have an analytically tractable approach to solving
for their overlap time.
The circular edge approximations are like the edges of circular ripples that result from dropping a pebble in a pond. They have the form
\begin{equation}
    (x-x_0)^2 + (y-y_0)^2 = (s t + c)^2
    \label{eq:circular-wavefront}
\end{equation}
where $s$ is the speed of propagation of the wavefront in the plane, $c$ determines the timing of its initiation, and $(x_0,y_0)$ is the point at which the wavefront originates. 

The linear edge approximations are more like the straight edges of waves one might find sweeping across the open ocean. They have the form
\begin{equation}
    x\cos\phi + y\sin\phi = (s t + c)
    \label{eq:plane-wavefront}
\end{equation}
where $c$ is the constant determining its position at specific times, and
$\phi$ is the direction of propagation.

The purpose of creating these partitions and approximations is to break apart
the conflict area $C$ into polygonal regions, so the timing expressions are no
longer piecewise in each polygon. Once these polygons are obtained, we can form
polynomial---and in some cases, linear---optimization problems to find the
collision interval.

These tailor-made approximations and refinements are each individually easier
and more accurate to reason about because the timing expression is no longer
piecewise inside each polygon. Table~\ref{wave-approx} shows which edge
approximations we use for each part of the timing equations, referring to both
a figure showing the domain, and the expression for the function's value being
approximated.

\begin{table}
\centering
\begin{tabular}{l|ll}
Region & Front Edge & Back Edge \\ \hline
A & Circular \eqref{eq:min1} \ref{mindist-turn} &  Circular \eqref{eq:max1} \ref{maxdist-turn}  \\
B & Linear \eqref{eq:min1} \ref{mindist-lin} & Circular \eqref{eq:max1} \ref{maxdist-turn}\\
C & Circular \eqref{eq:min1} \ref{mindist-circ} & Circular \eqref{eq:max1} \ref{maxdist-turn}\\
D & Linear \eqref{eq:min1} \ref{mindist-lin}  & Linear \eqref{eq:max1} \ref{maxdist-lin} \\
E & Linear \eqref{eq:min1} \ref{mindist-lin}  & Circular \eqref{eq:max1} \ref{maxdist-circ} \\
F & Circular \eqref{eq:min1} \ref{mindist-circ} & Circular \eqref{eq:max1} \ref{maxdist-circ} \\
G & Circular \eqref{eq:min1} \ref{mindist-circ} & Linear \eqref{eq:max1} \ref{maxdist-lin}\\
\end{tabular}
\caption{Each region has a combination of edge shapes that describe the front and back edges of the waves.}
\label{wave-approx}
\end{table}

Whenever these waves overlap, it means there exist trajectories that bring both
aircraft into those positions at that moment, and thus create a collision. 
We can consider the timing intervals of each polygon independently, which gives us some localization of where the collision may occur during that interval, or we can combine them together.  
The supremum and infimum of the union of the time intervals of wave intersection for each polygonal domain is a sound overapproximation of $t_e$ and $t_l$, giving us the time interval in which collisions may occur during the encounter.

Using appropriate approximations and the approach we describe,
we can analytically solve for the
intersection of these envelopes and waves in an efficient manner,
eliminating the quantification over time, and can establish the future
safety of different horizontal maneuvers in unbounded time. In effect,
our analysis allows us to ask and accurately answer: ``If the aircraft find themselves in
this configuration and the pilots restrict themselves to these turns,
bearings, and speed limits during the encounter, can we guarantee they
definitively do not collide?'' 
Section \ref{sec:solveifcollision} 
describes how we solve for the timing interval during which the waves overlap using these approximations.

%% file: happrox-approx.tex
\label{sec:soundapproxoftiming}

This section develops an approximation for path length whose accuracy can be
controlled that allows analytical solution for and efficient computation
of collision timing under non-deterministic, turn-to-bearing motion. Because of
the approximation's simpler form, we will be able to use it to develop a solution for collision timing that is quantified over the envelope of future positions possible for that maneuver in each polygonal domain.

\subsection{Fixed-radius turn-to-bearing approximation}

Points in the shaded area in Fig. \ref{mindist-circ} (or
Fig.~\ref{maxdist-circ}) are reachable by left-turning paths whose radii span
the full range of non-deterministic possibilities. The shortest (or longest)
path to a point in the shaded area is the one with the minimum (or maximum)
radius \ra (or \rb), so the function that expresses that bound matches the path
length for vehicles that turn at exactly that radius, and leave at an
appropriate tangent to reach the desired point. The expression is given by the
second piece of Eq.~\eqref{eq:min1} (or \eqref{eq:max1}).

\begin{figure}[htb]
    \centering
    \includegraphics[width=0.4\textwidth]{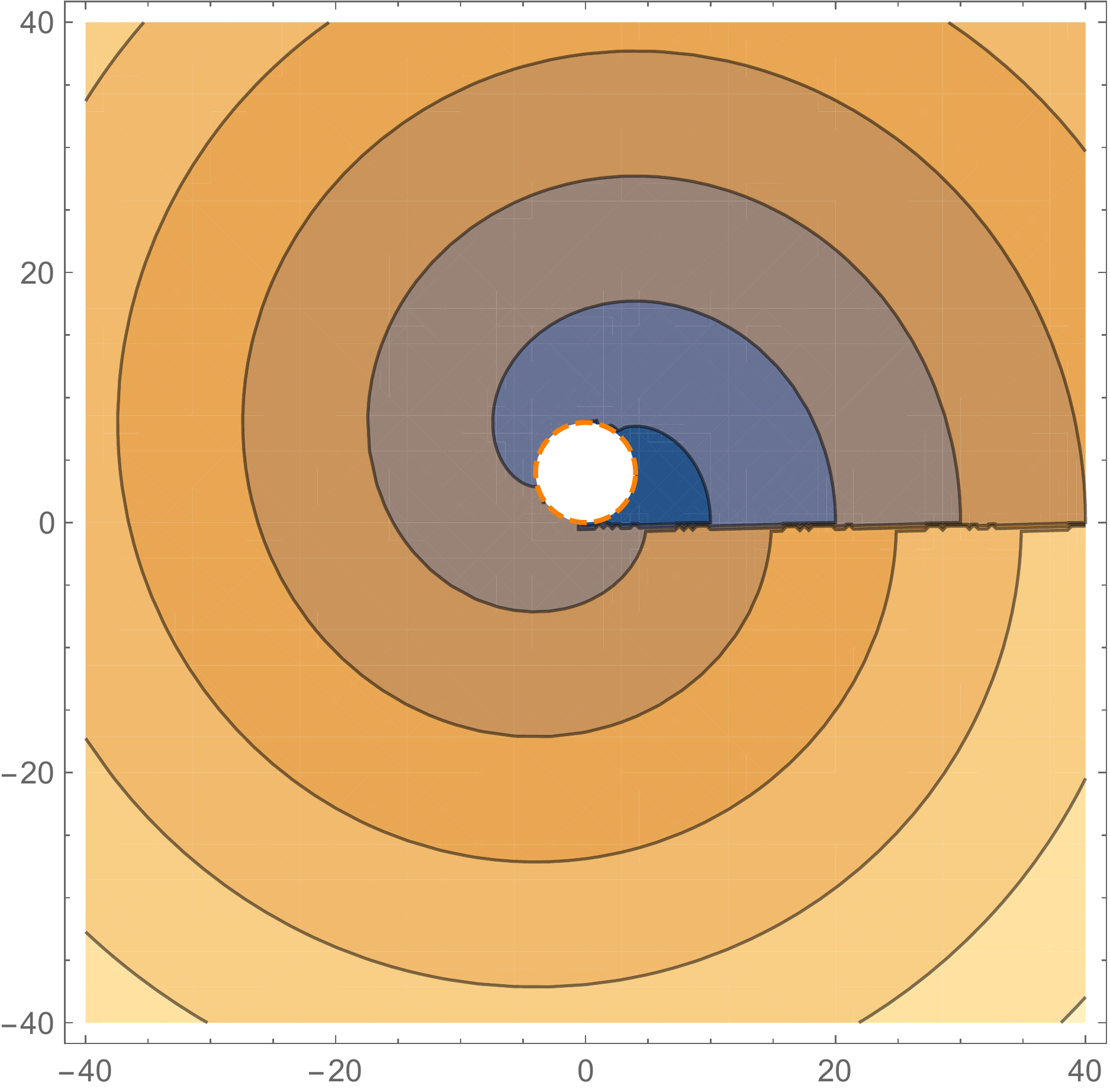}
    \includegraphics[width=0.05\textwidth]{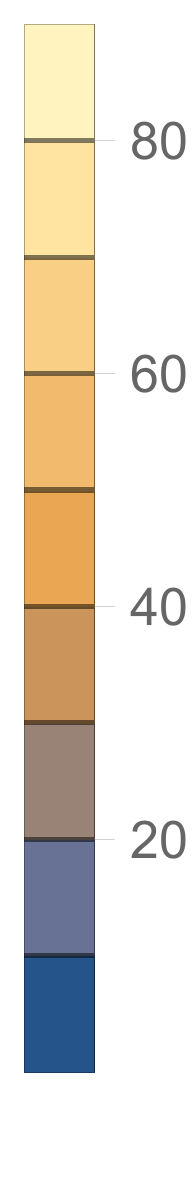}%
    \raisebox{107pt}[00pt][0pt]{
  \makebox[0pt][l]{
  \hspace*{-138pt}
  \includegraphics[angle=-147,scale=.01]{figs/airplane_topview_CC.png}}}%
    \caption{Trajectory length for fixed-radius turn-to-bearing
      motion, for left turns with $r=4$. Level sets of length are circle involutes.}
    \label{frttb}
\end{figure}

Fig.~\ref{frttb} is a contour plot where the color indicates the path length. Each path follows a circular arc of a particular radius, and then leaves the turn at a tangent to reach the destination point following a straight path thereafter. The plot has a cut, a discontinuity on the positive $x$-axis. The level sets of path length for these pieces is a circle involute, which we will approximate in a limited area using a circular arc.

We prove:
\begin{theorem}
For left turn-to-bearing motion with $0<\phi_2<2\pi$, the length of a path starting from the origin initially oriented in the direction of the positive x-axis
and arriving at a point $(x,y)$ following one-turn-to-bearing motion with a fixed radius is bounded by
\begin{multline}
\sqrt{x^2 + y^2}    \le L(x,y,\Theta(x,y,r),r) \le \\
r\phi_2 + \sqrt{(x-w_x)^2 + (y-w_y)^2} 
\end{multline}
where $w_x = r\sin(\phi_2)$ and $w_y = r(1-\cos(\phi_2))$.
\end{theorem}

We know that the shortest path between two points in a Cartesian plane is a line segment between those points. We can use this to create both an underapproximation of the shortest path and an overapproximation of the longest path for this region.

\begin{figure*}[htb]
    \centering
    \begin{subfigure}[t]{0.47\textwidth}\centering
    \includegraphics[width=.96\textwidth]{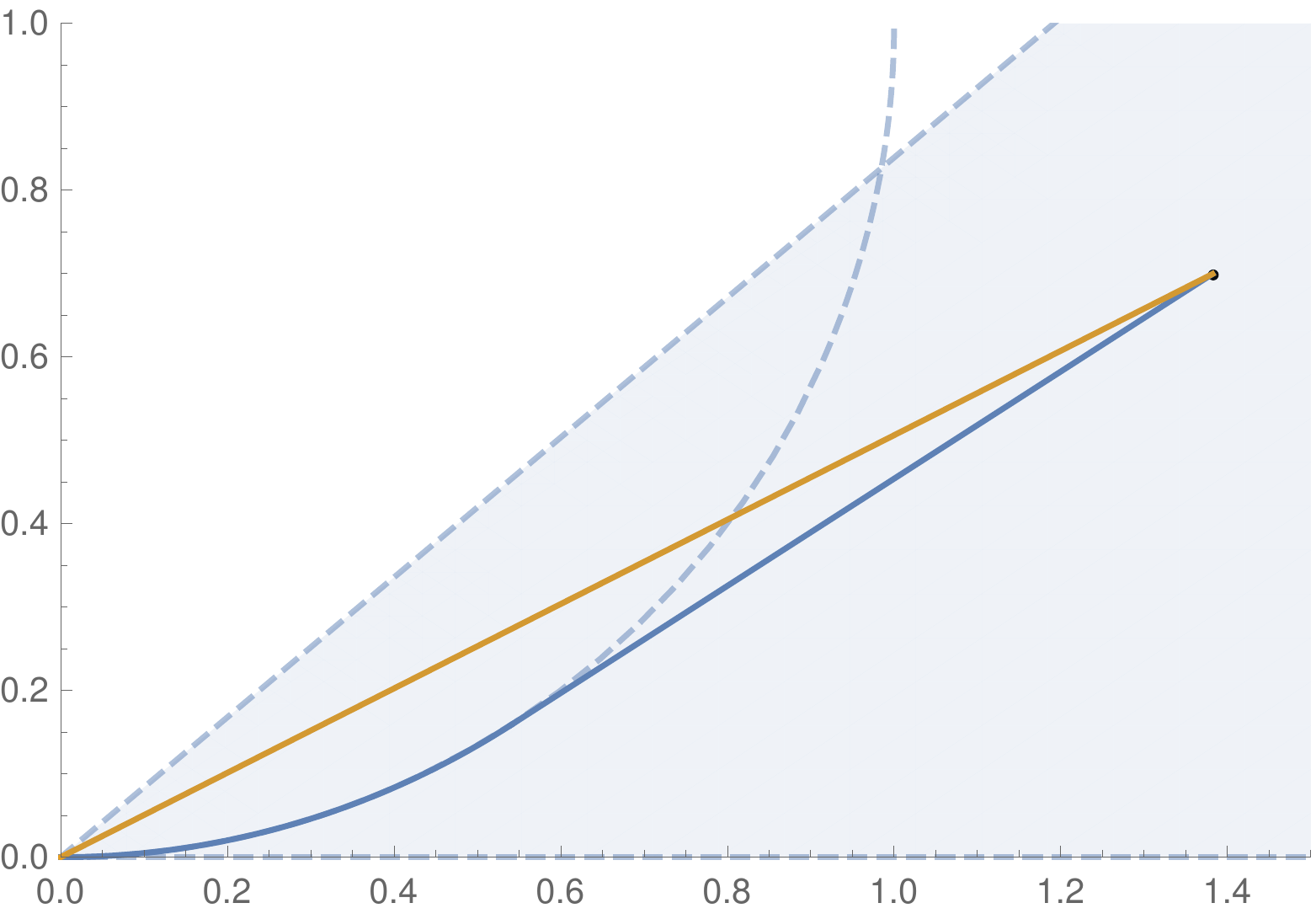}
    \caption{Path length underapproximation}
    \label{underapprox}
    \end{subfigure}
\begin{subfigure}[t]{0.47\textwidth}\centering
    \includegraphics[width=.96\textwidth]{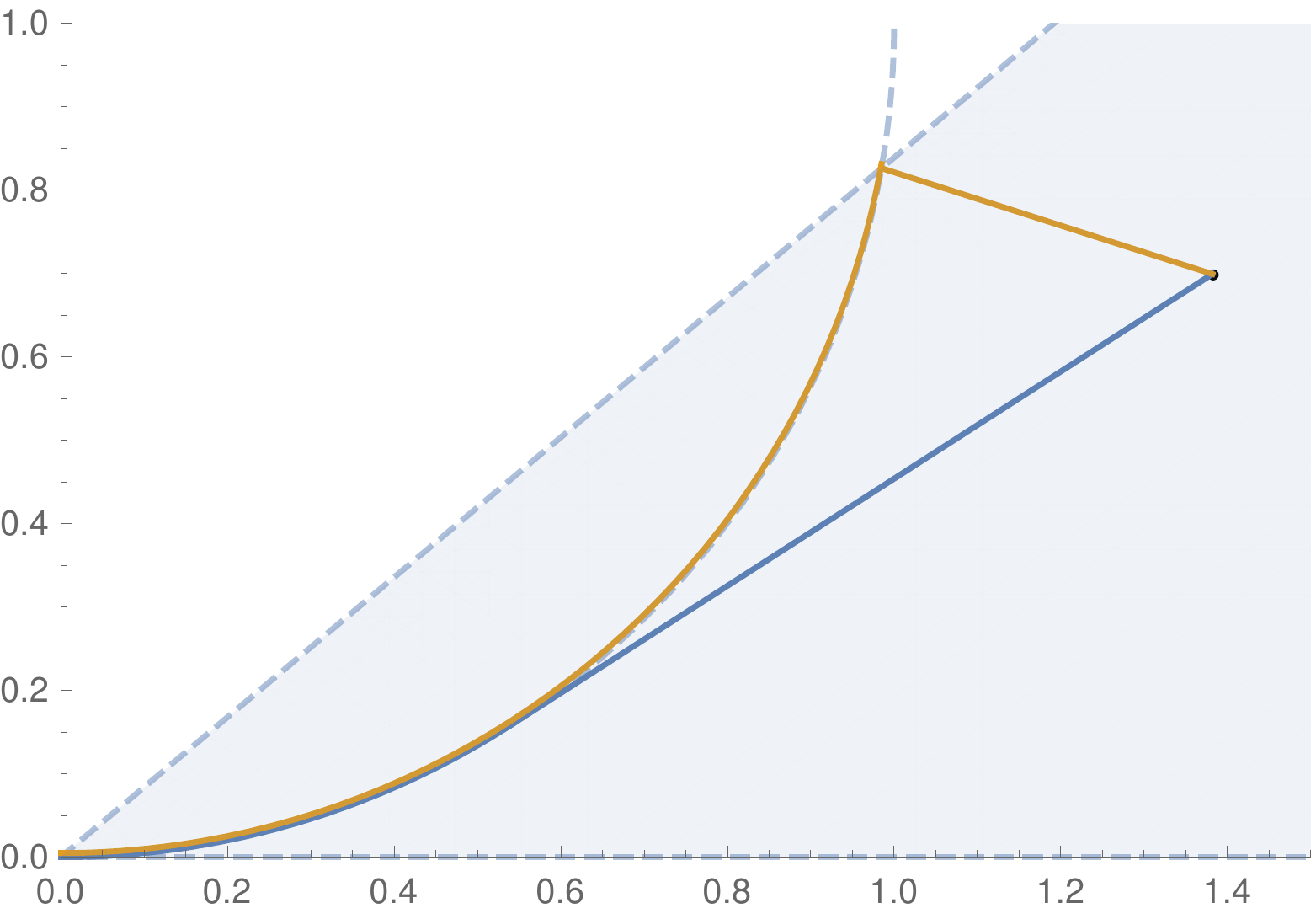}
    \caption{Path length overapproximation}
    \label{overapprox}
    \end{subfigure}
\caption{Turn-to-bearing paths (in blue) have length approximated by the length of the yellow paths. Path length for yellow paths can be expressed as functions of the length of their linear components, leading to contour plots with circular level sets. This approximation is valid within the domain indicated by the shaded wedge, whose angle in the approximating equations is $\phi_2$.}
\end{figure*}

Consider a turn-to-bearing path starting from the origin, traveling in a circular arc, leaving the arc at a tangent at point $T$, and traveling in a straight line thereafter to reach $(x,y)$.
The length of the line segment from the origin to $(x,y)$ is a lower bound for the path distance, since the true path does not follow a straight line to reach $(x,y)$. For an upper bound, we can use the length of a path that follows the circular arc but continues past $T$ to another point $U$ on the arc, and then follows a straight (non-tangent) line from $U$ to $(x,y)$. This is longer than the turn-to-bearing path, because both paths stay together until point $T$, and from that point the turn-to-bearing path follows a straight line to $(x,y)$. Every path that diverges at or after the tangent point must thus be longer. Figs.~\ref{underapprox} and \ref{overapprox} illustrate this strategy.

The domain of Figs. \ref{mindist-circ} and \ref{maxdist-circ} is nearly a polygonal boundary, but not quite. We create a polygonal domain for our approximation of path distance in the second pieces of Eqs. \ref{eq:min1} and \ref{eq:max1} for a vehicle starting at the origin using simple linear boundaries. We define an area
\begin{equation}
    S=\{(x,y) \mid y\ge 0\land \atan(y,x)\le\phi_2\}
\end{equation}
This area is equivalent to a domain
\begin{align}
  \begin{split}
  G^R_v(x,y) = & y \ge 0 \land
    \left(\cos(\phi_2) y \leq \sin(\phi_2) x \right)
  \end{split}
\end{align}

This wedge-shaped domain can be rotated and translated to allow us to approximate other parts of the circle. If we want to adjust it so that the initial bearing is $2 \phi_1$, we can translate the domain so the vertex is at $(r\sin(\phi_1),r(1-\cos(\phi_1))$, and rotate it so the clockwise-most linear boundary is tangent to the circle at that point.

\subsection{Fixed-bearing turn-to-bearing approximation}

There are a range of possible turn-to-bearing trajectories that reach from the origin to each point in the shaded area in Fig. \ref{mindist-lin}. The minimum path length for each point--given by the second piece of Eq.~\eqref{eq:min1} -- corresponds uniformly to a trajectory of whose final bearing $\thb$ is at the end of the allowable range, and the radius that achieves that bearing, which varies depending on the point. 

Turn-to-bearing trajectories reaching shaded points in Fig. \ref{maxdist-lin} have a similar property. The maximum possible path length to reach these points---given by the second piece of Eq.~\eqref{eq:max1}---also corresponds uniformly to trajectories of fixed final bearing (this time $\tha$) at the other end of the allowable range, and the radius that achieves that bearing, which differs depending on the point.

The level set of path length in the plane for these pieces is a line. Thus the boundary of the area that contains the non-deterministic possibilities of our motion is a linear wave, an isoline in the plane with a fixed orientation that is moving over time.

\begin{figure}[htb]
    \centering
    \includegraphics[width=0.4\textwidth]{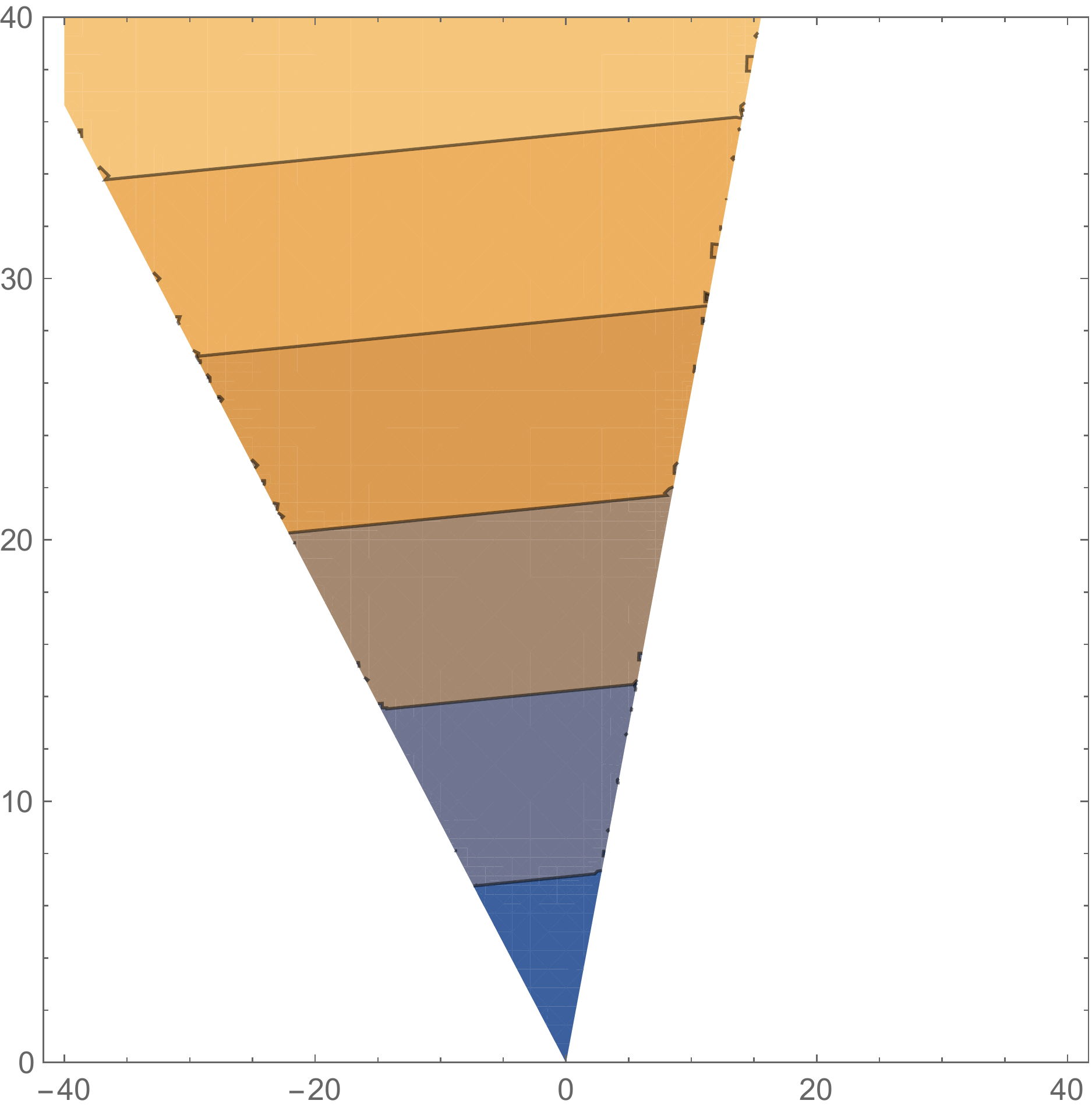}
    \includegraphics[width=0.05\textwidth]{figs/circle-involute-legend.pdf}%
\raisebox{18pt}[00pt][0pt]{
  \makebox[0pt][l]{
  \hspace*{-142pt}
  \includegraphics[angle=-147,scale=.015]{figs/airplane_topview_CC.png}}}%
    \caption{Trajectory length for fixed-bearing turn-to-bearing
      motion, for left turns with final bearing $\theta=2.4$. Level sets of length are lines.}
    \label{fig:fbttb}
\end{figure}

Fig. \ref{fig:fbttb} shows the exact length of the turn-to-bearing path
reaching each each point in the plane of motion when those paths are constrained to end
with uniform orientation. There is only one such path
that reaches each point for a particular choice of bearing, and each
point is colored according to the path length. We do not consider
paths that end during the initial turn unless their bearing matches
the motion we are analyzing, so although some points to the right of this
wedge are traversed by these paths, the paths are circular at that
stage, and their lengths are not shown in this figure. Points on the left side of the wedge are not reachable or
traversable by this type of motion. For this section, each contour is
linear and this type of motion does not require an approximation to
express the path lengths at each point. We prove:

\begin{theorem}
For left turn-to-bearing motion, 
path distance starting from the origin with orientation $0$
(facing the direction of the positive x-axis)
and arriving at a point $(x,y)$ with orientation $\theta$ is given by 
\begin{equation}
\label{eqn:linearwavefront}
\begin{split}
    & L(x,y,\theta,R(x,y,\theta)) = \\
    & x\left( \cot{\left(\frac{\theta}{2}\right)}\theta - 1\right) + 
     y \left( \cot{\left(\frac{\theta}{2}\right)} - \frac{ \cos{\left(\theta\right)}}{1-\cos(\theta)}\theta \right)
\end{split}
\end{equation}

\end{theorem}
The key insight here is that for each region, the angle $\theta$ required to find the minimum or maximum path length is constant, and thus the expression for path distance has the form of a plane wave as given by Eq.~\eqref{eq:plane-wavefront}. A plot of path lengths produce linear level sets, which can be interpreted as a linear wavefront in the plane.

The domains of the second pieces of Eqs.~\eqref{eq:min1} and \eqref{eq:max1} are each convex, open sets with linear boundaries. They need no approximation to ensure convexity and can be subdivided into convex polygons to localize the timing of potential future collisions within them.

\subsection{Turning approximation}

Each turn-to-bearing maneuver begins with a turn that follows a circular arc. There is only one path that leads from the origin to each of the points in the shaded area in Fig. \ref{mindist-turn}, so that path is the shortest length path possible. The path length is given by the third piece of Eq.~\eqref{eq:min1}. The domain of Fig. \ref{maxdist-turn} is a superset of Fig. \ref{mindist-turn}, where some of the points have more than one way to approach them. The longest possible path to reach each of these points also follows a circular arc, given by the same expression shown in the second piece of Eq.~\eqref{eq:max1}.

The level set of path length for these pieces is a cardioid, which we will approximate using a circular wave.
\begin{figure}[htb]
    \centering
    \includegraphics[width=0.4\textwidth]{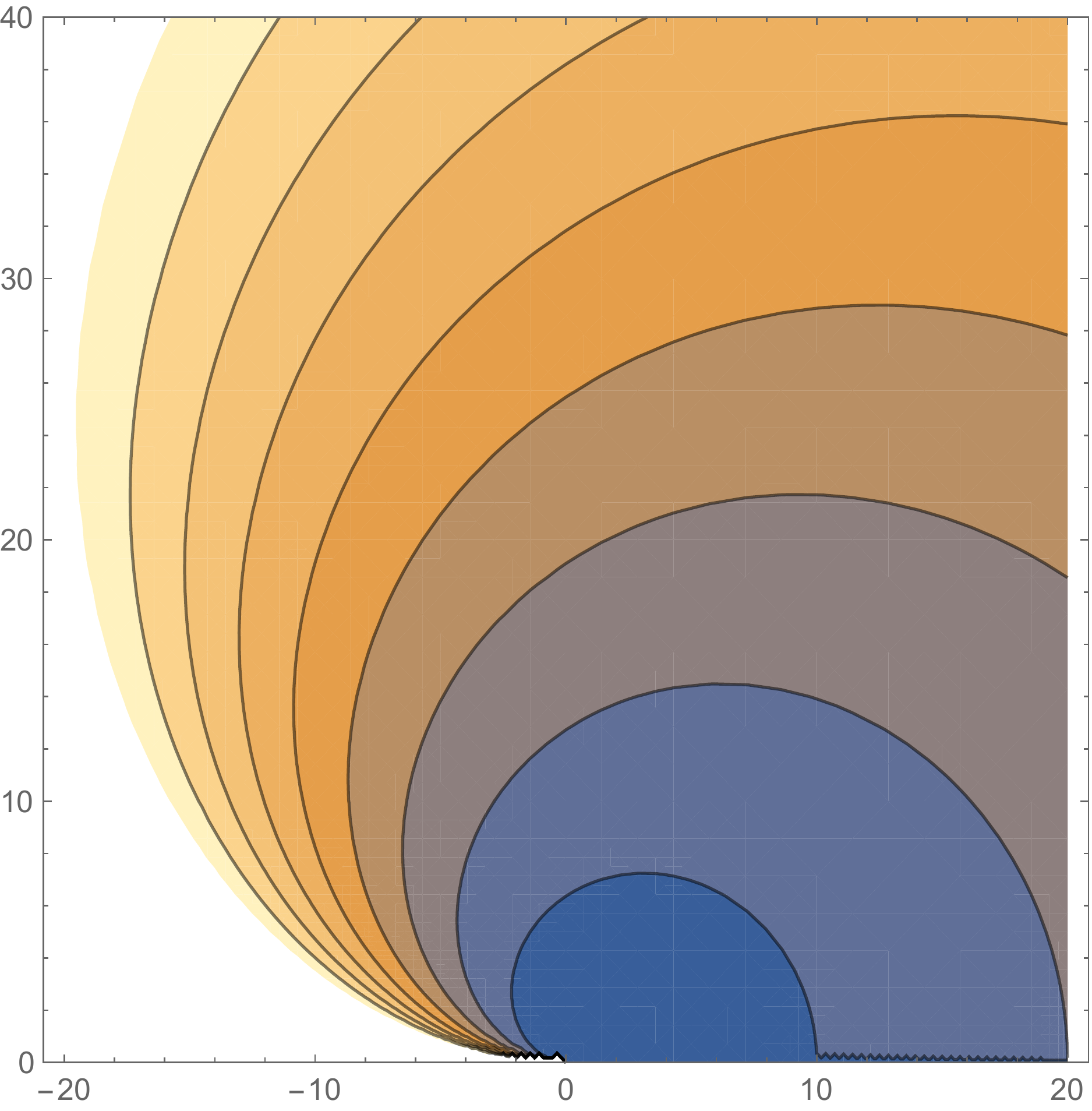}
    \includegraphics[width=0.05\textwidth]{figs/circle-involute-legend.pdf}%
    \raisebox{18pt}[00pt][0pt]{
  \makebox[0pt][l]{
  \hspace*{-142pt}
  \includegraphics[angle=-147,scale=.015]{figs/airplane_topview_CC.png}}}%
    \caption{Trajectory length for a left-turning circular arc. Level sets of length are cardioids.}
    \label{fig:circ}
\end{figure}

Fig.~\ref{fig:circ} shows the lengths of turn-to-bearing paths starting at the origin as shown, reaching each point in the plane of motion, when each path is following a circular arc from beginning to end.

The key insight here is that if we examine the length of circular paths that end at points whose location is at a constant angle in the plane (and thus have a constant final orientation), we can create a function that has the form of a circular wavefront that matches the path lengths along that ray. We prove:
\begin{lemma}[Circular path length]
The length of a left-turning circular path equals the distance from the
origin scaled by a factor of $\sinc(x)=\sin(x)/x$:
\begin{equation}
r_m\theta_m=\frac{\sqrt{x^2 + y^2}}{\sinc(\frac{\theta_m}{2})}.
\end{equation}
\end{lemma}
If we solve the implicit equation of a circle for the radius, we will find that it is proportional to $\sqrt{x^2 + y^2}$. We are using the radius to compute the path lengths, so the bounds will have this term in them as well.
\begin{figure*}[htb]
    \centering
    \begin{subfigure}[t]{0.47\textwidth}\centering
    \includegraphics[width=.7\textwidth]{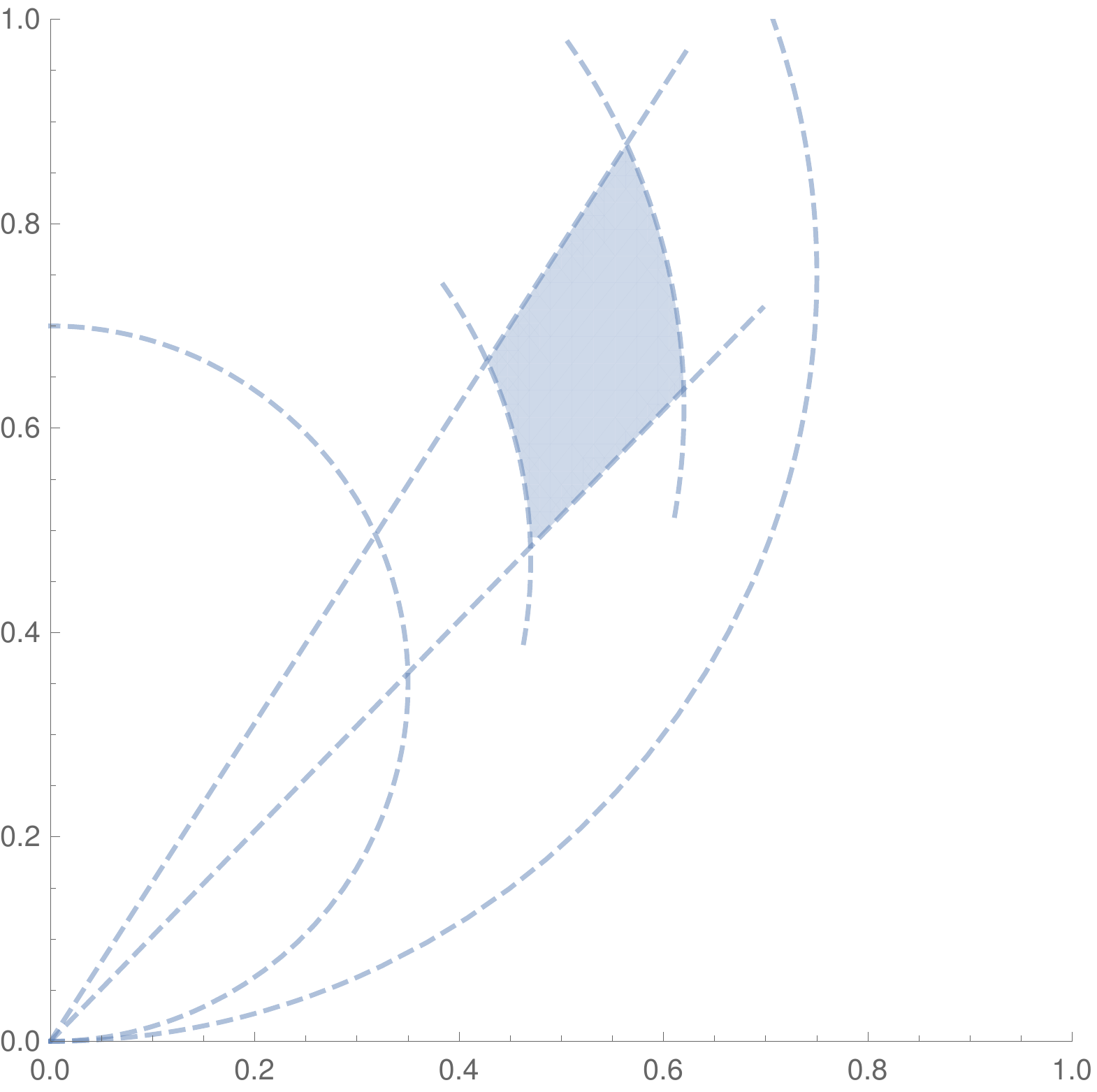}
    \caption{}
    \label{fig:constant-wave}
    \end{subfigure}
\begin{subfigure}[t]{0.47\textwidth}\centering
    \includegraphics[width=.7\textwidth]{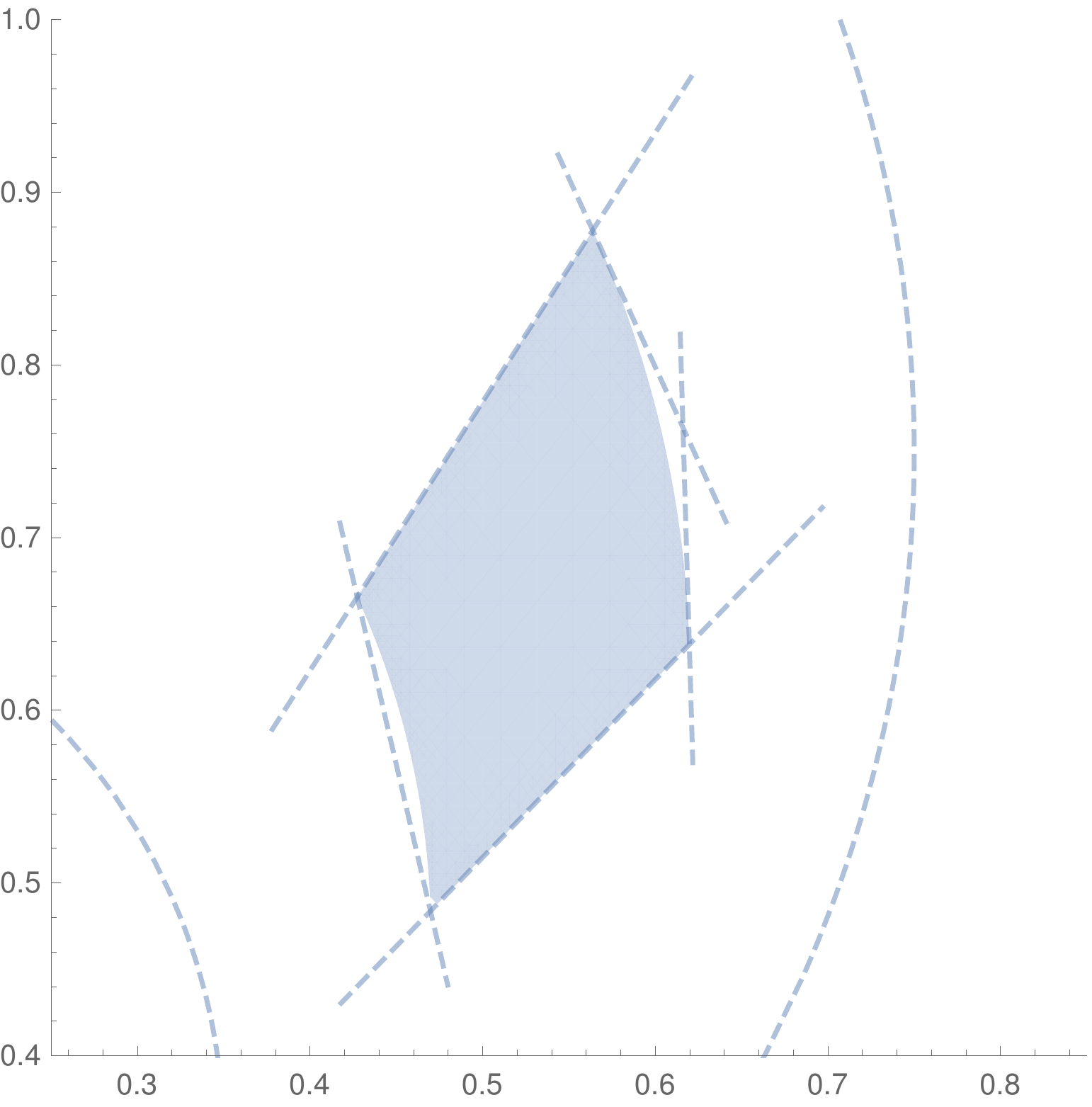}
    \caption{}
    \label{fig:constant-wave-approx}
    \end{subfigure}
\caption{Boundaries of a pixel that can be used to tile the reachable area for motion in a circular turn.}
\end{figure*}
Because $\sinc(\cdot)$ decreases monotonically over the interval
$(0, \pi)$, 
we know that a circular function that matches one angle is a lower bound for
angles with the same radii above it.

We define an area
\begin{multline}
S = \left\{
(x,y) \mid \exists\ r\in[r_1,r_2], \theta\in[\theta_1,\theta_2], \right. \\
   \left. x = r\sin(\theta) \land 
   y = r(1- \cos(\theta)) \right\}
   \end{multline}
whose boundary is limited by circular turning arcs with different radii and straight lines radiating from the starting point at different angles, as shown in Fig. \ref{fig:constant-wave}. We can create circular wavefronts centered at the origin that match the path lengths at the straight edges, and serve as upper and lower bounds for all of the points within the shaded area. 

These boundaries define a closed region, and the reachable area of circular turning motion can be tiled by these sets. In Thm.~\ref{thm:circularbounds}, we prove upper and lower bounds for path length in this type of region.

\begin{theorem}\label{thm:circularbounds}
For left turn-to-bearing motion with $0<\theta_1<\theta_m<\theta_2<2\pi$, the length of a path starting from the origin initially oriented in the direction of the positive x-axis
and arriving at a point $(x,y)$ following a circular arc is bounded by
\begin{equation}
\frac{\sqrt{x^2 + y^2}} {\sinc\left(\frac{\theta_1}{2}\right)}
    \le L(x,y,\theta_m,r_m) \le 
\frac{\sqrt{x^2 + y^2}} {\sinc\left(\frac{\theta_2}{2}\right)}.
\label{eqn:circleapprox}
\end{equation}
\end{theorem}

The partition has elements of the form of $S$, but with different parameters $r_1$, $r_2$, $\theta_1$, and $\theta_2$. Elements have curved boundaries and are not convex, so we create a polygonal overapproximation of these elements which serves as the domain. Fig.~\ref{fig:constant-wave-approx} shows this approximation in the first quadrant, where the region is defined by
\begin{equation}
\begin{split}
G^C(x,y) = &
    \left( \cos(\theta_1/2) (y-\nu_y) \geq \sin(\theta_1/2) (x-\nu_x) \land \right. \\
   &        \cos(\theta_2/2) (y-\iota_y) \leq \sin(\theta_2/2) (x-\iota_x) \land  \\
   &       (\iota_x - \nu_x) (y-\nu_y) \leq (\iota_y - \nu_y) (x-\nu_x) \land \\
   &        \cos(\theta_1) (y-\upsilon_y) \geq \sin(\theta_1) (x-\upsilon_x) \land  \\
   & \left. \cos(\theta_2) (y-\omega_y) \geq \sin(\theta_2) (x-\omega_x) \right), \end{split}
  \label{eq:linvalidity}
\end{equation}
where
\begin{align*}
    (\nu_x,\nu_y) &= r_1 (\sin(\theta_1), (1-\cos(\theta_1))) \\
    (\iota_x,\iota_y) &= r_1 (\sin(\theta_2), (1-\cos(\theta_2))) \\
    (\upsilon_x,\upsilon_y) &= r_2 (\sin(\theta_1), (1-\cos(\theta_1))) \\
    (\omega_x,\omega_y) &=r_2 (\sin(\theta_2), (1-\cos(\theta_2))).
\end{align*}

When this approximation is used instead of a partition for refinement, the other areas that also are part of the refinement may be duplicated twice, and need to be evaluated with more than one wavefront combination.

This approximation can be used to safely represent positions
for the highly non-linear area at the beginning of the circular turn.


\section{Sound Solution for Collision Timing between Two Turning Vehicles}
\label{sec:solveifcollision}


In Secs.~\ref{sec:happrox-intro} and \ref{sec:soundapproxoftiming}, we showed how
to subdivide the reachable region of turn-to-bearing maneuvers into a covering
with polygonal regions.  Recall that the polygonal regions are sound
overapproximations of refinements of the conflict area $C$, the region where two
vehicles could both potentially reach. 

In each of these polygonal regions, the vehicle location over time is bounded
by the region, which is fixed in time, and linear or circular curve segments
that propagate in time. These boundaries move and form the front and back edges
of propagating waves---the front edge bounds the location of the vehicle in the longest
paths that could be followed for a given time and the back edge bounds the
location by the shortest paths that could be followed.

Guaranteeing the absence of collisions now equates to ensuring that, for each
of these polygonal regions, there are no locations simultaneously contained by
the front and back edges of the waves of position possibilities for both vehicles at the
same time. 

We represent each region as a polygon, which is an intersection of finitely
many half-planes. A compact notation for the region is
\begin{equation}
    \mathcal{P} = \left\{ {\bf x} \in \mathbb{R}^2 \mid A{\bf x} \leq {\bf b} \right\} 
    \label{def:polyhedron}
\end{equation}
where ${\bf x} = \begin{bmatrix}x & y\end{bmatrix}^T \in \mathbb{R}^2$ and 
${\bf b} \in \mathbb{R}^k$. We do not require the region be bounded. 

\paragraph{The overlap property}
Evaluating timing for positions to overlap is necessary to establish collision possibilities. This was first discussed for individual points in Thm. \ref{thm:collision_timing}.
A necessary condition for there to be an overlap between the time that ownship
and intruder are in a particular point in $x,y$ space is that the waves that
constrain where each vehicle is must overlap at that point for some time $t$.
If there exists a point and time
that is within the polygon and within the wavefronts of both the ownship and
the intruder at a common time, then a collision is possible at that place and future time.

In Sec.~\ref{ss:ptwisecolltime}, Eqs.~\eqref{eq:te_p} and \eqref{eq:tx_p}
define the earliest and latest times that the vehicle could reach point $p$.
Our objective is to compute over the polygonal region, defined in
\eqref{def:polyhedron}, the earliest and latest times where the intersections
of the ownship and intruder waves can meet. 
The intersection of the ownship and
intruder waves form another wave, a set of points at each instant in time that for that future instant might produce a collision. We will call the wave resulting from the intersection the conflict wave.
Since the over- and
under-approximations for vehicle locations are sound in the polygonal region of
interest, if the conflict wave intersects any part of our polygon, then that
denotes a potential collision. The overall earliest and latest times for
collision in this region will be denoted $t_e(\mathcal{P})$ and
$t_l(\mathcal{P})$.

\subsection{The case with purely linear wavefronts}\label{ss:linearwaves}

We first treat the case where the front and back edges of both ownship and
intruder position waves are all described with the linear boundary given in
Eq.~\eqref{eqn:linearwavefront}. This case happens in region D
(see Fig.~\ref{fig:configA} and Table~\ref{wave-approx}).
The front and back of the linear edges are
formed using the upper and lower bounds on the vehicle speed, $\sb$ and $\sa$. 
Each ``wavefront'' is defined by a tangent and a velocity of propagation. 
The ownship front wave is
\begin{eqnarray}
\begin{bmatrix}f_1^o & f_2^o\end{bmatrix}\begin{bmatrix}x \\ y\end{bmatrix} 
= \sb^o t,  
\end{eqnarray}
and the ownship back wave is
\begin{eqnarray}
\begin{bmatrix}b_1^o & b_2^o\end{bmatrix}\begin{bmatrix}x \\ y\end{bmatrix} 
= \sa^o t.  
\end{eqnarray}
The intruder's front wave, which starts at point ${\bf q} = \begin{bmatrix}q_1 & q_2\end{bmatrix}^T$, is
\begin{eqnarray}
\begin{bmatrix}f_1^i & f_2^i\end{bmatrix}\left(\begin{bmatrix}x \\ y\end{bmatrix} - {\bf q}\right) = \sb^i t,
\end{eqnarray}
and the intruder back wave is
\begin{eqnarray}
\begin{bmatrix}b_1^i & b_2^i\end{bmatrix}\left(\begin{bmatrix}x \\ y\end{bmatrix} - {\bf q}\right)
= \sa^i t.
\end{eqnarray}

To relate these coefficients to those in Eq.~\eqref{eqn:linearwavefront}, we define 
\begin{align}
    f_1^o &= \cot{\left(\tha/2\right)}\tha - 1 \\
    f_2^o &= \cot{\left(\tha/2\right)} - \frac{ \cos{\left(\tha\right)}}{1-\cos{\tha}}\tha  \\
    b_1^o &= \cot{\left(\thb/2\right)}\thb - 1 \\
    b_2^o &= \cot{\left(\thb/2\right)} - \frac{ \cos{\left(\thb\right)}}{1-\cos{\thb}}\thb
\end{align}
where the $\tha$ and $\thb$ parameters correspond to the bounds on angle that apply to this particular ownship refinement. 
We then do the same for the intruder front and back wave coefficients. 

We now write the full set of constraints over the search space of feasible, $x$, $y$, and $t$. 
The joint set of constraints are:
\begin{eqnarray}
\begin{bmatrix}f_1^o & f_2^o\end{bmatrix}\begin{bmatrix}x \\ y\end{bmatrix} 
    \leq \sb^o t  \label{eq:firstlincons} 
\\
\begin{bmatrix}b_1^o & b_2^o\end{bmatrix}\begin{bmatrix}x \\ y\end{bmatrix} 
\geq \sa^o t  
\\
\begin{bmatrix}f_1^i & f_2^i\end{bmatrix}\left(\begin{bmatrix}x \\ y\end{bmatrix} - {\bf q}\right) \leq \sb^i t \\
\begin{bmatrix}b_1^i & b_2^i\end{bmatrix}\left(\begin{bmatrix}x \\ y\end{bmatrix} - {\bf q}\right)
\geq \sa^i t \\
A {\bf x} - {\bf b} \leq 0. \label{eq:lastlincons}
\end{eqnarray}
With purely affine inequality constraints, the problem of finding the earliest
and latest times of a potential collision can be solved by minimizing or
maximizing $t$ over these constraints. For example, $t_e(\mathcal{P})$ is the
solution to the optimization problem
\begin{equation}
    \begin{array}{ll}
        \mbox{minimize} 
            & t \\
        \mbox{subject to} 
            & \eqref{eq:firstlincons}\text{--}\eqref{eq:lastlincons}, \\
            & t \geq 0,
    \end{array}
    \label{eq:linoptmin}
\end{equation}
over the variables $x$, $y$, and $t$. If the problem~\eqref{eq:linoptmin} is
infeasible, then collision is not possible in $\mathcal{P}$.  Similarly, we can
find $t_l(\mathcal{P})$ by forming a maximization problem.

\subsection{Case including only circular wavefronts}\label{ss:withcircwaves}

We now consider the cases where the front
and back waves of ownship and intruder involve only circular waves. These would
correspond to intersections of polygons where both the ownship and the intruder
are in regions A, C or F (see Fig.~\ref{fig:configA} and Table~\ref{wave-approx}). 
As described in Sec.~\ref{sec:soundapproxoftiming},
these turning sections are also subdivided (or tiled) into smaller regions and
then overapproximated as polygons (Fig.~\ref{fig:constant-wave-approx}). 

The timing bounds for these circular wavefronts are given in Thm.~\ref{thm:circularbounds}, Eq.~\eqref{eqn:circleapprox}. 
They have the form
\begin{equation}
    (x-x_0)^2 + (y-y_0)^2 = (v t + c_0)^2
\end{equation}
where $v$ is the speed of propagation of the wavefront in the plane, $c_0$ determines the timing of its initiation, and $(x_0,y_0)$ is the point at which the wavefront originates. We note that the front and back waves may have different parameters, $x_0$, $y_0$, and $c_0$, and will be denoted with subscript $f$ for the front wave and with $b$ for the back wave. 

With appropriate coefficients for the ownship and intruder front and back wave edges, the constraints for a collision become the following.
\begin{eqnarray}
(x - x_f^o)^2 + (y - y_f^o)^2 \leq (\sb^o t + c_f^o)^2 \label{eq:own-circfront}\\
(x - x_b^o)^2 + (y - y_b^o)^2 \geq (\sa^o t + c_b^o)^2 \label{eq:own-circback}\\
(x - x_f^i)^2 + (y - y_f^i)^2 \leq (\sb^i t + c_f^i)^2 \label{eq:int-circfront}\\
(x - x_b^i)^2 + (y - y_b^i)^2 \geq (\sa^i t + c_b^i)^2 \label{eq:int-circback}\\
A {\bf x} - {\bf b} \leq 0 
\end{eqnarray}
The four constraints~\eqref{eq:own-circfront}--\eqref{eq:int-circback} are
quadratic in the decision variables $x$, $y$, and
$t$, but they are not convex constraints. In fact, only the linear case results
in a convex linear program. All other cases can be solved, in general, using
polynomial optimization. Exact algorithms based on the Cylindrical Algebraic
Decomposition (CAD)~\cite{collins1975quantifier}, such
as~\cite{fotiou2006,lavalle2006planning}, as well as approximation techniques
involving moments~\cite{lasserre2001global}, or sums-of-squares and
semidefinite relaxations~\cite{prajna2002,parrilo2003semidefinite} can be used
to determine the timing boundaries on each region.

\subsection{Visualizing the solutions of wave intersections}\label{ss:locimethod} 
Using informal, geometric, and CAD-inspired arguments, we present an approach
to evaluating collision safety and solving for collision timing. We leave it as
future work to formalize this proof.
In Fig.~\ref{o} we saw that at
each future moment in time, each vehicle has an area in which it will be
found, encompassing the uncertainty in its motion between the present and that moment.
This area is a propagating wave of position possibilities that changes shape and moves forward as time progresses. It 
is bounded by the curves that define the reachable area, and
irregularly-shaped front and back edges that move orthogonally to the reachable area
boundaries, expanding over time. We use the path distance 
approximations from Sec.~\ref{sec:soundapproxoftiming} to 
represent the front and back edges of the wave. Each
encounter has two waves and four different edges, i.e. the leading
and trailing edges of the area describing the possible positions for each vehicle.

\begin{figure}[htbp]
    \centering
\includegraphics[width=0.47\textwidth]{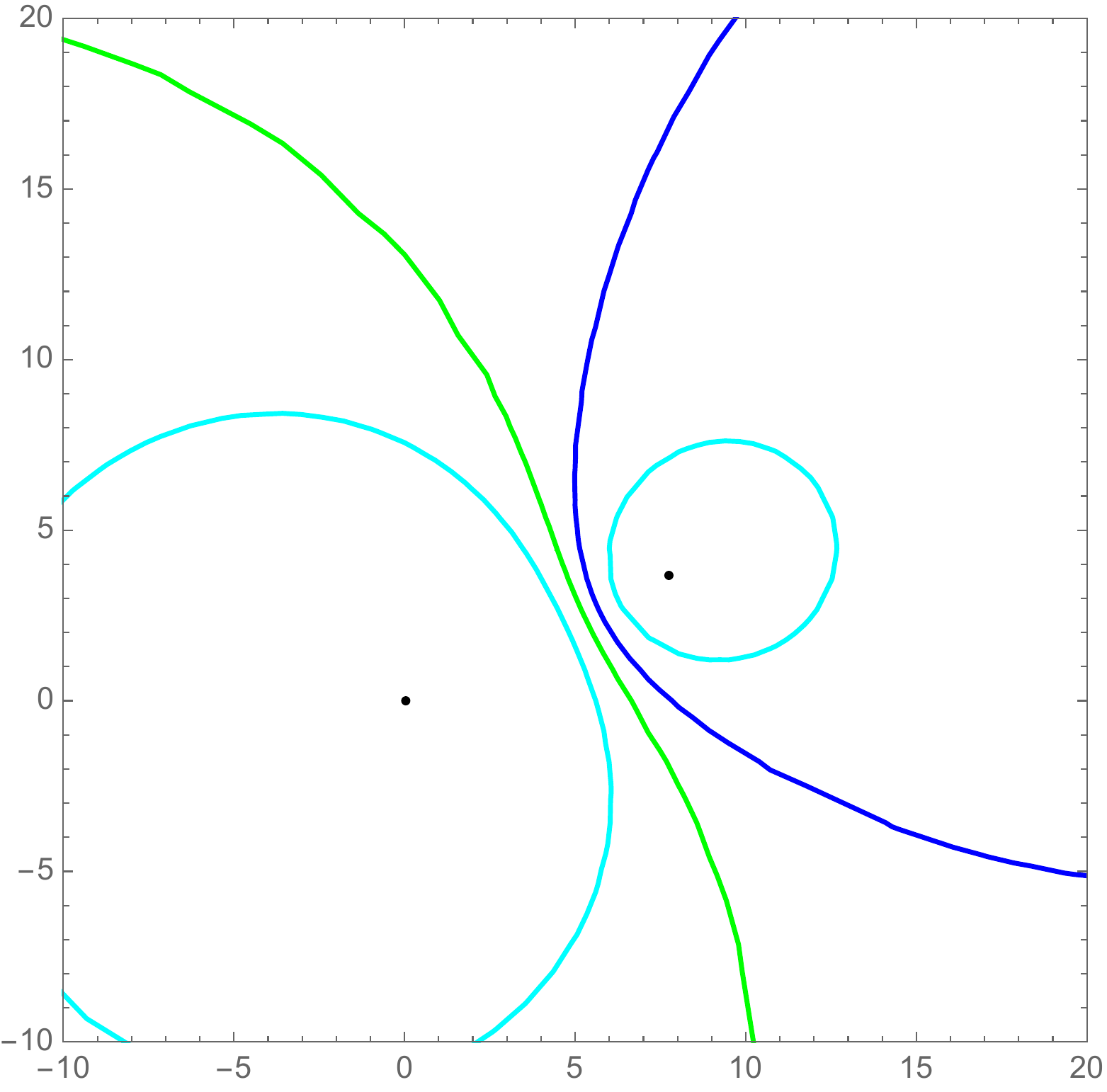}%
\raisebox{75pt}[00pt][0pt]{\makebox[0pt][l]{\hspace*{-165pt}
Ownship}}%
\raisebox{100pt}[00pt][0pt]{\makebox[0pt][l]{\hspace*{-100pt}
Intruder}}%
\raisebox{100pt}[00pt][0pt]{\makebox[0pt][l]{\hspace*{-200pt}
A: $t^o_l < t^i_e$}}%
\raisebox{120pt}[00pt][0pt]{\makebox[0pt][l]{\hspace*{-100pt}
E: $t^i_l<t^o_e$}}%
\raisebox{150pt}[00pt][0pt]{\makebox[0pt][l]{\hspace*{-215pt}
B: $t^o_e<t^i_e<t^o_l<t^i_l$}}%
\raisebox{210pt}[00pt][0pt]{\makebox[0pt][l]{\hspace*{-180pt}
C: $t^i_e<t^o_e<t^o_l<t^i_l$}}%
\raisebox{150pt}[00pt][0pt]{\makebox[0pt][l]{\hspace*{-105pt}
D: $t^i_e<t^o_e<t^i_l<t^o_l$}}%
\caption{Loci of intersections between leading and trailing edges
of circular wavefronts are plotted together for a specific, example
geometry. The loci impose an ordering of earliest and latest arrival times for each vehicle for the points in each region of the plane.}
\label{fig:loci}
\end{figure}

In the previous section, we created a covering of convex polygons, each with a positional wave whose edges (level sets of path length) are represented by simple polynomials of at most order two. This breaks the problem up into a set of simpler problems. In this section, we will focus on the problem of solving for wave intersection within one polygonal region.

For now, we eliminate the polygonal boundaries so that we can more clearly see the geometry and timing of overlap of different position possibility waves. Later we will add the polygonal boundaries back into the problem.

In Sec.~\ref{ss:ptwisecolltime} we developed the overlap criteria for pointwise collision, namely that the time intervals when two vehicles might arrive at a point must overlap for there to be a collision. We can now apply that criteria to the geometry of uncertain position waves. 

One key insight is that we can order the arrival of wave edges at each point by creating loci of the moving edges of the position waves. Each locus is a curve consisting of all the points of intersection over time between
two different position wave edges, one from each vehicle. The loci divide
the horizontal plane into regions that identify the order the earliest
arrival and latest departure times for each vehicle at each point in the region.

\begin{figure}[htbp]
    \centering
    \begin{subfigure}[t]{0.3\textwidth}
        \centering
        \includegraphics[width=\textwidth]{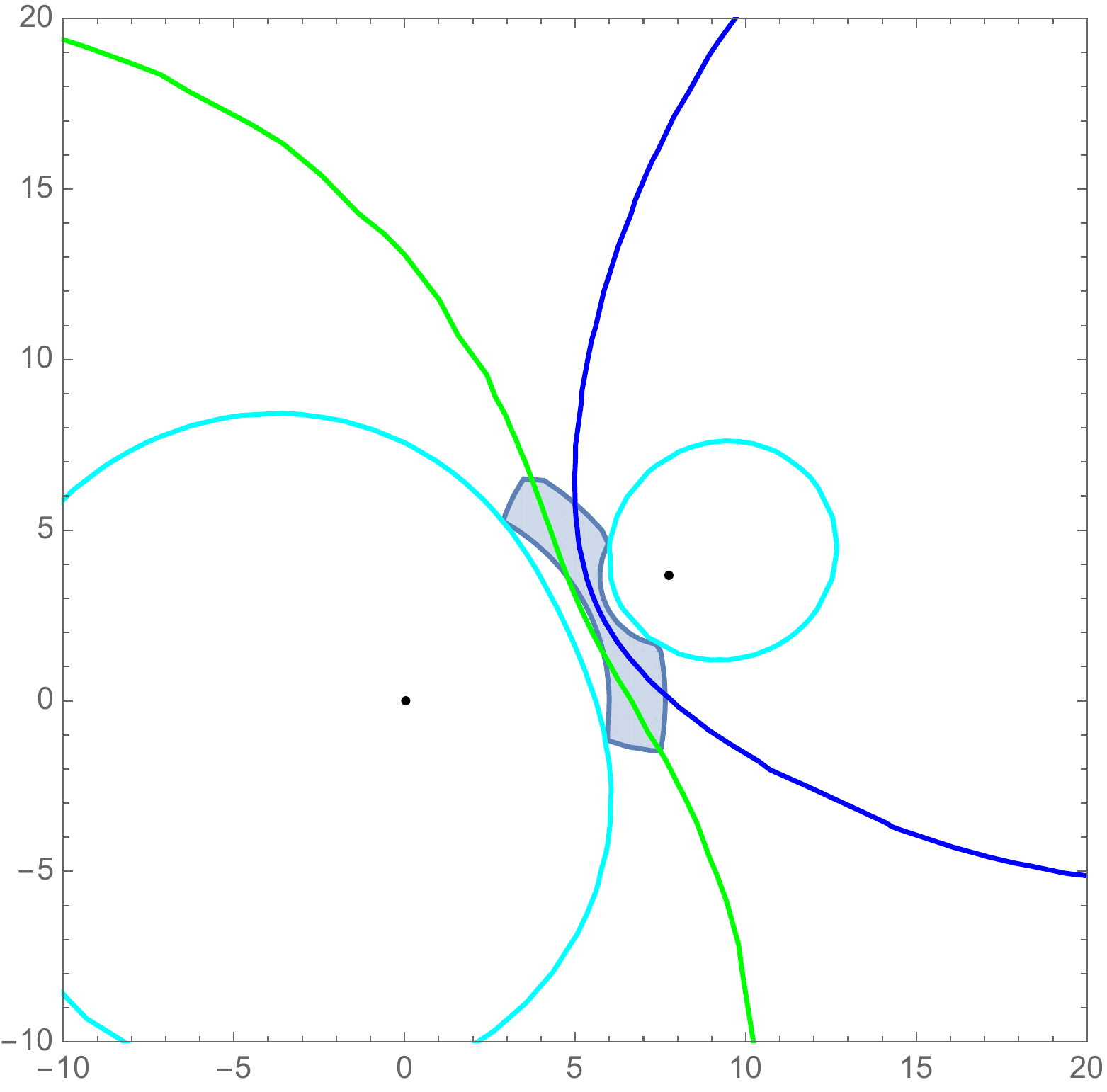}
        \caption{Conflict wave at $t=6.1$}
    \end{subfigure}
    \begin{subfigure}[t]{0.3\textwidth}
        \centering
        \includegraphics[width=\textwidth]{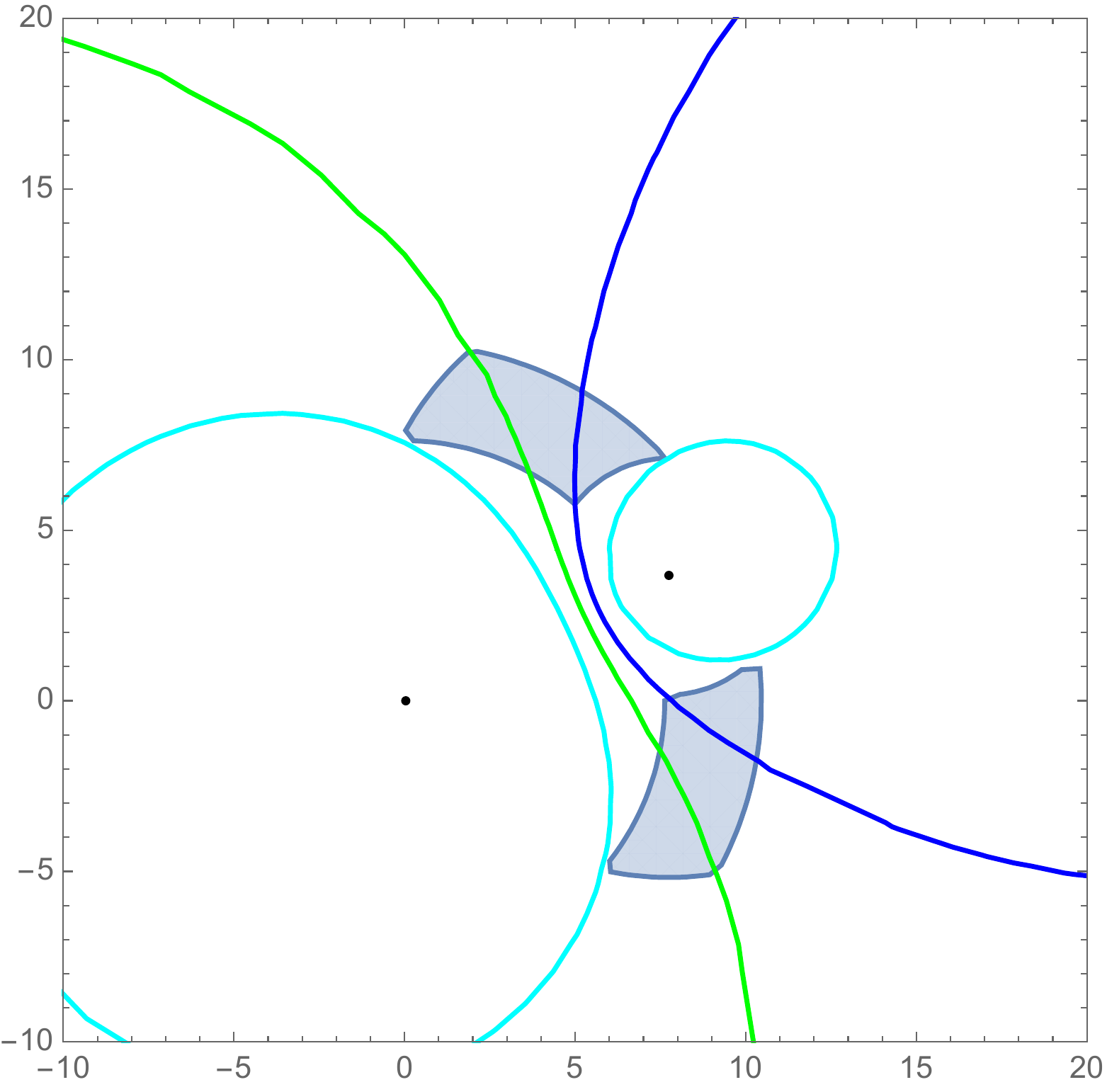}
        \caption{Conflict wave at $t=10.5$}
    \end{subfigure} \\
    \begin{subfigure}[t]{0.3\textwidth}
        \centering
        \includegraphics[width=\textwidth]{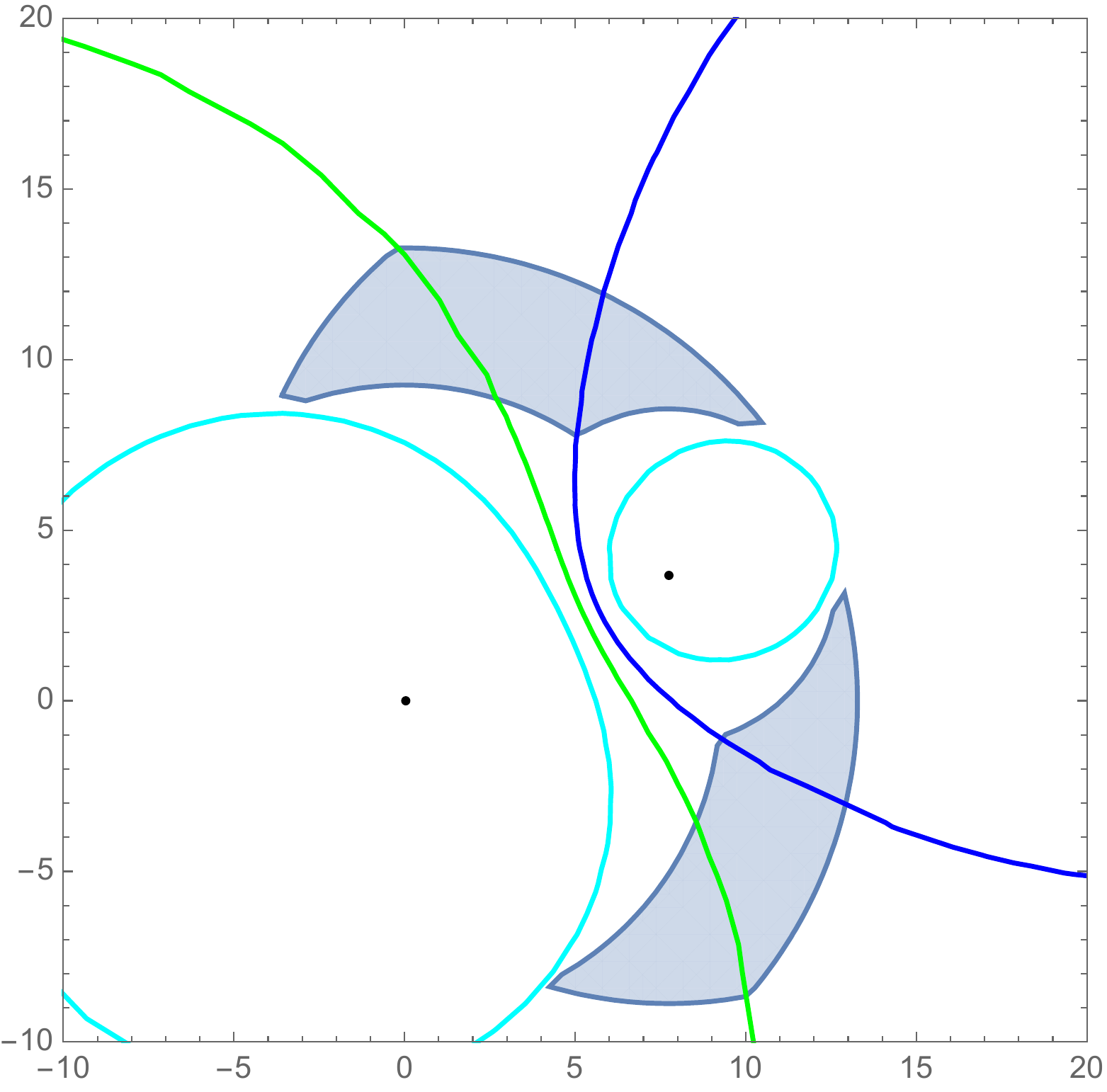}
        \caption{Conflict wave at $t=14.9$}
    \end{subfigure}
    \begin{subfigure}[t]{0.3\textwidth}
        \centering
        \includegraphics[width=\textwidth]{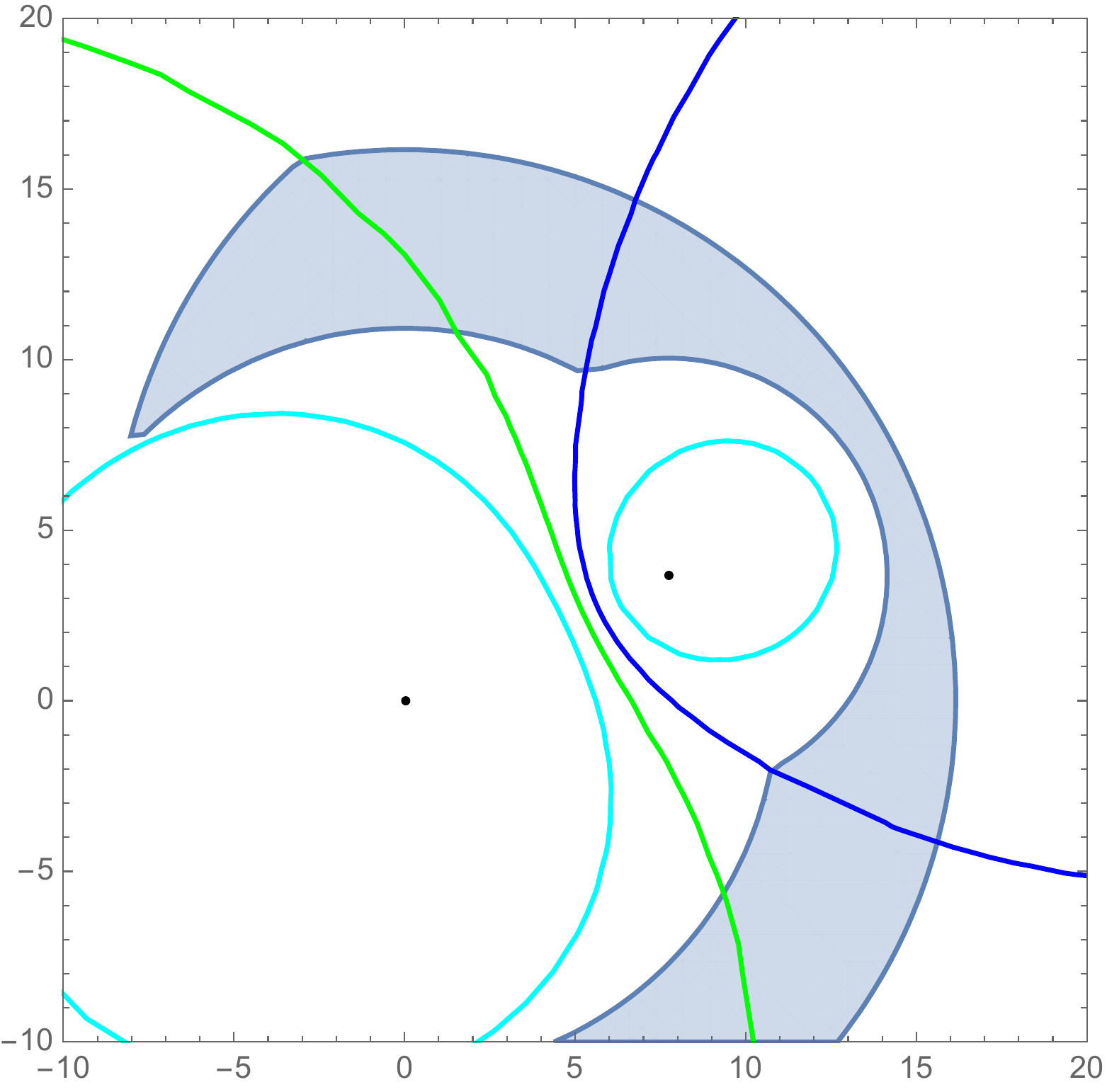}
        \caption{Conflict wave at $t=19.3$}
    \end{subfigure}
\caption{Using the example geometry, we plot areas in which the aircraft can
collide at different instants in time--the conflict wave.}
\label{fig:loci-collision-wavefront}
\end{figure}

We consider an example encounter shown in Fig. \ref{fig:loci} between
two aircraft whose positions are localized by annuli whose outer and
inner curves are expanding circular waves. The green curve is the locus of the two
leading edges; the blue curve is the locus between trailing edges; the
cyan curves are the loci of one leading and one trailing edge from
each vehicle. Points labeled with vehicle designations mark the center
of the circular waves. The regions are labeled and the ordering for
points in each region imposed by the annuli are indicated.

In regions A and E of Fig.~\ref{fig:loci},
there is no possibility of
collision; the conflict wave never enters these regions. For regions
B, C, and D, we can choose any point, and identify the earliest and
latest times that the vehicles can collide at that point by looking at the ordering, and calculating the beginning and ending of the collision timing interval using
Eqs.~\eqref{eq:te_p}--\eqref{eq:tx_p} and
Eqs.~\eqref{eq:min1}--\eqref{eq:max1}.

The intersection of the position waves for two vehicles is an area that moves over time and represents where collisions may occur at each moment; we call this the conflict wave. Fig.~\ref{fig:loci-collision-wavefront} shows the conflict wave for our example geometry at different snapshots in time by shading the area of overlap between the areas of position possibilities. Our position wave edges, and thus the loci are based on the circular edge approximations from Section 6, so the shaded conflict waves are also approximations. This approach also works if one or more of the position-wave edges is a linear boundary.

With a visual understanding of conflict waves, we can reintroduce
the convex polygonal boundaries we created for our approximations.

To compute the earliest or latest time that a collision is possible,
we do not need to evaluate every single point in a polygonal area; we
can compute earliest and latest collision times, i.e. 
Eq.~\eqref{eq:te1}--\eqref{eq:tx1}, by looking at the timing of the collision
wave with respect to a finite set of points that we call critical points.

This approach corresponds to something akin to Cylindrical Algebraic
Decomposition, which can be used to find extrema of polynomials on sets whose
boundaries are polynomial equations and inequalities. CAD can be applied
algorithmically, using techniques such as~\cite{fotiou2006}, however the order
of decomposition is often important, and it may be necessary to ensure that
the polynomial coefficients are rational. 

\subsubsection{Computing the earliest conflict time}
\label{ss:comp-te}

\begin{figure}[htbp]
    \centering
    \begin{subfigure}[t]{0.3\textwidth}
        \centering
        \includegraphics[width=\textwidth]{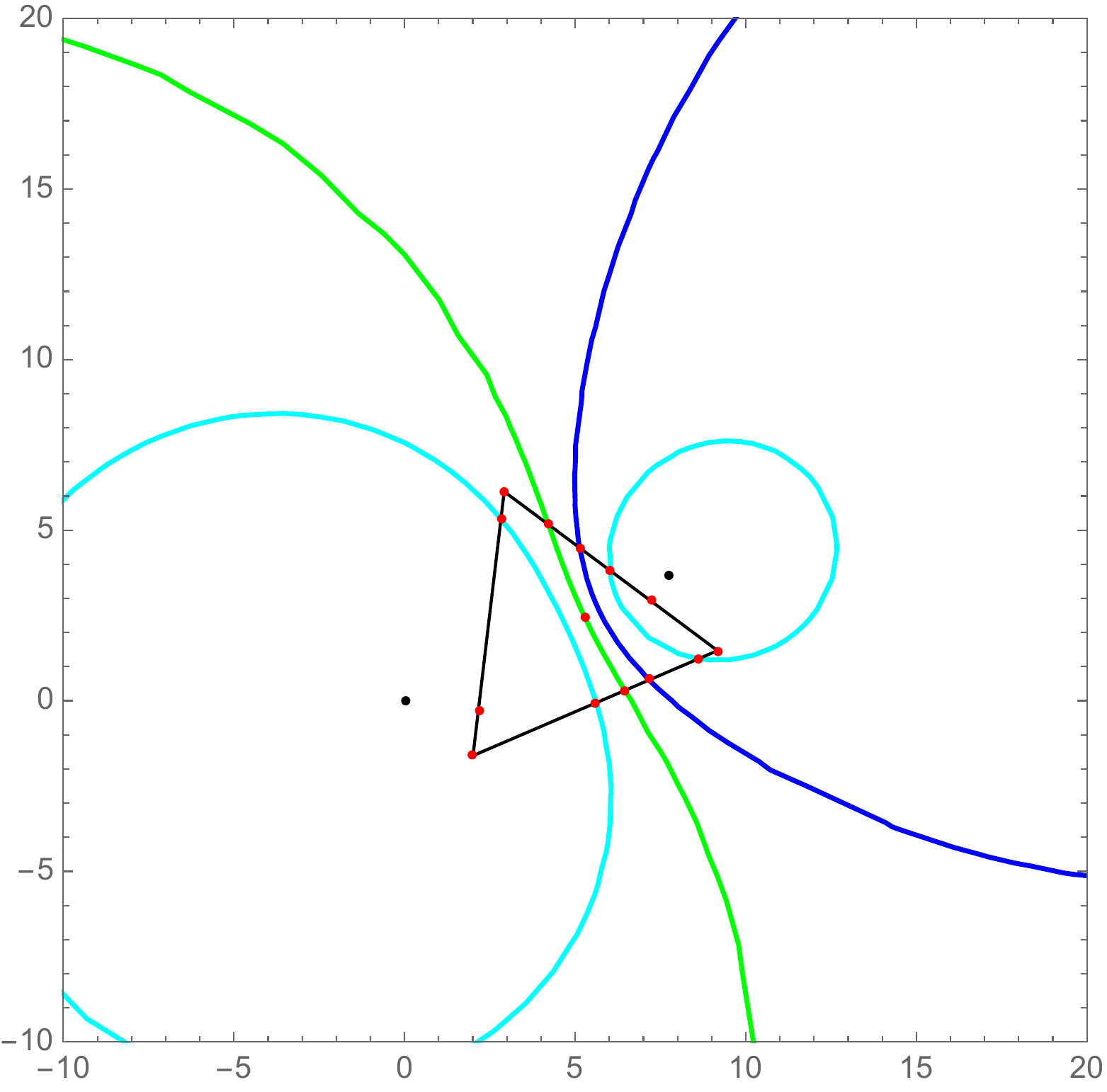}
        \caption{Conflict wave at $t=19.3$}
    \end{subfigure}
    \begin{subfigure}[t]{0.3\textwidth}
        \centering
        \includegraphics[width=\textwidth]{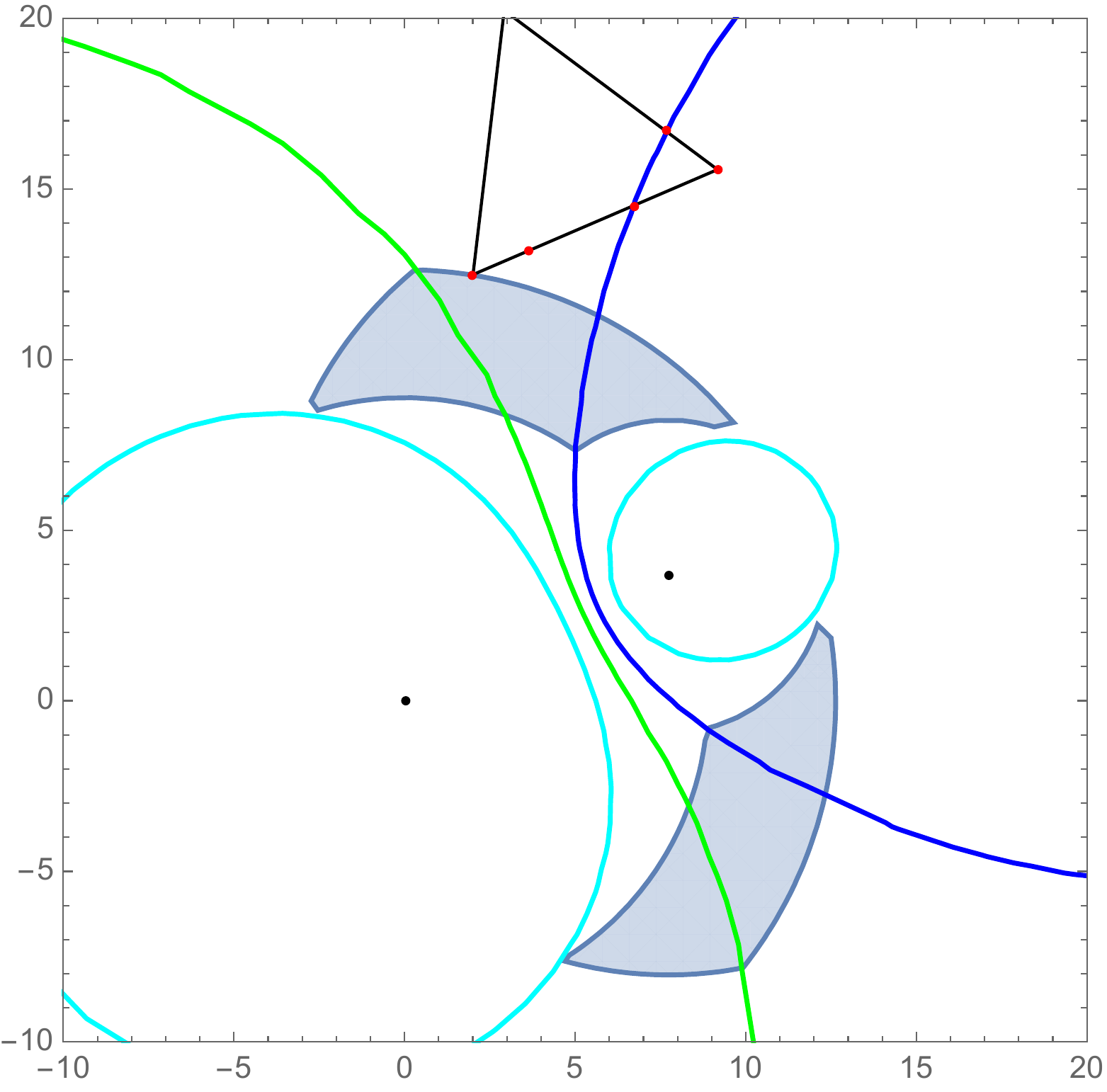}
        \caption{Conflict wave at $t=14.9$}
    \end{subfigure} \\
    \begin{subfigure}[t]{0.3\textwidth}
        \centering
        \includegraphics[width=\textwidth]{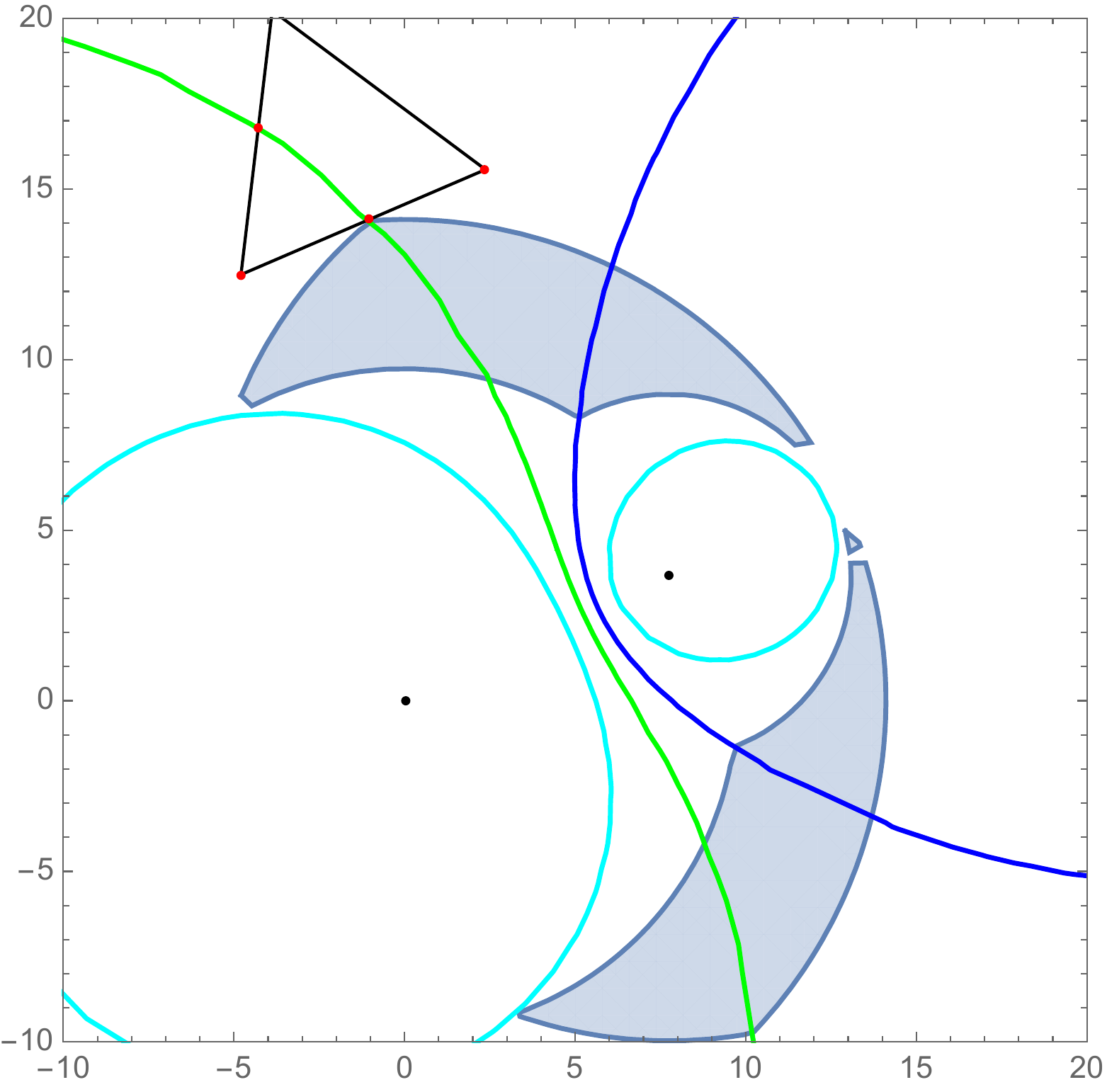}
        \caption{Conflict wave at $t=10.5$}
    \end{subfigure}
    \begin{subfigure}[t]{0.3\textwidth}
        \centering
        \includegraphics[width=\textwidth]{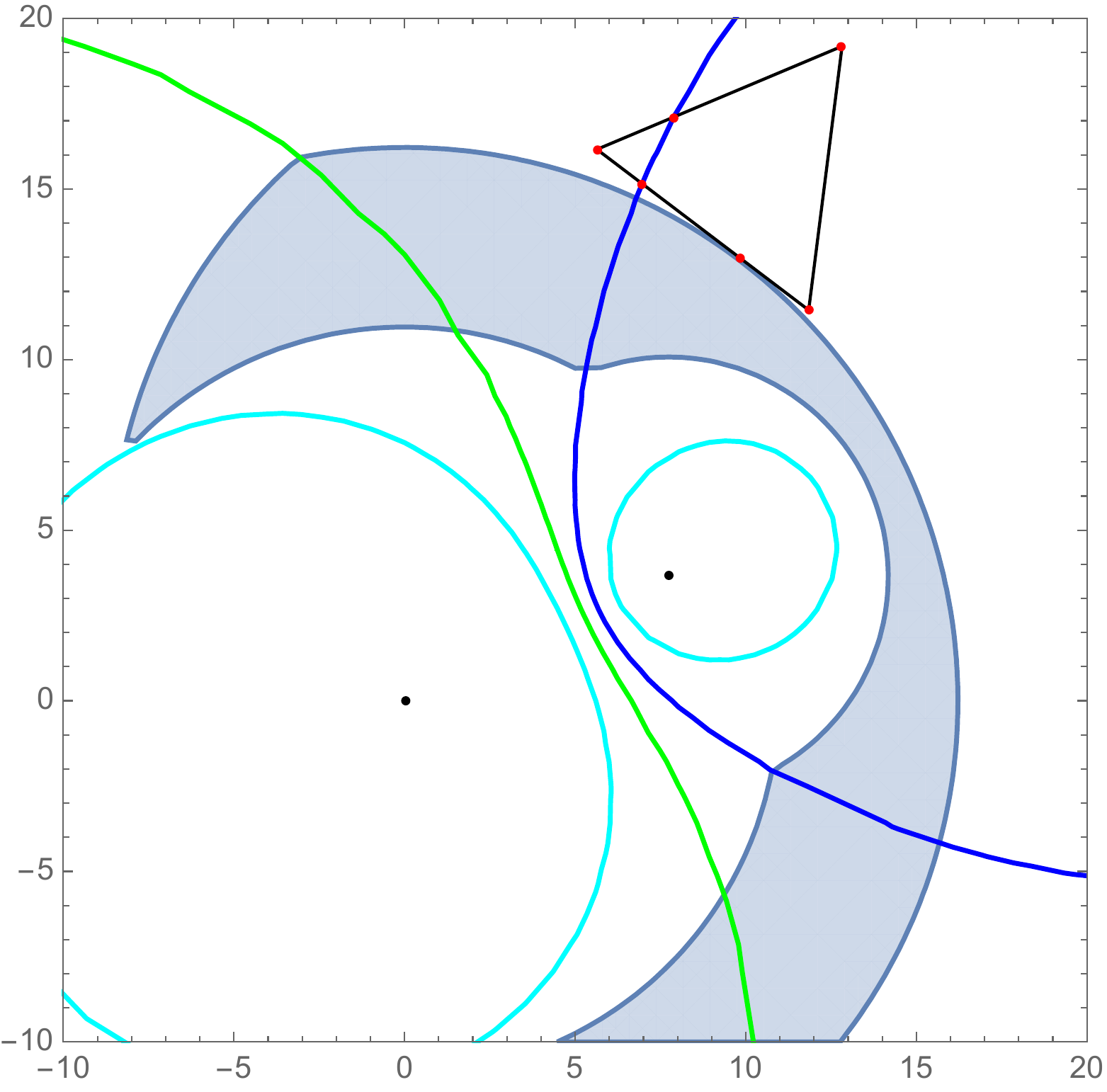}
        \caption{Conflict wave at $t=6.1$}
    \end{subfigure}
\caption{Types of critical points for computing $t_e$}
\label{fig:critical-points-te}
\end{figure}

To find the earliest possible collision time (and the location of that
potential collision) we need to evaluate when and where the collision
wave first touches the area of the polygon.
Fig. \ref{fig:critical-points-te}, uses the example geometry from the
previous section, adding a triangle to represent the convex polygon that
defines the boundary of this area, and plotting the critical points for
computing $t_e$ in red. One of these critical points will represent the
point where the earliest possible collision may occur.

Each panel illustrates a type of critical point by moving the
triangle to show a geometry and earliest moment at which that critical
point is the first one in the polygon that comes in contact with
the conflict wave.
Fig. \ref{fig:critical-points-te}a shows that there is a special case
of a critical point that is not on the edges of the polygon at the
location of the point where the leading wavefronts of each of the
vehicles first touch, if that point is in the collision
area. This is where the conflict wave first ``appears.'' 
Fig. \ref{fig:critical-points-te}b shows that the vertices of
the polygon are also critical 
points--this is a consequence of the convexity of the polygon and front
edge of the conflict wave.
Fig. \ref{fig:critical-points-te}c shows that points where the
loci intersect the edges of the polygon are
also critical points and may be the earliest locations at which 
the conflict wave comes into contact with the polygon.
Fig. \ref{fig:critical-points-te}d shows that
points on a segment of the polygon that are tangent to the
leading wavefront are also critical points.

Critical points can be identified automatically, and once this is done, evaluation of the earliest
time of collision is a matter of evaluating the earliest time of
collision for each point that contacts the conflict wave, 
and choosing the minimum time overall.

\subsubsection{Computing the latest conflict time}
\label{ss:comp-tx}

\begin{figure}[htbp]
    \centering
    \begin{subfigure}[t]{0.3\textwidth}
        \centering
        \includegraphics[width=\textwidth]{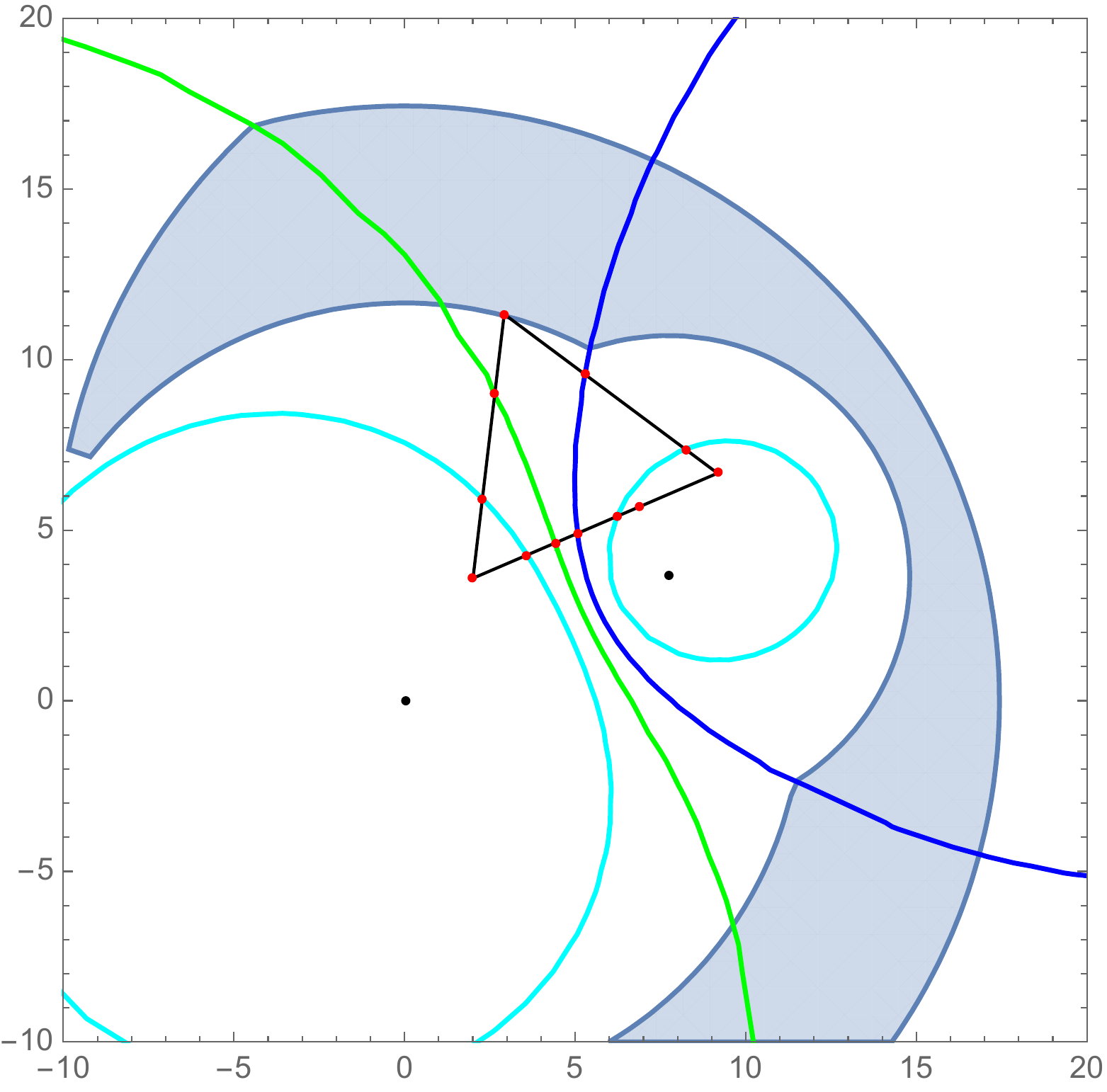}
        \caption{Conflict wave at $t=19.3$}
    \end{subfigure}
    \begin{subfigure}[t]{0.3\textwidth}
        \centering
         \includegraphics[width=\textwidth]{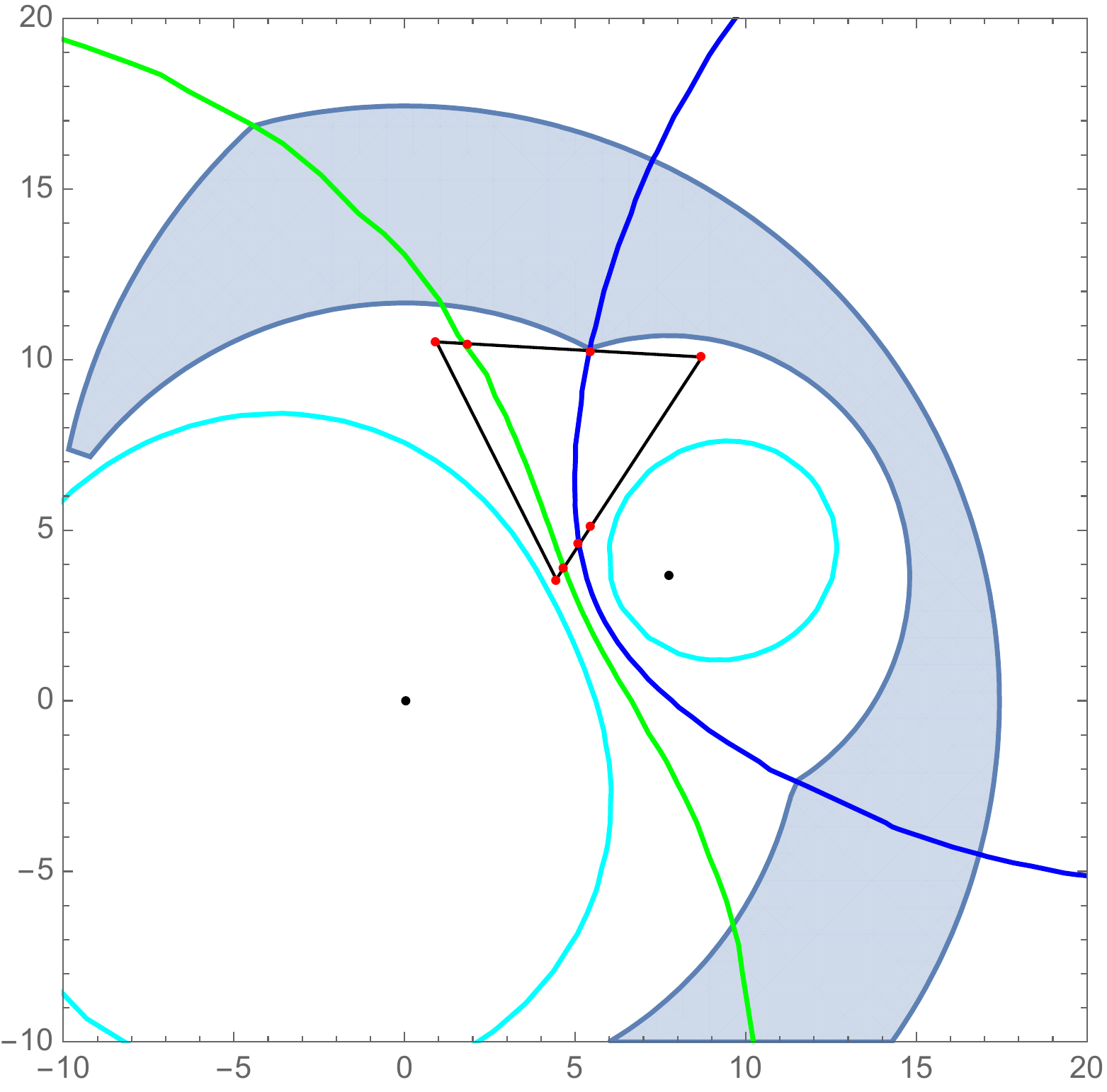}
         \caption{Conflict wave at $t=14.9$}
    \end{subfigure}
\caption{Types of critical points for computing $t_l$}
\label{fig:critical-points-tx}
\end{figure}

To find the latest possible collision time  (after which there can be no further possibility of collision) and location, we need to calculate when and where the conflict wave last touches the area of the polygon. 
Fig.~\ref{fig:critical-points-tx} uses the example geometry from the previous
section, again adding a triangle to represent the convex polygonal that defines the domain of our approximations, and plotting critical points in red. One of these critical points will be the point where the latest possible collision may occur.

As before, each subfigure illustrates a type of critical point by moving the triangle to show a geometry and timing where that critical point is the last one the polygon comes into contact with, before the conflict waves moves away entirely.
The types of critical points used in computing $t_l$ are a subset of
the types used for computing $t_e$. Fig.~\ref{fig:critical-points-tx}a
shows that the vertices of the polygon are critical points. Fig.~\ref{fig:critical-points-tx}b shows
that points where the loci intersect the line segments that make up
the polygon are also critical points. 
In some cases, the polygon that contains the approximation is open, i.e. the edges do not create a closed circuit. In these cases, the polygonal boundary has rays at the end of it extending to
infinity. We use the angles of these rays to identify a range of
angular directions by which we can approach infinity and still remain
in the polygon. We then consider the angular position of the
loci in the limit as we approach infinity, and the ordering of arrival
and departure times at each angle as we approach infinity. By
comparing these angular intervals, we can identify angular intervals
and angular points in the region of validity where collisions may or
may not occur. If a collision may not occur, then we can safely
guarantee that $t_l$ must have some finite value. If there exist
angular directions where a collision may occur, then we cannot
guarantee safety, there can be collisions that occur in unbounded time. In this case, these angles indicate that there is no upper bound to the interval in which collisions may occur; we can say the latest collision time is at infinity. 

Once all critical points are identified, evaluation of the latest collision time is a matter of evaluating the latest collision time for each point that contacts the conflict wave, and choosing the maximum time overall. If we have an unbounded region with a latest collision time that is also unbounded, then the correct latest collision time is infinity.

%% file: horiz-conclusion.tex
\section{Future Work}
There are many ways to improve on this work, some of which we have begun to explore.

The most straightforward continuation of this work is to move it closer to practical application. To this end, we have implemented the algorithm in the Julia language to compute horizontal conflict intervals and plan to synthesize a safety controller that uses it based on \cite{kouskoulas2019}. We are also developing a formalization and proofs of correctness specifically for the controller synthesis, which would allow us to extract a correct-by-construction controller implementation. We have started to experiment with the calculation in a few different simulations: to evaluate the safety of maneuvers made by autonomous boats; and also as part of a mixed vertical/horizontal collision avoidance system for aircraft, computing horizontal conflict intervals for \cite{kouskoulas2017a}.

Another enhancement would be to explore the tradeoff in our approximations between accuracy and computational efficiency. The calculations and maneuvers based on them are provably safe, but coarse approximations lead to maneuvers that are more conservative and might restrict the system unnecessarily, and tight approximations increase the computational burden. We feel that the approximations we have described strike a good balance, but different applications may require more accuracy or more computational speed from the analysis. The approach in this paper could be used to adjust the fineness/coarseness of the covering polygons and thus set the tradeoff according to requirements of a particular application.

To make the system more useful in application, we have experimented with representing position uncertainty in the vehicles -- which could represent sensor error or unexpected variations in future trajectories -- by expanding polygons to contain shapes created by convolving a circular disk with each of the position waves, and by extension, the envelopes and collision waves. This can be done with minimal additional computational effort, but needs further proof and formal verification. 

In addition, we are pursuing methods for constraining the learned policies of safe neural network controllers using the safety predicates from this work, such as in \cite{geninNSV}. The safety predicates developed here may be used to guide the training of such networks, verify the correctness of the network policies, and could someday be used directly in the optimization of such neural controllers. 

While the model developed in this work applies only to turn-to-bearing kinematics, due to the non-determinism that we incorporate there is a family of trajectories that are also encompassed by these proofs. The extension representing position uncertainty allows additional flexibility. Future research could include characterizing this family of trajectories.

\section{Conclusion}

In this work, we have created a formally verified library that describes
uncertain turn-to-bearing kinematics and allows us to reason about the timing
of such maneuvers without approximation. The representation allows
non-determinism in all turn parameters by quantifying over state variables.

We have applied the library to compute timing intervals during which the 
intersecting turns of two vehicles might collide. These timing computations can
be used to determine whether two aircraft will ever travel close enough to each
other (under the range of assumed kinematics) to be in horizontal conflict, and,
if so, what the earliest and latest times of the horizontal conflict can be. By
combining horizontal conflict timing with reasoning about the vertical
separation of aircraft, we can ensure that the aircraft are not simultaneously
in horizontal and vertical conflict and guarantee the absence of collisions. 
To find the horizontal conflict time range, we first developed expressions of
time intervals without approximation, for a given point 
accounting for non-deterministic horizontal maneuvers for each aircraft. 
We then applied the library to create approximations of the position waves
that are useful for calculating the intersection between two position waves in
subregions of the envelope in which collisions may occur. 
Finally, we showed a method for tiling and fitting sound polygonal
approximations of each subregion, resulting in computationally efficient
methods for solving for the earliest and latest horizontal conflict times.